\documentclass[prd,nofootinbib,showpacs,twocolumn]{revtex4-1}
\usepackage{color}
\usepackage{bm}
\usepackage{graphicx}
\usepackage{amsmath}
\usepackage{amssymb}
\usepackage{enumitem}
\usepackage{hyperref}
\usepackage{subfigure}
\usepackage{array}
\usepackage{multirow}
\usepackage[english]{babel}
\usepackage{tensor}
\usepackage{comment}
\usepackage[normalem]{ulem}
\usepackage[dvipsnames]{xcolor}

\def\be{\begin{equation}}
\def\ee{\end{equation}}
\def\ba{\begin{eqnarray}}
\def\ea{\end{eqnarray}}
\def\l{\left}
\def\r{\right}
\def\f{\frac}

\allowdisplaybreaks

\def\nn{\nonumber}

\def\l{\left}
\def\r{\right}
\def\f{\frac}

\def\nn{\nonumber}

\def\be{\begin{equation}}
\def\ee{\end{equation}}
\def\ba{\begin{eqnarray}}
\def\ea{\end{eqnarray}}
\def\gn{G_\mathrm{N}}

\def\ak{\alpha_K}
\def\ab{\alpha_B}

\def\mnras{MNRAS}

\begin{document}

\title{Spherical collapse and halo abundance in shift-symmetric Galileon theory}

\author{In\^es S.~Albuquerque$^1$, Noemi Frusciante$^2$, Francesco Pace$^{3,4,5}$, Carlo Schimd$^6$}
\affiliation{
\smallskip
$^1$ Instituto de Astrofis\'ica e Ci\^{e}ncias do Espa\c{c}o, Faculdade de Ci\^{e}ncias da Universidade de Lisboa, Edificio C8, Campo Grande, P-1749016, Lisboa, Portugal\\ 
$^2$ Dipartimento di Fisica ``E. Pancini", Universit\`a degli Studi  di Napoli  ``Federico II", Compl. Univ. di Monte S. Angelo, Edificio G, Via Cinthia, I-80126, Napoli, Italy\\
$^3$ Dipartimento di Fisica, Universit\`a degli Studi di Torino, Via P. Giuria 1, I-10125 Torino, Italy\\
$^4$ INFN-Sezione di Torino, Via P. Giuria 1, I-10125 Torino, Italy\\
$^5$ INAF-Istituto Nazionale di Astrofisica, Osservatorio Astrofisico di Torino, strada Osservatorio 20, 10025, Pino Torinese, Italy\\
$^6$ Aix Marseille Univ, CNRS, CNES, LAM, IPhU, Marseille, France}

\begin{abstract}
We present the nonlinear growth of bound cosmological structures using the spherical collapse approach in the shift-symmetric Galileon theories. In particular, we focus on the class of models belonging to the Kinetic Gravity Braiding by adopting a general parametrization of the action encoding a large set of models by means of four free parameters: two defining the background evolution and two affecting the perturbations. For the latter we identify their specific signatures on the linearised critical density contrast, nonlinear effective gravitational coupling and the virial overdensity and how they drive their predictions away from $\Lambda$CDM. We then use the results of the spherical collapse model to predict the evolution of the halo mass function. We find that the shift-symmetric model predicts a larger number of objects compared to $\Lambda$CDM for masses $M \gtrsim 10^{14} h^{-1} \mathrm{M}_\odot$ and such number increases for larger deviations from the standard model. Therefore, the shift-symmetric model shows detectable signatures which can be used to distinguish it from the standard scenario.
\end{abstract}

\date{\today}

\maketitle

\section{Introduction}

Since the first observational evidences of the late-time accelerated expansion of the Universe, an enormous effort has been put forward towards its theoretical modelling. The current standard cosmological model, dubbed $\Lambda$-Cold-Dark-Matter ($\Lambda$CDM), is the simplest model to account for the cosmic acceleration in the framework of General Relativity (GR). This is realized through the introduction of the cosmological constant, $\Lambda$, to describe a negative-pressure energy component boosting the acceleration. However, the $\Lambda$CDM model is plagued by some theoretical problems \cite{Weinberg:1988cp,Carroll:2000fy,Velten:2014nra,Joyce:2014kja} as well as some mild observational tensions \cite{Riess:2019cxk,Wong:2019kwg,Freedman:2019jwv,DiValentino:2020zio,Kuijken:2015vca,deJong:2015wca,Hildebrandt:2016iqg,DiValentino:2020vvd}, which might hint to new physics. One appealing proposal is to modify the gravitational interactions on large cosmological scales, e.g., through the introduction of new degrees-of-freedom \cite{Joyce:2014kja,Lue:2004rj,Copeland:2006wr,Silvestri:2009hh,Nojiri:2010wj,Tsujikawa:2010zza,Capozziello:2011et,Clifton:2011jh,Kobayashi:2019hrl,Frusciante:2019xia,CANTATA:2021ktz}. This is the realm of modified theories of gravity (MG) \cite{CANTATA:2021ktz}.

Among the many proposals of MG, scalar-tensor theories \textit{à la} Galileon \cite{Horndeski:1974wa,Nicolis:2008in,Deffayet:2009wt,Deffayet:2009mn,Kobayashi:2011nu} are characterized by a modified gravitational interaction through the introduction of a scalar field $\phi$. Their attractiveness results from their capability to account for self-accelerating solutions adopted for both primordial inflation
\cite{Burrage:2010cu,Creminelli:2010qf,Renaux-Petel:2011rmu,Kamada:2010qe,Kobayashi:2011pc,Frusciante:2013haa} and late-time cosmic acceleration \cite{Deffayet:2010qz,Kobayashi:2010cm,Nesseris:2010pc,Charmousis:2011bf,Barreira:2013xea,Barreira:2014jha,Renk:2017rzu,Peirone:2017vcq,Kase:2018iwp,Frusciante:2018aew,Albuquerque:2018ymr,Frusciante:2019puu,Peirone:2019aua,Albuquerque:2021grl}. Additionally, the generality of the Galileon theory relies on an action that features four free functions of the scalar field $\phi$ and its kinetic energy $X\equiv \partial^\mu\phi\partial_\mu\phi$, which allow the design of different models. Interestingly, the observation of the gravitational wave (GW) event GW170817 together with its electromagnetic counterpart GRB170817A has set a tight constraint on the speed of propagation of GWs \cite{LIGOScientific:2017zic}, restricting the form of the Galileon action \cite{Creminelli:2017sry,Baker:2017hug,Ezquiaga:2017ekz}.  

A class of surviving Galileon models, known as Kinetic Gravity Braiding (KGB) \cite{Deffayet:2010qz}, is characterised by a specific combination of the scalar and metric fields that arises from the 
term $G_3(\phi, X) \Box \phi$ in the action. In KGB models the scalar field can exhibit a phantom behaviour and cross the so-called phantom divide without the appearance of ghosts or gradient instabilities \cite{Deffayet:2010qz,Kase:2018iwp}. Moreover, KGB models admit scaling solutions \cite{Gomes:2013ema,Frusciante:2018aew,Albuquerque:2018ymr}. Among KGB models, we will be interested in studying shift-symmetric Galileon models, which satisfy the shift symmetry $\phi\rightarrow \phi+\tilde{c}$, where $\tilde{c}$ is a constant \cite{Deffayet:2010qz}. For these models, cosmological perturbations have been explored at both linear \cite{Nesseris:2010pc,Barreira:2013xea,Renk:2017rzu,Peirone:2017vcq,Frusciante:2019puu,Peirone:2019aua,Albuquerque:2021grl} and nonlinear scales \cite{Kimura:2010di,Kimura:2011td,Bellini:2012qn,Barreira:2013eea,Frusciante:2020zfs}, showing detectable signatures to be used to distinguish shift-symmetric models from the standard $\Lambda$CDM scenario. In particular, on smaller nonlinear scales Galileon theories show a \textit{screening mechanism}, known as Vainshtein mechanism \cite{Vainshtein:1972sx,Kimura:2011dc,Babichev:2013usa,Koyama:2013paa}, able to suppress any modification to the gravitational interaction in high-density environments, acting through the second derivative of the scalar field. The KGB models can therefore accommodate the Solar System constraints, according to which no new degrees of freedom have been detected so far, proving the exquisite validity of GR on these scales \cite{Baessler:1999iv,Will:2005va,Uzan:2010pm}. It is then clear how such mechanism is crucial in the study of the formation of gravitationally bound structures, especially on extragalactic scales. 

The spherical collapse model \cite{Gunn:1972sv} provides the simplest semi-analytical framework to follow the formation of nonlinear gravitationally bounded structures, supposed to be spherically symmetric. This technique has already been applied to some shift-symmetric Galileon models \cite{Kimura:2010di,Bellini:2012qn,Barreira:2013eea,Frusciante:2020zfs}. To mention a few, the investigation of the covariant Galileon \cite{Bellini:2012qn} and Galileon Ghost Condensate \cite{Frusciante:2020zfs} models showed that the linearised critical density contrast, the nonlinear effective gravitational coupling and the virial overdensity have peculiar features with respect to the standard $\Lambda$CDM model, non-trivially affecting the phenomenology of the nonlinear matter and lensing power spectra. As a concrete application, the theoretical predictions of the spherical collapse method in MG scenarios are one of the ingredients employed in software such as \texttt{ReACT} \cite{Cataneo:2018cic}, adopting analytical methods to compute the nonlinear power spectra. These solutions are required to exploit data from upcoming weak lensing and galaxy clustering surveys \cite{LSST:2008ijt,DESI:2016fyo,EUCLID:2011zbd,Weltman:2018zrl}.

The studies cited above adopted specific functional forms of the free functions that define  the shift-symmetric Galileon action. In this paper we investigate this class of models in a more general way, by employing the model-independent framework of the Effective Field Theory (EFT) of dark energy (DE) and MG \cite{Gubitosi:2012hu,Bloomfield:2012ff} (see \cite{Frusciante:2019xia} for a review), which encompasses the theories of gravity with a single extra scalar degree-of-freedom by constructing the action with a set of time-dependent functions. Firstly, we develop the spherical collapse approach in terms of these general functions; then, aiming at exploring some theoretical predictions, we adopt a simple but accurate parameterization of these functions that fit more than 98\% of randomly generated shift-symmetric models \cite{Traykova:2021hbr}. This way we gauge the general class of shift-symmetric models in terms of few free functions of time. We are then able to make predictions and interpret the observations directly in the space of this class of models and not within a single specific paradigm. The results presented in this study are therefore expected to be of large applicability for the construction of data-analysis pipelines of upcoming and future data.

The paper is organized as follows. In Sec.~\ref{Sec:II} we detail the equations describing the evolution of linear and nonlinear perturbations in shift-symmetric Galileon models. The spherical collapse model is reviewed in Sec.~\ref{Sec:III}, while its specific application to shift-symmetric Galileon models and the analysis of its results is shown in Sec.~\ref{sec:IV}. In Sec.~\ref{Sec:V} we compute theoretical predictions for the halo mass function and discuss the impact of the shift-symmetric Galileon models on the abundance of halos. Finally, we conclude in Sec.~\ref{Sec:VI}. We adopt units such that $c=1$, and indicate by $\partial_\mu$ and $\nabla_\mu$ the partial and covariant derivatives and by $\Box\equiv\nabla^\mu\nabla_\mu$ the d'Alambert operator.

\section{Perturbations in Shift-Symmetric Galileon theory} \label{Sec:II}

\subsection{Action and field equations}\label{Sec:IIA}

Shift-symmetric Galileon models are a sub-class of Galileon theories invariant under the linear transformation of the scalar field, $\phi \rightarrow \phi + \tilde{c}$, for some constant $\tilde{c}$ \cite{Deffayet:2010qz}. For the particular case in which shift-symmetric theories satisfy also the GWs constraint on the speed of propagation of tensor modes, i.e., $c_{\rm T}^2=1$ \cite{LIGOScientific:2017vwq}, their general action can be written as \cite{Deffayet:2010qz,Kobayashi:2010cm}
\be\label{eq:action}
S=\int{}\mathrm{d}^4x\sqrt{-g}\l[ G_2(X)+G_3(X)\Box\phi+\f{M_{\rm Pl}^2}{2}R+L_{\rm M}\r]\,,
\ee
where $g$ is the determinant of the metric $g_{\mu\nu}$, $M_{\rm Pl} = (8 \pi G_{\rm N})^{-1/2}$ is the Planck mass with $G_{\rm N}$ the Newtonian gravitational constant, $R$ is the Ricci scalar, $G_{2}$ and $G_{3}$ are free functions of $X\equiv \partial^\mu\phi\partial_\mu\phi$, and $L_{\rm M}$ is the Lagrangian of matter and radiation fields. 

From the above action, one can obtain the field equations which read \cite{Kobayashi:2011nu}
\begin{align}
    &G_{2X} \nabla_{\mu}\phi \nabla_{\nu}\phi - \frac{1}{2} G_2 g_{\mu \nu} + G_{3X} \Box \phi \nabla_{\mu} \phi \nabla_{\nu} \phi \nonumber \\ 
    &- \nabla_{( \mu} G_3 \nabla_{\nu )} \phi + \frac{1}{2} g_{\mu \nu} \nabla_{\lambda} G_3 \nabla^{\lambda} \phi + \frac{M^2_{\rm Pl}}{2} G_{\mu \nu} = T_{\mu\nu}\,, \label{eq:eq1}\\
    & \nabla^{\mu} \left[ 2 G_{2X} \nabla_{\mu} \phi + 2 G_{3X} \Box \phi \nabla_{\mu} \phi - G_{3X} \nabla_{\mu} X \right] = 0\,, \label{eq:eq2}
\end{align}
where $G_{iX} \equiv \partial G_i / \partial X$, $G_{\mu \nu}$ is the Einstein tensor and $T_{\mu\nu}$ is the stress energy tensor. The system is closed by the continuity equation
\begin{equation} \label{eq:continuityequation}
 \nabla^\mu T_{\mu\nu}=0\,.
\end{equation}
We consider barotropic fluids with pressure $p$ and density $\rho$ related by $p=w\rho$, with $w=0$ for baryons, cold dark matter, or any non-relativistic component, and $w=1/3$ for radiation or any relativistic component.

\subsection{Linear perturbations}

The evolution of perturbations is studied adopting a perturbed Friedmann-Lema\^{i}tre-Robertson-Walker (FLRW) metric in the Newtonian gauge,
\begin{equation}
 \mathrm{d}s^2 = -(1+2\Psi)\mathrm{d}t^2 + a^2(t)(1-2\Phi)\delta_{ij}\mathrm{d}x^i\mathrm{d}x^j\,,
\end{equation}
where $a(t)$ is the scale factor and $\Phi(t,x^i)$ and $\Psi(t,x^i)$ are the two gravitational potentials, the dependence on the cosmic time $t$ and comoving position $x^i$ being omitted for simplicity. 

At linear order and in Fourier space, the field equations yield the modified Poisson and lensing equations \cite{Bean:2010zq,Silvestri:2013ne,Pogosian:2010tj}, which in the absence of matter anisotropic stress read
\begin{eqnarray}
 -k^2\Psi & = & 4\pi G_{\rm N} a^2\mu^{\rm L}(a,k) \Bar{\rho}_{\rm m}\delta_{\rm m}\,, \label{mudef} \\
 -k^2(\Psi+\Phi) & = & 8\pi G_{\rm N} a^2\Sigma^{\rm L}(a,k) \Bar{\rho}_{\rm m} \delta_{\rm m}\,,\label{sigmadef}
\end{eqnarray}
Here the same symbols are used for the Fourier transforms of $\Psi$ and $\Phi$, which are functions of the comoving number $k$ and time (omitted for clarity). The matter density perturbation $\delta_{\rm m}\equiv \delta\rho_{\rm m}/\Bar{\rho}_{\rm m}$ is relative to the background matter density $\Bar{\rho}_{\rm m}$, and $\mu^{\rm L}$ and $\Sigma^{\rm L}$ are, respectively, the linear effective gravitational coupling and the light deflection parameter. In the GR limit $\mu^{\rm L} = \Sigma^{\rm L} = 1$ and $\Phi=\Psi$. For generic MG the two potentials differ and are routinely considered as proportional,
\begin{equation}
 \Phi = \eta^\mathrm{L}\Psi\,.
\end{equation}
Equations~\eqref{mudef}-\eqref{sigmadef} yield
\begin{equation}
 \Sigma^{\rm L} = \frac{\mu^{L}}{2}(1+\eta^{\rm L})\,.
\end{equation}
For values of $\mu^{\rm L}>1$ ($\mu^{\rm L}<1$), gravitational interactions are stronger (weaker) than in GR. Equations \eqref{mudef} and \eqref{sigmadef} are valid for the general class of MG models. However, the quasi-static approximation (QSA) is necessary to obtain analytical expressions for $\mu^{\rm L}$ and $\Sigma^{\rm L}$ in a specific theory \cite{Boisseau:2000pr,DeFelice:2011hq}. In QSA, the time derivatives of the perturbed quantities can be neglected when compared with their spatial derivatives.

For Galileon models, the QSA for perturbations inside the sound horizon of scalar field fluctuations is a valid assumption for $k\gtrsim 10^{-3}h~\mathrm{Mpc}^{-1}$ \cite{Sawicki:2015zya,Pogosian:2016pwr,Frusciante:2019xia}. Applied to Eqs.~\eqref{eq:eq1}-\eqref{eq:eq2} it gives
\begin{eqnarray}\label{eq:musigmaLgalileon}
  \mu^{\rm L} = 1+\dfrac{2\ab^2}{\alpha c_\mathrm{s}^2}\,, \qquad \Sigma^{\rm L}=\mu^{\rm L}\,,
\end{eqnarray}
where $\alpha \equiv \ak +6 \ab^2$ is written in terms of the kineticity and the braiding function, respectively defined by \cite{Bellini:2014fua}
\begin{eqnarray}
H^2 M_{\rm Pl}^2 \ak &=& 2X \left( G_{2X} + 2XG_{2XX} \right)  \nonumber \\
&&\; - 12 \dot{\phi} X H \left( G_{3X} + XG_{3XX} \right)\,, \label{eq:aK} \\
H M_{\rm Pl}^2 \ab &=& \dot{\phi} X G_{3X} \,, \label{eq:ab}
\end{eqnarray}
and where the speed-of-sound of scalar modes is given by
\begin{equation}\label{eqn:cs2}
 \alpha c_\mathrm{s}^2 =
 -2(1+\alpha_{\rm B})\l(\frac{H^\prime}{H}+\ab\r)-2\ab^{\prime}-3\Omega_{\rm m}.
\end{equation}
Here the prime denotes the derivative with respect to the $e$-fold time $\ln{a}$, $\Omega_{\rm m}\equiv\rho_{\rm m}/(3M_{\rm Pl}^2H^2)$ is the matter density parameter and $H=\dot{a}/a$ is the Hubble function with a dot representing a time derivative. Equations~\eqref{eq:aK}-\eqref{eq:ab} fully define the so-called Kinetic Gravity Braiding (KGB) models \cite{Deffayet:2010qz}. Note that $\alpha$ and $c_\mathrm{s}^2$ must be positive to avoid, respectively, ghost and gradient instabilities.

From these requirements it follows that $\mu^{\rm L}>1$, i.e., KGB models are characterized by stronger gravitational interactions than GR. Moreover, according to Eq.~\eqref{eq:musigmaLgalileon} one obtains $\Phi=\Psi$, i.e., there is no anisotropic stress.
 
The evolution of the matter density perturbations follows from Eq.~\eqref{eq:continuityequation}. On linear scales, it yields
\begin{equation}
 \delta_{\rm m}^{\prime\prime} + \l(2+\f{H^\prime}{H}\r)\delta_{\rm m}^{\prime} - \f{\nabla^2\Psi}{a^2H^2} = 0\,,
\end{equation}
which combined with the Poisson equation gives
\be\label{eqn:linpert}
\delta_{\rm m}^{\prime\prime} + \l(2+\f{H^\prime}{H}\r) \delta_{\rm m}^\prime - \f{3}{2}\Omega_{\rm m} \mu^{\rm L}(a) \delta_{\rm m}=0\,.
\ee

Eqs.~\eqref{mudef}, \eqref{sigmadef} and \eqref{eqn:linpert} constitute the full system of equations that define the linear evolution of gravitational and matter fields.

\subsection{Nonlinear perturbations}

On small scales, the evolution of perturbations becomes nonlinear. Assuming the validity of the QSA, the equations presented in Sec.~\ref{Sec:IIA} are
\begin{align}
&\nabla^2\Psi=\frac{ \Bar{\rho}_{\rm m} \delta_{\rm m}}{2M_{\rm Pl}^2} +\ab H\nabla^2\chi \,, \label{PoissonNL}\\
&\nabla^2\chi+\lambda^2\left[(\nabla_i\nabla_j\chi)^2 - \left(\nabla^2\chi\right)^2\right]=-\frac{\lambda^2}{2M_{\rm Pl}^2} \Bar{\rho}_{\rm m}\delta_{\rm m }\,, \label{secondfield}
\end{align}
where $\chi\equiv\delta\phi/\dot{\phi}$ is an auxiliary field proportional to the scalar field perturbations $\delta \phi$, $(\nabla_i\nabla_j\chi)^2$ indicates the norm of $\nabla_i\nabla_j\chi$, and
\be
\lambda^2\equiv-\f{2\ab}{H\alpha c_\mathrm{s}^2}\,.
\ee
On nonlinear scales the absence of anisotropic stress is still valid, hence $\Phi=\Psi$.

This system of equations is solved in spherical symmetry. Introducing the mass enclosed in a sphere of density $\Bar{\rho}_{\rm m}\delta_{\rm m}$ and radius $r$ as
\be
 m(r) = 4\pi \int_0^r {r^{\prime}}^2 \Bar{\rho}_{\rm m} \delta_{\rm m}(r^\prime)\mathrm{d}r^\prime\,,
\ee
a first integration of Eq.~\eqref{secondfield} gives
\be
r^2\f{\mathrm{d}\chi}{\mathrm{d}r} - 2\lambda^2r\l(\f{\mathrm{d}\chi}{\mathrm{d}r}\r)^2 = -\lambda^2G_{\rm N}m(r)\,,
\ee
with algebraic solution
\be\label{fieldsol}
\f{\mathrm{d}\chi}{\mathrm{d}r} = \f{r}{4\lambda^2}
\l(1-\sqrt{1+\f{r_{\rm V}^3}{r^3}}\r)\,,
\ee
written in terms of the Vainshtein radius of the enclosed mass perturbation,
\be
 r_{\rm V} = [8\lambda^4G_{\rm N} m(r)]^{1/3}\,. \label{eq:vainsrad}
\ee

For a spherical halo with total mass $M$ and constant density up to a scale $R$, the Vainshtein radius increases linearly with $r$ and $\mathrm{d}\chi/\mathrm{d}r$ as well. Equation~\eqref{secondfield} then simplifies to
\be
\nabla^2\chi = 8\pi G_{\rm N} a^2\Bar{\rho}_{\rm m}\lambda^2
\l(\f{R}{R_{\rm V}}\r)^3\l[1-\sqrt{1+\f{R_{\rm V}^3}{R^3}}\r]\delta_{\rm m}\,,
\ee
with $R_{\rm V}^3=8\lambda^4G_{\rm N}M$. Accordingly, Eq.~\eqref{PoissonNL} becomes
\be\label{nonlinearpoisson}
\nabla^2 \Psi = 4\pi G_{\rm N} \mu^{\rm NL}(a,R)a^2
\Bar{\rho}_{\rm m}\delta_{\rm m}\,,
\ee
with the nonlinear effective gravitational coupling defined by
\begin{equation}\label{muNL}
\mu^{\rm NL}=1 + 2\l(\mu^{\rm L}-1\r)\l(\f{R}{R_{\rm V}}\r)^3 \l(\sqrt{1+\f{R_{\rm V}^3}{R^3}}-1\r)\,.
\end{equation}

GR is recovered in the limit $R\rightarrow 0$, while for $R\gg R_{\rm V}$ one obtains $\mu^{\rm NL}\rightarrow \mu^{\rm L}$ as expected. Note also that, since in the nonlinear regime the relation $\Phi=\Psi$ is still valid, $\Sigma^{\rm NL}=\mu^{\rm NL}$. Finally, it is worth to stress that the nonlinear interaction described by Eq.~\eqref{muNL} is general to any KGB model.

\section{Spherical collapse model in KGB models} \label{Sec:III}

The nonlinear evolution of the scalar field described in the previous section is crucial in the process of formation of bounded structures. In this section we consider the spherical collapse model \cite{Weinberg:2002rd}, which captures the main MG effects on bounded structures in the weak-field limit, affecting their abundance in a Press-Schechter like approach.

As routinely done in this kind of calculations \cite{Gunn:1972sv}, one considers the nonlinear continuity equation \eqref{eq:continuityequation} around a flat FLRW background and obtains the nonlinear evolution equation for the matter density,
\be\label{nonlinearoverdensityeq}
\ddot{\delta}_{\rm m} + 2H\dot{\delta}_{\rm m} - \f{4}{3}\f{\dot{\delta}_{\rm m}^2}{1+\delta_{\rm m}} = \l(1+\delta_{\rm m}\r)\f{\nabla^2\Psi}{a^2}\,,
\ee
which, combined with Eq.~\eqref{nonlinearpoisson} yields
\be
\ddot{\delta}_{\rm m} + 2H\dot{\delta}_{\rm m} - \f{4}{3}\f{\dot{\delta}_{\rm m}^2}{1+\delta_{\rm m}} = 4\pi G_{\rm N} \mu^{\rm NL}\,
\Bar{\rho}_{\rm m}\l(1+\delta_{\rm m}\r)\delta_{\rm m}\,.
\ee
This equation will be considered at second order in perturbations. 

We suppose as usual that the density profile is spherical \cite{Gunn:1972sv} and that the total mass inside the radius $R$ is conserved during the collapse phase, which translates to
\be\label{eq:M}
M = \f{4\pi}{3}R^3\Bar{\rho}_{\rm m}(1+\delta_{\rm m})=\text{const} \,.
\ee
Then, its second derivative w.r.t.\ time combined with Eq.~\eqref{nonlinearoverdensityeq} yields
\be\label{collapseR}
 \f{\ddot{R}}{R} = H^2 + \dot{H} - 
 \f{4\pi G_{\rm N}}{3}\mu^{\rm NL}\,\Bar{\rho}_{\rm m}\delta_{\rm m}\,.
\ee
The change of variable
\be\label{defy}
 y=\f{R}{R_i} - \f{a}{a_{i}}\,,
\ee
with $R_{i}$ and $a_{i}$ the initial values of perturbation radius and scale factor, is instrumental to rewrite Eq.~\eqref{collapseR} in a more convenient way for numerical solutions, i.e.
\be\label{yeq}
y^{\prime\prime} = -\f{H^\prime}{H}y^\prime + \l(1+\f{H^\prime}{H}\r)y - \f{\Omega_{\rm m}}{2}\mu^{\rm NL}\delta_{\rm m}\l(y +\f{a}{a_{i}}\r)\,.
\ee
In this equation one needs to specify the evolution of $\mu^{\rm NL}(a,R)$ and of the ratio $R/R_{\rm V}$. Considering the mass conservation and the definition of the Vainshtein radius, we obtain
\ba\label{RRV3}
\l(\f{R}{R_{\rm V}}\r)^3 & = & \f{1}{4\Omega_{\rm m}H^2\lambda^4 }\f{1}{\delta_{\rm m}}\,.
\ea
Note that $(R/R_{\rm V})^3 \propto \ab^2$. Thus, only in presence of a cubic term ($G_3$) in the Lagrangian will a relation between the Vainshtein radius and the collapsing overdensity exist. In the following we will numerically solve Eq.~\eqref{yeq} with initial conditions such that the collapse occurs today, $a_{\rm coll}=1$. Then we have $a_i=6.66 \times 10^{-6}$, $y_i = 0$ and $y^\prime_i = -\delta_{{\rm m},i}/3$, where $\delta_{{\rm m},i}$ is the initial density obtained from linear theory in the matter dominated era assuming the collapse ($R=0$) at $a_{\rm coll}=1$ \cite{Bellini:2012qn}. Because of mass conservation, overdensity can be written as
\be
 \delta_{\rm m} = (1+\delta_{{\rm m},i})\l(1+\f{a_{i}}{a}y\r)^{-3}-1\,.
\ee
After the maximum expansion and approaching collapse, which mathematically would correspond to a singularity, the system enters in the virialization stage: a stable, self-gravitating, spherical distribution reaches the equilibrium described by the virial theorem, i.e., the total kinetic energy of the object ($T$) and the total gravitational potential energy ($U$) satisfy the relation
\be\label{virialtheorem}
 T + \f{1}{2}U = 0\,.
\ee
For a top-hat profile one has
\be
 T \equiv \f{1}{2}\int\mathrm{d}^3\mathbf{x} \,\rho_{\rm m} \mathbf{v}^2 = \f{3}{10}M\dot{R}^2\,,
\ee
and \cite{Kimura:2010di}
\ba
 U & \equiv & -\int \mathrm{d}^3x \,\rho_{\rm m}({\bf x}) \,{\bf x} \cdot \nabla \Psi \nn \\
 & = & \f{3}{5}\l(\dot{H}+H^2\r)MR^2 - 
       \f{3}{5}G_{\rm N}\mu^{\rm NL}\f{M}{R}\delta M\,.
\ea
Since energy conservation is not strictly satisfied for a time-dependent dark energy or modified gravity model, instead of assuming it to define the virial overdensity, we evaluate the condition given by Eq.~\eqref{virialtheorem} during collapse in order to find the virialization time $a_{\rm vir}$ at which it is satisfied as suggested in Ref.~\cite{Schmidt:2009yj}. We then define the virial overdensity as
\be \label{eq:viroverdens}
 \Delta_{\rm vir} \equiv \f{\rho_{\rm vir}}{\rho_{\rm coll}} = 
 \l[1+\delta_{\rm m}(R_{\rm vir})\r]\l(\f{a_{\rm coll}}{a_{\rm vir}}\r)^3\,.
\ee

The virial overdensity determines the mean square velocity of particles in virialized halos. Using Eq.~\eqref{eq:M} and the virial theorem, one obtains
\begin{align}
\overline{V_\mathrm{vir}^2} = &\, \frac{3}{5}\l[\gn MH\sqrt{\Omega_{\rm m}\Delta_{\rm vir}} \r]^{2/3} \nonumber \\
\times & \l\{\frac{1}{\Delta_{\rm vir}}\l[1 + \frac{\Omega_{\rm DE}}{\Omega_{\rm m}}(1+3w_{\rm DE})\r] + \mu^{\rm NL}\l(1-\frac{1}{\Delta_{\rm vir}}\r)\r\}\,,
\end{align}
where $\Omega_{\rm DE}=1-\Omega_{\rm m}-\Omega_{\rm r}$ and $w_{\rm DE}$ are the time-dependent density parameter and equation of state of the DE scalar field and the mass $M$ does not include any contribution from scalar field fluctuations. The known $\Lambda$CDM expression \cite{WangSteinhardt1998} is recovered for $\mu^{\rm NL} = 1$ and $w_{\rm DE}=-1$. One can eventually deduce the  kinetic energy of a virialized halo, $T_\mathrm{vir}=\frac{1}{2}M\overline{V_\mathrm{vir}^2}$, and model its gas temperature, which for a uniform isothermal cloud of monatomic gas with vanishing external pressure is given by $T=\mu m_\mathrm{p}\sigma^2/3\beta=\mu m_\mathrm{p}\overline{V_\mathrm{vir}^2}/3\beta$, where $\mu m_\mathrm{p}$ is the mean mass of particles, $\sigma^2$ is the isotropic velocity dispersion, $\beta$ is the ratio of the kinetic energy to the temperature, and $k_\mathrm{B}$ is the Boltzmann constant.

\section{Application of the spherical collapse model to a KGB-mimic model} \label{sec:IV}

In previous sections we have defined a general framework for studying KGB models. In literature there are many applications of the spherical collapse for KGB models \cite{Kimura:2010di,Bellini:2012qn,Barreira:2013eea,Frusciante:2020zfs} that require the specification of the forms for the $G_{2,3}$ functions, so that $\ab$ and $\ak$ are fixed. In this study we will pursue the most general approach possible, not focusing on any specific form of the $G_i$-functions. To this purpose, we will consider the general parametrization for $\ak$ and $\ab$ presented in \cite{Traykova:2021hbr} encompassing a general class of shift-symmetric Galileon models. One more reason for this general approach also lies in the future availability of $N$-body simulations, which might be based on spherical collapse computations \cite{Hassani:2020rxd}, or semi-analytic approaches such as the reaction method \cite{Bose:2022vwi}, capturing in a general way the nonlinear phenomenology of KGB models.

KGB features are encoded in the equation-of-state for the effective DE component, $w_{\rm DE}$, and in $\ab$, each of them with only two parameters, while leaving $\ak$ constant ~\cite{Traykova:2021hbr}. Such modelling fits incredibly well more than 98\% of the randomly generated models. We will assume $\ab(a)$ to be described by~\cite{Traykova:2021hbr}
\begin{equation}
    \alpha_B(a)=\alpha_{B,0}\l(\frac{H_0}{H(a)}\r)^\frac{4}{m} \,, \label{eq:alphaKGB}
\end{equation}
where $\alpha_{B,0}$ and $m$ are constants and $H_0$ is the Hubble constant, while the DE equation-of-state takes the Chevallier-Polarski-Linder (CPL) form \cite{Chevallier:2000qy,Linder:2002et}:
\begin{equation}
    w_{\rm DE}(a) = w_0+w_a(1-a) \,,
\end{equation}
with $w_0$ and $w_a$ constants. The latter fully determines the Hubble parameter $H(a)$ and the DE density parameter $\Omega_{\rm DE}(a)$. We have four free parameters, $\{ \alpha_{B,0}, m, w_0, w_a \}$. 
Following \cite{Traykova:2021hbr} we fix $\alpha_{K,0}=10$; indeed, this parameter does not enter the equations we are considering, moreover it cannot be constrained by data \cite{Bellini:2015xja,Frusciante:2018jzw}.  
Throughout this analysis the DE parameters $w_0$ and $w_a$ are fixed and we focus on the parameters controlling $\ab(a)$, i.e.\ $\alpha_{B,0}$ and $m$. For this purpose, we use as \textit{baseline} the best fit values of the KGB-mimic parameters found in \cite{Traykova:2021hbr} for a combined analysis of Cosmic Microwave Background (CMB), Baryon Acoustic Oscillations (BAO), Redshift Space Distortions (RSD) and type Ia supernovae (SnIa) data, together with some theoretical priors. This corresponds to $\alpha_{B,0} = -0.30$\footnote{Note that the $\alpha_{B,0}$ used here differs from the one reported in \cite{Traykova:2021hbr} by a factor of $-1/2$. This is due to a different sign choice for the Lagrangian term $\mathcal{L}_3$ together with a different definition of $X$, which consequently change the definition of the braiding function.}, $m = 2.4, w_0=-0.97, w_a=-0.11$, hereafter referred as Best Fit KGB model. The other KGB models that we will explore correspond to variations around this baseline, letting $\alpha_{B,0}$ or $m$ vary while keeping all the other parameters fixed. We have checked that the DE parameters mainly introduce minor modifications in the amplitudes of physical quantities such as $\delta_{\rm c}$ and $\Delta_{\rm vir}$ \cite{Pace:2010sn,Pace:2011kb} (see also \cite{Pace:2013pea,Nazari-Pooya:2016bra} for a comparison between dark energy and modified gravity models). Additionally, we have checked that no ghost or gradient instabilities occur when the test values of $\alpha_{B,0}$ and $m$ are used.

Beside the baseline Best Fit KGB model, we have explored the parameter space by varying $\alpha_{B,0}$ and $m$ individually, considering two values for $\alpha_{B,0}$ and two for $m$. The other cosmological parameters are fixed to $H_0=70$~km~s$^{-1}$~Mpc$^{-1}$, $\Omega_{\rm r,0} = 8.516 \times 10^{-5}$, $\Omega_{\rm DE,0} = 0.69$.

\begin{figure}[t!]
\includegraphics[width=.46\textwidth]{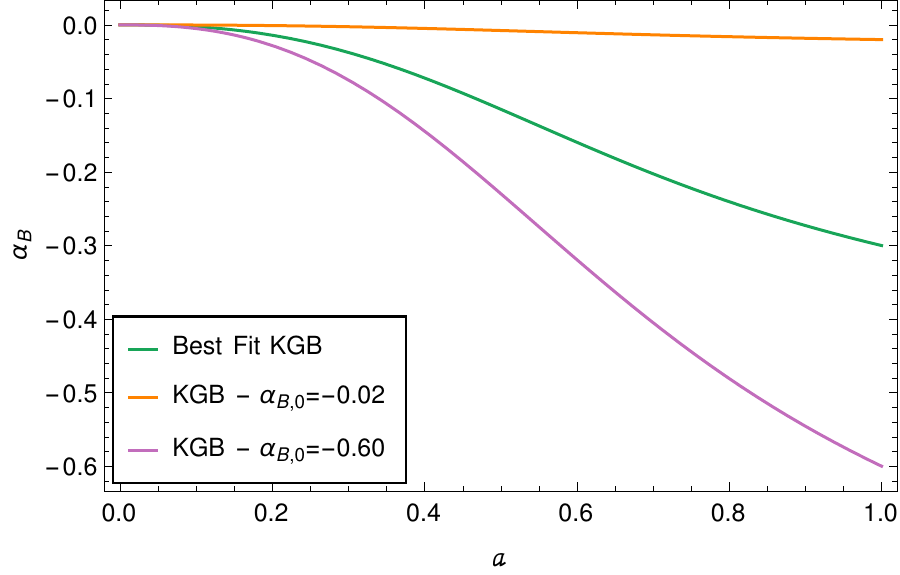}
\caption{Time evolution of the braiding function $\ab$ for the three KGB-mimic models with varying $\alpha_{B,0}$: the Best Fit KGB having $\alpha_{B,0} = -0.30$ (green line), $\alpha_{B,0} =-0.02$ (orange line) and $\alpha_{B,0} = -0.60$ (purple line).}
\label{fig:alphaBevolalpha} 
\end{figure}

\begin{figure}[h!]
\includegraphics[width=.46\textwidth]{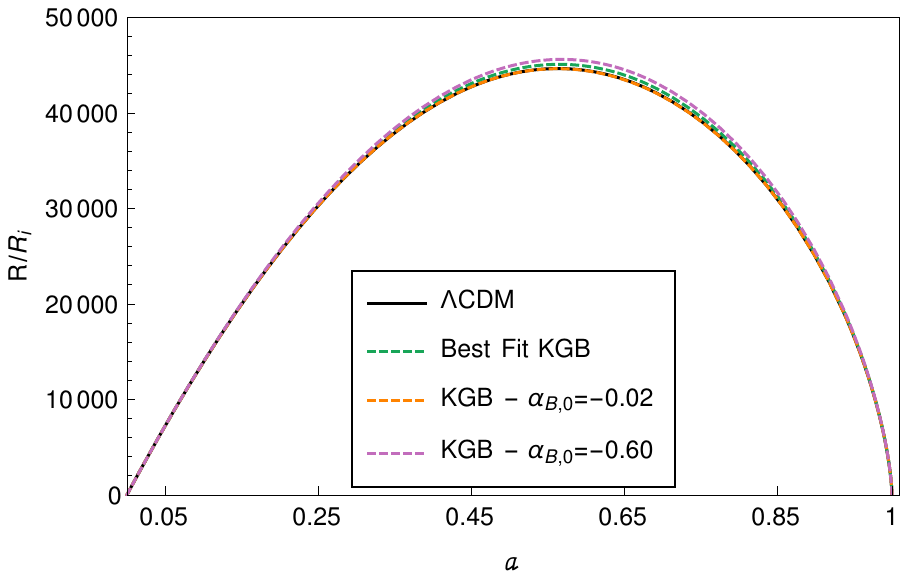}
\caption{Time evolution of $R/R_{i}$ for KGB-mimic models with varying $\alpha_{B,0}$. The $\Lambda$CDM (black solid line) is shown for comparison. The two KGB-mimic models with varying $\alpha_{B,0}$ share the values of the Best Fit KGB model for the remaining parameters.}
\label{fig:RRiKGBalpha} 
\end{figure}

\subsection{Varying \texorpdfstring{$\alpha_{B,0}$}{aB0}} \label{sec:KGBVaralpha}

\autoref{fig:alphaBevolalpha} shows the time evolution of the braiding function for $\alpha_{B,0}=-0.02$ and $\alpha_{B,0}=-0.60$, with fixed $m=2.4$. As evident from the functional form of Eq.~\eqref{eq:alphaKGB}, the largest the value of $\lvert \alpha_{B,0} \rvert$, the largest the departure from zero of $\alpha_B(a)$, regardless the value of $m$.

The time evolution of the radius of spherical perturbations is shown in \autoref{fig:RRiKGBalpha}. Non-vanishing values of $\alpha_{B,0}$ lead to non-trivial deviations from the $\Lambda$CDM behaviour; see~\autoref{tab:KGBalphaquantities}. The Best Fit KGB model and the model with $\alpha_{B,0}=-0.6$ have the largest turnaround radius ($R_{\rm ta}$) with respect to the $\Lambda$CDM reference model. This behaviour reflects the values of the initial overdensities, $\delta_{{\rm m},i}$, which follow the order $\delta_{{\rm m},i} (\alpha_{B,0} = -0.60) < \delta_{{\rm m},i} (\text{Best Fit}) < \delta_{{\rm m},i} (\Lambda\text{CDM})\simeq \delta_{{\rm m},i} (\alpha_{B,0} = -0.02)$. A smaller initial overdensity determines a delay of the onset of collapse phase, leaving more time for evolution so that the radius can reach larger values at turnaround. Note that the evolution of $R/R_i$ for the best Fit KGB and the $\alpha_{B,0}=-0.6$ are very similar to the one of the Cubic Galileon, while the one of $\alpha_{B,0}=-0.02$ follows the one of the Galileon Ghost Condensate \cite{Frusciante:2020zfs}. Interestingly enough, the turnaround occurs at very similar times; see the third line in \autoref{tab:KGBalphaquantities}.

\begin{figure}[t!]
\includegraphics[width=.46\textwidth]{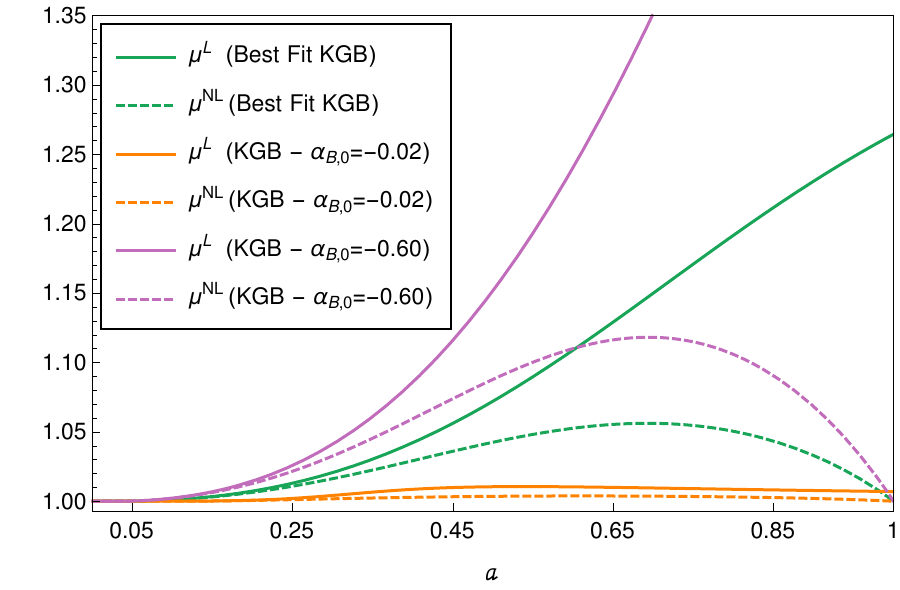}
\caption{Time evolution of $\mu^{\rm L}$ (solid lines) and $\mu^{\rm NL}$ (dashed lines) for for KGB-mimic models with varying $\alpha_{B,0}$. Collapse is set at present time.}
\label{fig:muKGBalpha} 
\end{figure}

The time evolution of the linear and nonlinear gravitational couplings are shown in \autoref{fig:muKGBalpha}. As expected $\mu^{\rm L}>1$, consistent with a gravitational interaction stronger than in $\Lambda$CDM. Additionally, larger negative values of $\alpha_B$ show a larger deviation from the GR limit, i.e.\ from $\mu^{\rm L} = 1$. Specifically, at present time $\mu^{\rm L}$ deviates from unity by about $0.6\%$, $26\%$ and $77\%$ for the $\alpha_{B,0} = -0.02$, Best Fit KGB, and $\alpha_{B,0} = -0.60$ models, respectively.

\begin{figure}[t!]
\includegraphics[width=.46\textwidth]{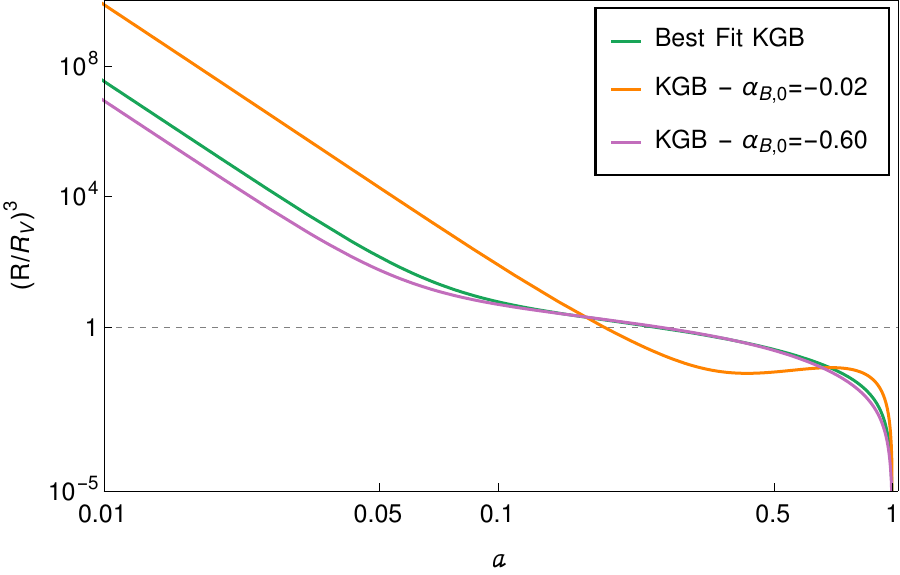}
\caption{Time evolution of $(R/R_{\rm V})^3$ for for KGB-mimic models with varying $\alpha_{B,0}$, for a matter overdensity collapsing at present time.}
\label{fig:RRV3KGBalpha} 
\end{figure}

The time evolution of the nonlinear effective gravitational coupling (dashed lines in \autoref{fig:muKGBalpha}) is determined by the Vainshtein mechanism, which crucially modifies the gravitational interactions on small scales. Depending on the value of $\alpha_{B,0}$ the radius of spherical shells crosses the Vainshtein radius at a different time, which can be extrapolated from \autoref{fig:RRV3KGBalpha}. Specifically, $R/R_{\rm V} = 1$ at $a=0.186$ for the KGB model with $\alpha_{B,0} = -0.02$, at $a=0.253$ for the Best Fit KGB model, and at $a=0.260$ for $\alpha_{B,0}=-0.60$. Namely, the Vainshtein mechanism becomes effective later for larger values of $|\alpha_{B,0}|$. Note also that the model with the smallest value, $\alpha_{B,0}=-0.02$, shows a rather different evolution of $(R/R_{\rm V})^3$. This order can be understood by estimating the Vainshtein radius for the different cases; using Eq.~(\ref{eq:vainsrad}) at present time for a point-like source, we find that the $\alpha_{B,0}=-0.02$ model has the smallest Vainshtein radius among the three KGB models, $R_{\rm V} = 0.95 \ (M/10^{12} \mathrm{M}_\odot)^{1/3}$~Mpc, which explains why $R/R_{\rm V}$ crosses unity first. The other two models have larger Vainshtein radii, following the growing order of $\alpha_{B,0}$. Specifically, for the Best Fit KGB model $R_{\rm V} = 1.76 \ (M/10^{12} \mathrm{M}_\odot)^{1/3}~\mathrm{Mpc}$, while for $\alpha_{B,0} = -0.60$ the Vainshtein radius is $R_{\rm V} = 2.26 \ (M/10^{12} \mathrm{M}_\odot)^{1/3}~\mathrm{Mpc}$.

\renewcommand{\arraystretch}{1.2}
\begin{table*}[t!]
  \caption{\label{tab:KGBalphaquantities} Physical quantities characterising the spherical collapse at present time for different KGB-mimic models with varying $\alpha_{B,0}$ in comparison with $\Lambda$CDM. The cosmological parameters are $H_0=70$ km~s$^{-1}$Mpc$^{-1}$, $\Omega_{\rm m,0}=0.31$, $\Omega_{\rm r,0}=8.51\times10^{-5}$. The initial scale factor is fixed deep into the radiation-dominated era. The KGB-mimic models (three rightmost columns) share all the parameter values ($w_0, w_a, m$) but $\alpha_{B,0}$ of the Best Fit KGB model ($\alpha_{B,0}=-0.30$).}
 \centering
 \begin{tabular}{|c|c|c|c|c|}
  \hline
  & $\Lambda$CDM & Best Fit KGB & $\alpha_{B,0} = -0.02$ & $\alpha_{B,0} = -0.60$ \\
  \hline \hline
  $\delta_{\rm c}$ & 1.676 & 1.697 & 1.679 & 1.726 \\
  $\Delta_{\rm vir}$ & 333.1 & 311.1 & 332.9 & 289.0 \\
  $a_{\rm ta}$
  & 0.5634
  & 0.5653
  & 0.5629 
  & 0.5680
  \\
  $a_{\rm vir}$ 
  & 0.9218  
  & 0.9199 
  & 0.9217
  & 0.9179
  \\
  $R_{\rm ta}/R_i$ ($\times 10^{-2}$) & 446.2 & 450.6 & 446.1 & 455.8 \\
  $R_{\rm vir}/R_i$ ($\times 10^{-2}$) & 216.4 & 221.4 & 216.5 & 226.9 \\
  $\delta_{{\rm m},i}$ ($\times 10^{4}$) & 1.23 & 1.21 & 1.23 & 1.19 \\
  \hline
 \end{tabular}
\end{table*}
\renewcommand{\arraystretch}{1}

\begin{figure}[t!]
\includegraphics[width=.46\textwidth]{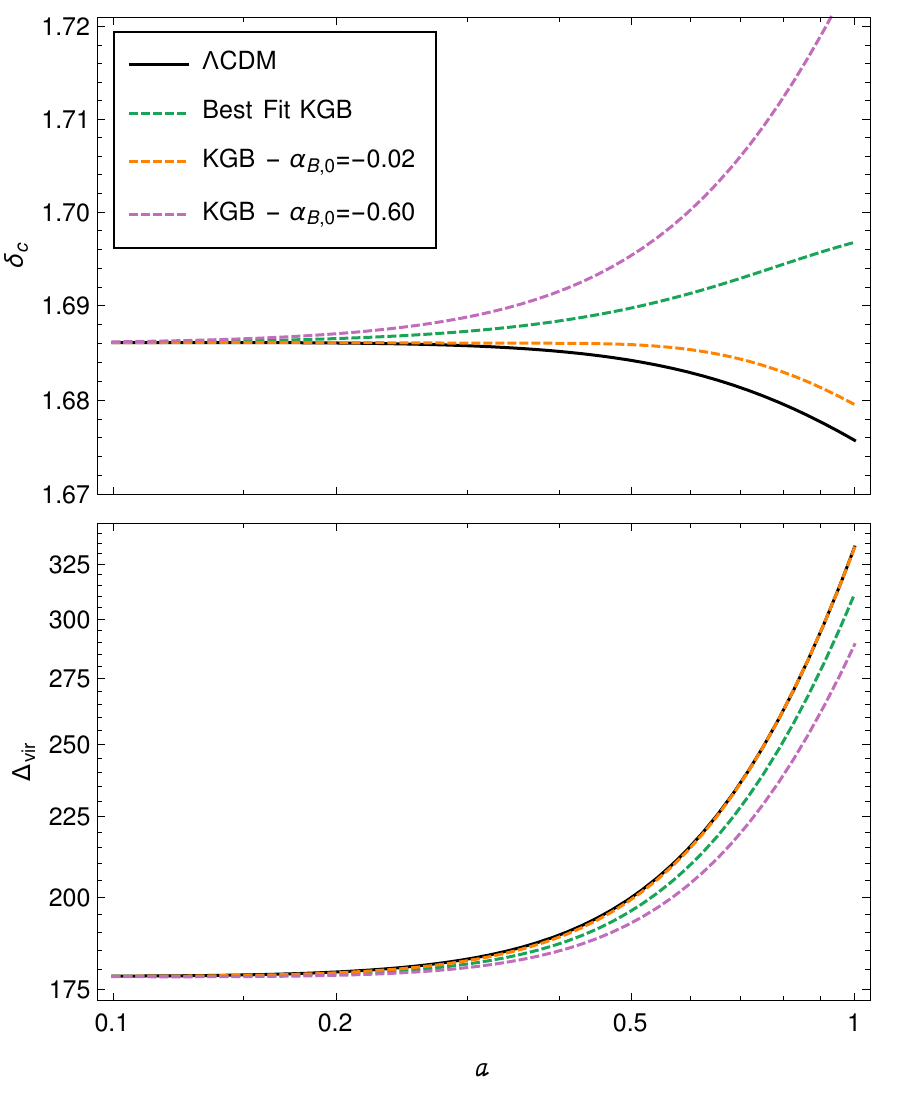}
\caption{Evolution of the critical density contrast $\delta_{\rm c}$ (top panel) and of the virial overdensity $\Delta_{\rm vir}$ (bottom panel) as a function of the scale factor $a$ for the three KGB-mimic models with varying $\alpha_{B,0}$. The $\Lambda$CDM model (black solid line) is shown for comparison.}
\label{fig:deltasKGBalpha}
\end{figure}

Regarding the last stages of collapse, \autoref{tab:KGBalphaquantities} reports the values of some important physical quantities such as $\delta_{\rm c}$, $\Delta_{\rm vir}$ and $a_{\rm vir}$. The critical density contrast $\delta_{\rm c}$ is the value of the linear matter density contrast $\delta_{\rm m}$ at the time of collapse, when the initial conditions have been set in order for the nonlinear equation to diverge at collapse. This quantity defines a linear scale at which one should expect collapse to have occurred. The virial overdensity $\Delta_{\rm vir}$ is defined in Eq.~\eqref{eq:viroverdens}, and occurs at time $a_{\rm vir}$. \autoref{fig:deltasKGBalpha} illustrates the time evolution of $\delta_{c}$ and $\Delta_{\rm vir}$. In the top panel, we note that $\delta_{\rm c}$ for the KGB model with $\alpha_{B,0} = -0.02$ follows the $\Lambda$CDM behaviour up to $a\simeq 0.3$, instead the models with larger values of $\lvert \alpha_{B,0} \rvert$ deviate from the Einstein-de Sitter value ($\delta_{\rm c}=1.686$) earlier than $a\simeq 0.2$. At later times, the critical density is larger compared to the $\Lambda$CDM model, with some distinctions: in models with more negative values of $\alpha_{B,0}$, $\delta_{\rm c}$ is monotonically increasing, instead the model with $\alpha_{B,0}$ close to zero has $\delta_{\rm c}$ monotonically decreasing. The evolution of $\delta_{\rm c}$ is explained by the time-dependence of $\mu^{\rm L}$, as modifications of gravity are larger and start earlier in time in the two models with larger values of $\lvert \alpha_{B,0} \rvert$. Furthermore, we note that the models with larger values of $\delta_{\rm c}$ at present time correspond to smaller initial overdensities, reaching a larger radius before the time of turnaround.

\begin{figure}[t!]
\includegraphics[width=.46\textwidth]{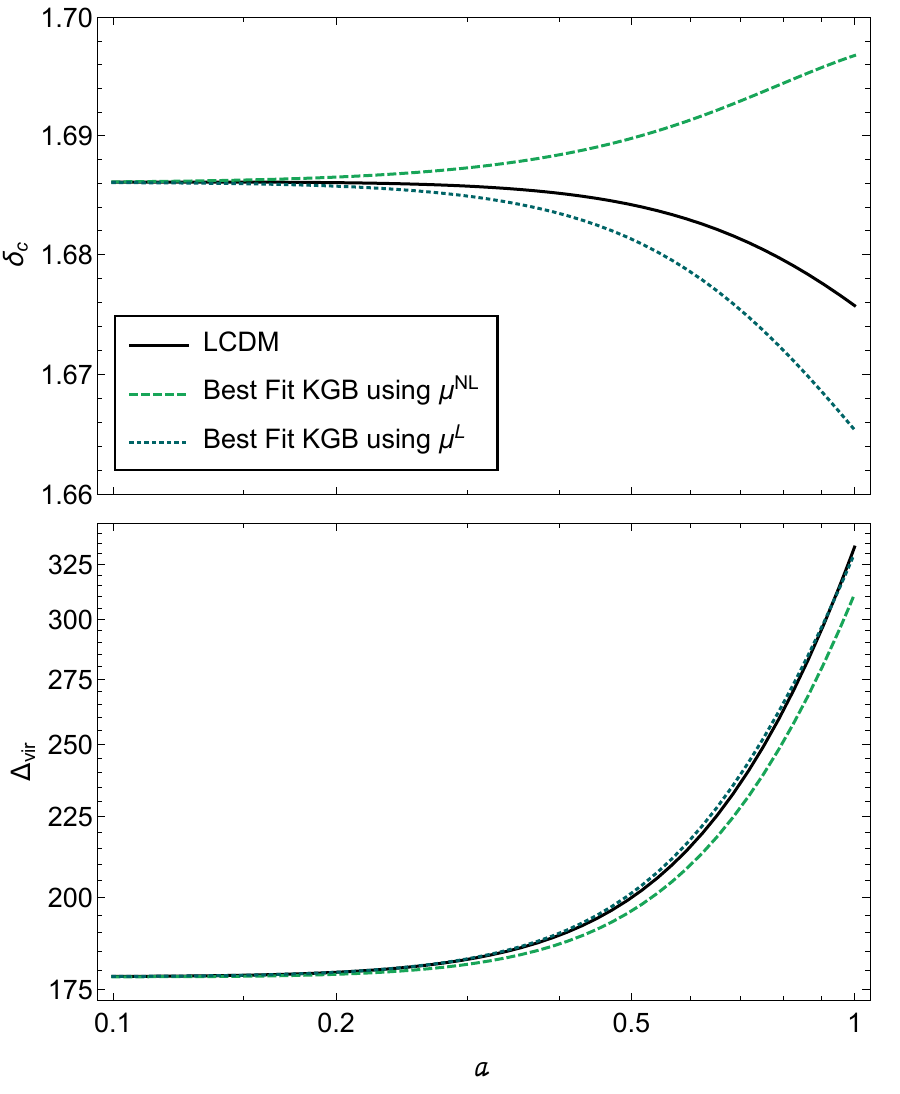}
\caption{Evolution of the critical density contrast $\delta_{\rm c}$ (top panel) and of the virial overdensity $\Delta_{\rm vir}$ (bottom panel) as a function of the scale factor $a$ for the Best Fit KGB model using $\mu^{\rm NL}$ (dashed lines) and $\mu^{\rm L}$ (dotted lines) to solve the nonlinear equations. The $\Lambda$CDM model (black solid line) is shown for comparison.}
\label{fig:VainshteinEffectonDeltas}
\end{figure}

\begin{figure}[t!]
\includegraphics[width=.48\textwidth]{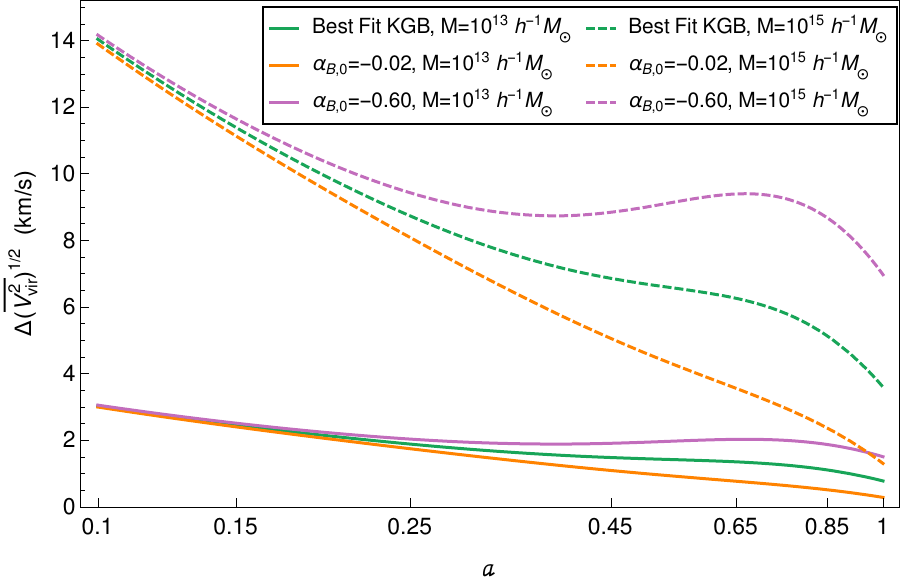}
\caption{Time evolution of the difference between the root-mean-square virial velocity $(\overline{V_\mathrm{vir}^2})^{1/2}$ of three KGB-mimic models with varying $\alpha_{B,0}$ and the corresponding value for the $\Lambda$CDM model, for halos with mass $M = 10^{13} \, h^{-1} \mathrm{M}_\odot$ (solid lines) and $M = 10^{15} \, h^{-1} \mathrm{M}_\odot$ (dashed lines).}
\label{fig:deltaVvirKGBalpha}
\end{figure}

The bottom panel of \autoref{fig:deltasKGBalpha} shows the time evolution of the virial overdensity. In all the models, $\Delta_{\rm vir}$ remains approximately constant close to the Einstein-de Sitter value ($\Delta_{\rm vir}\approx 177.8$) until $a \simeq 0.2$, then it is suppressed compared to the $\Lambda$CDM model as collapse approaches present time. Note that the larger $|\alpha_{B,0}|$ is, the smaller the present day value of $\Delta_{\rm vir}$; see \autoref{tab:KGBalphaquantities}. Indeed, the model with the largest $\lvert \alpha_{B,0} \rvert$, i.e.\ $\alpha_{B,0} = -0.6$, has $\Delta_{\rm vir}(a=1)=289.0$, while the model with the smallest absolute value of this parameter, $\alpha_{B,0}=-0.02$, remains very close to $\Lambda$CDM with $\Delta_{\rm vir}(a=1)=332.9$.

The importance of including the effect of the Vainshtein mechanism in the effective gravitational coupling can be assessed from the evolution of $\delta_{\rm c}$ and $\Delta_{\rm vir}$. By comparing the previous results for the Best Fit KGB model with the case where nonlinear equations use $\mu^{\rm L}$ instead of $\mu^{\rm NL}$, the critical density results are underestimated while the virial overdensity becomes overestimated at late collapse times, as shown respectively in the top and bottom panels of \autoref{fig:VainshteinEffectonDeltas}. Specifically, for a collapse at present time, $\delta_{\rm c}$ would be underestimated by $\sim 1.8\%$ and $\Delta_{\rm vir}$ would be overestimated by $\sim 6.1\%$. This behaviour is consistent with the difference between the time evolution of the linear and nonlinear effective gravitational couplings shown in \autoref{fig:muKGBalpha}. Indeed, by using $\mu^{\rm L}$ rather than $\mu^{\rm NL}$ we are considering a stronger gravitational interaction, specially at late times, since the linear coupling is not limited by the Vainshtein mechanism. This will accelerate the collapse of the overdensity, which will translate into a smaller value of $\delta_{\rm c}$ and a larger $\Delta_{\rm vir}$.

In summary, the amplitude of the critical and virial overdensities is non-trivially affected as long as $\alpha_{B,0}$ is allowed to vary while keeping all other parameters fixed, with more negative values of $\alpha_{B,0}$ corresponding to larger deviations from the $\Lambda$CDM scenario. Note also that deviations from the standard cosmological model become more evident as the time of collapse approaches present time. The reason is that we are considering late-time deviations in $\alpha_B$. Additionally, in this case for values of $\alpha_{B,0}$ very close to zero the KGB models clearly mimic the $\Lambda$CDM behaviour. It is important to stress that the values at present time and the time evolution of $\delta_c$ and $\Delta_{\rm vir}$ are clearly different from those in $\Lambda$CDM and, as such, modifications of gravity can be distinguishable from GR.

The small deviation of $\mu^{\rm NL}$ from unity and the very small variation of $\Delta_{\rm vir}$ from the $\Lambda$CDM values yield a tiny variation of the root-mean-square virial velocity $(\overline{V_\mathrm{vir}^2})^{1/2}$ from $\Lambda$CDM values, approximately decreasing with time unless very massive halos are considered; see \autoref{fig:deltaVvirKGBalpha}. For the more extreme KGB-mimic model considered in this section, with $\alpha_{B,0}=-0.6$, the 10\% variation in $\mu^{\rm NL}$ from unity occurring at $a\approx0.7$, or redshift $z=0.43$, jointly with the $\sim1\%$ variation of $\Delta_{\rm vir}$ at the same epoch lead to $\sim5\%$ or 10~km~s$^{-1}$ difference in the rms-velocity compared to $\Lambda$CDM for halos with mass $M=10^{15}h^{-1} \mathrm{M}_\odot$. Less massive halos have virial velocity almost indistinguishable from that in $\Lambda$CDM.

\subsection{Varying \texorpdfstring{$m$}{m}}

\begin{figure}[t!]
\includegraphics[width=.46\textwidth]{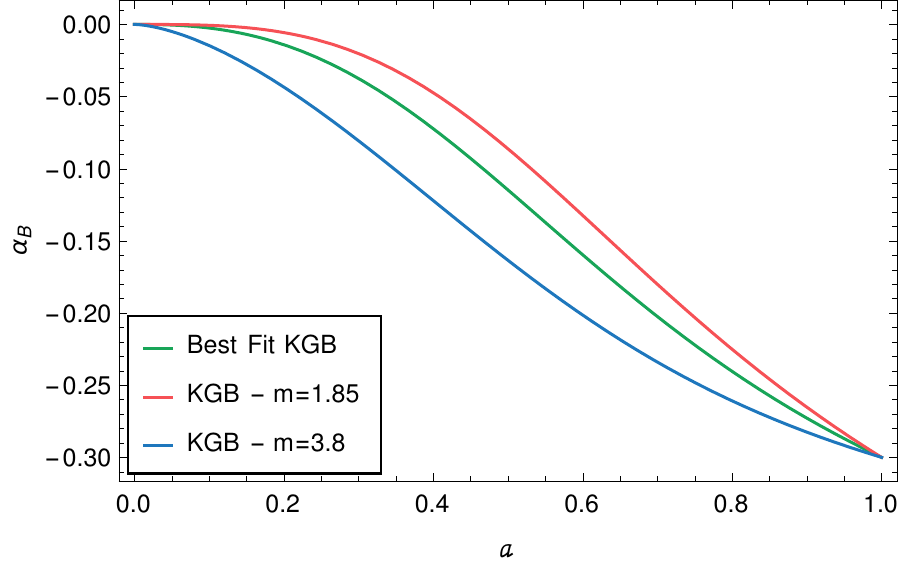}
\caption{Time evolution of the braiding function $\alpha_B$ for the three KGB-mimic models with varying $m$: the Best Fit KGB model with $m=2.4$ (green line), $m=1.85$ (red line) and $m=3.8$ (blue line).}
\label{fig:alphaBevolm} 
\end{figure}

\begin{figure}[t!]
\includegraphics[width=.46\textwidth]{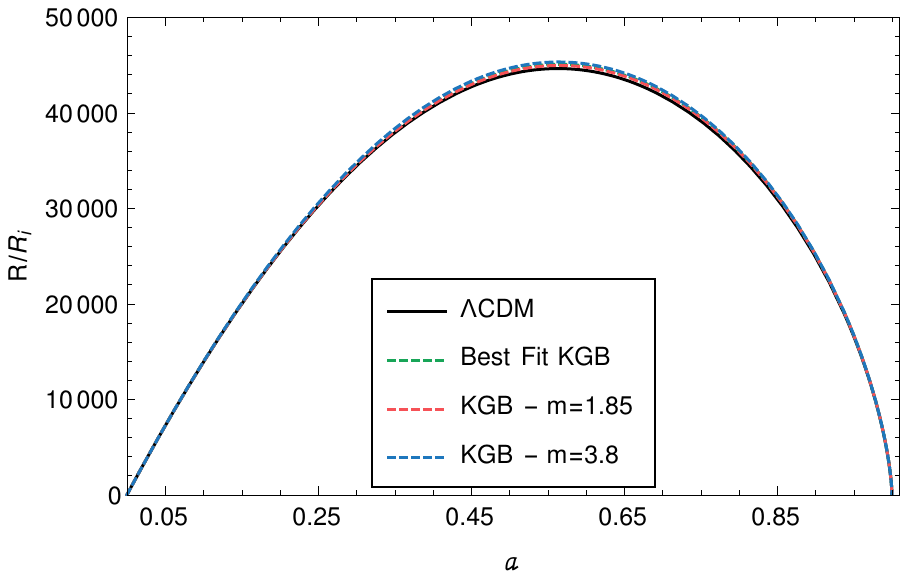}
\caption{Time evolution of $R/R_{i}$ for the KGB-mimic models with varying $m$. The $\Lambda$CDM (black solid line) is shown for comparison. The two KGB-mimic models with varying $m$ share the values of the Best Fit KGB model for the remaining parameters.}
\label{fig:RRiKGBm} 
\end{figure}

\begin{figure}[t!]
\includegraphics[width=.46\textwidth]{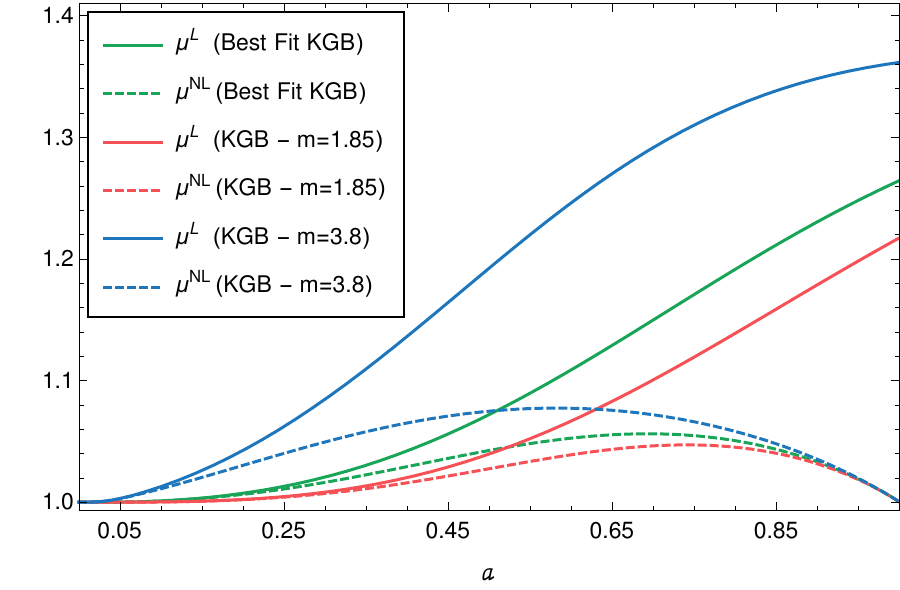}
\caption{Time evolution of $\mu^{\rm L}$ (solid lines) and $\mu^{\rm NL}$ (dashed lines) for the KGB-mimic models with varying $m$. Collapse is set at present time.}
\label{fig:muKGBm} 
\end{figure}

The time variation of the braiding function $\alpha_B (a)$ for varying $m$ and fixed $\alpha_{B,0}$ is qualitatively different but quantitatively comparable with the results discussed in the previous Section; see \autoref{fig:alphaBevolm}, in which the Best Fit KGB model with $m = 2.4$ is compared to two KGB-mimic models with $m = 1.85$ and $m = 3.8$. Since all models share the same value of $\alpha_{B,0}$, by definition they reach the same value of the braiding function at present time. Moreover, since the braiding function vanishes at early times, any change of $m$ translates into a different rate of change of $\alpha_B$ for $a\approx0$; the larger the value of $m$, the earlier the braiding function departs from zero.

The time evolution of $R/R_i$ is shown in \autoref{fig:RRiKGBm} for the Best Fit KGB and $m$-varying models, including the curve for the reference $\Lambda$CDM for comparison. The time of collapse is again fixed at $a_{\rm coll} = 1$. \autoref{tab:KGBmquantities} quotes the same quantities as \autoref{tab:KGBalphaquantities}. As for the turn-around radius (fifth line in the table), all the KGB-mimic models attain a larger value than in $\Lambda$CDM, the increase in $R_{\rm ta}/R_i$ following the growing order of $m$. Contrarily, the values of the initial overdensity $\delta_{{\rm m},i}$ (bottom line) are smaller for larger $m$. This is consistent with a later collapse of the smaller overdensities, which therefore attain larger radii. In this case, the modifications are due to the different time evolution of the braiding function $\alpha_B (a)$, which changes the evolution of the matter density perturbations $\delta_{\rm m}$. 

\renewcommand{\arraystretch}{1.2}
\begin{table*}[t!]
 \caption{Physical quantities characterising the spherical collapse at the present time for different KGB-mimic models with varying $m$ in comparison with $\Lambda$CDM. The cosmological parameters are: $H_0=70$ km~s$^{-1}$~Mpc$^{-1}$, $\Omega_{\rm m,0}^0=0.31$, $\Omega_{\rm r,0}^0=8.51\times10^{-5}$. The initial scale factor is fixed deep into the radiation-dominated era. The KGB-mimic models (three rightmost columns) share all the parameter values $(w_0,w_a,\alpha_{B,0})$ but $m$ of the Best Fit KGB ($m=2.4$).}
 \centering
 \begin{tabular}{|c|c|c|c|c|}
  \hline
  & $\Lambda$CDM & Best Fit KGB & $m=1.85$ & $m=3.8$ \\
  \hline \hline
  $\delta_{\rm c}$ & 1.676 & 1.697 & 1.685 & 1.757 \\
  $\Delta_{\rm vir}$ & 333.1 & 311.1 & 316.0 & 302.8 \\
  $a_{\rm ta}$ 
  & 0.5634
  & 0.5653
  & 0.5654
  & 0.5645
  \\
  $a_{\rm vir}$ 
  & 0.9218
  & 0.9199
  & 0.9205
  & 0.9189
  \\
  $R_{\rm ta}/R_i$ ($\times 10^{-2}$) & 446.2 & 450.6 & 449.4 & 453.1 \\
  $R_{\rm vir}/R_i$ ($\times 10^{-2}$) & 216.4 & 221.4 & 220.2 & 223.4 \\
  $\delta_{{\rm m},i}$ ($\times 10^{4}$) & 1.23 & 1.21 & 1.22 & 1.18 \\
  \hline
 \end{tabular}
 \label{tab:KGBmquantities}
\end{table*}
\renewcommand{\arraystretch}{1}

\begin{figure}[t!]
\includegraphics[width=.46\textwidth]{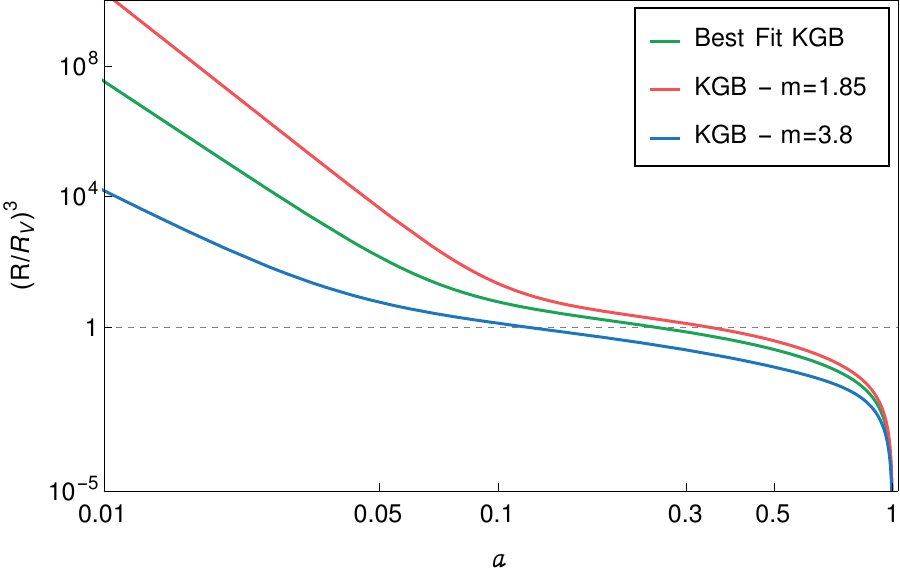}
\caption{Time evolution of $(R/R_{\rm V})^3$ for the KGB-mimic models with varying $m$, for a matter overdensity collapsing at present time.}
\label{fig:RRV3KGBm} 
\end{figure}

The parameter $m$ has a clear impact on the evolution of the linear and nonlinear gravitational couplings, as shown in \autoref{fig:muKGBm}. The model with the largest $m$ shows the largest modifications of the gravitational interaction w.r.t.\ GR. Departures from GR also occur at earlier times for larger $m$. Note that depending on the value of $m$, the radius of spherical shells enters the Vainshtein radius at different times. Using the evolution of $(R/R_{\rm V})^3$ plotted in \autoref{fig:RRV3KGBm} to estimate the scale factor at which $R/R_{\rm V}=1$, we obtain the following order: $a(m=3.8) = 0.112$, $a(\rm Best \ Fit)=0.253$, $a(m=1.85)=0.345$. The sizes of the Vainshtein radii in the limit of a point source are respectively $R_{\rm V} = 1.54 \ (M/10^{12} \mathrm{M}_\odot)^{1/3}$~Mpc for $m=1.85$, $R_{\rm V} = 1.76 \ (M/10^{12} \mathrm{M}_\odot)^{1/3}$~Mpc for Best Fit KGB, and $R_{\rm V} = 2.17 \ (M/10^{12} \mathrm{M}_\odot)^{1/3}$~Mpc for $m=3.8$. Note also that the model with the largest Vainshtein radius has the smallest $\delta_i$. However, while for varying $\alpha_{B,0}$ the model with the largest Vainshtein radius is the \emph{last} for which $R$ becomes smaller than $R_{\rm V}$, in the case with varying $m$ it is the \emph{first}. This can be explained because the effects of $\alpha_{B,0}$ start affecting the gravitational coupling at later times, while the ones of $m$ start earlier.

\begin{figure}[t!]
\includegraphics[width=.46\textwidth]{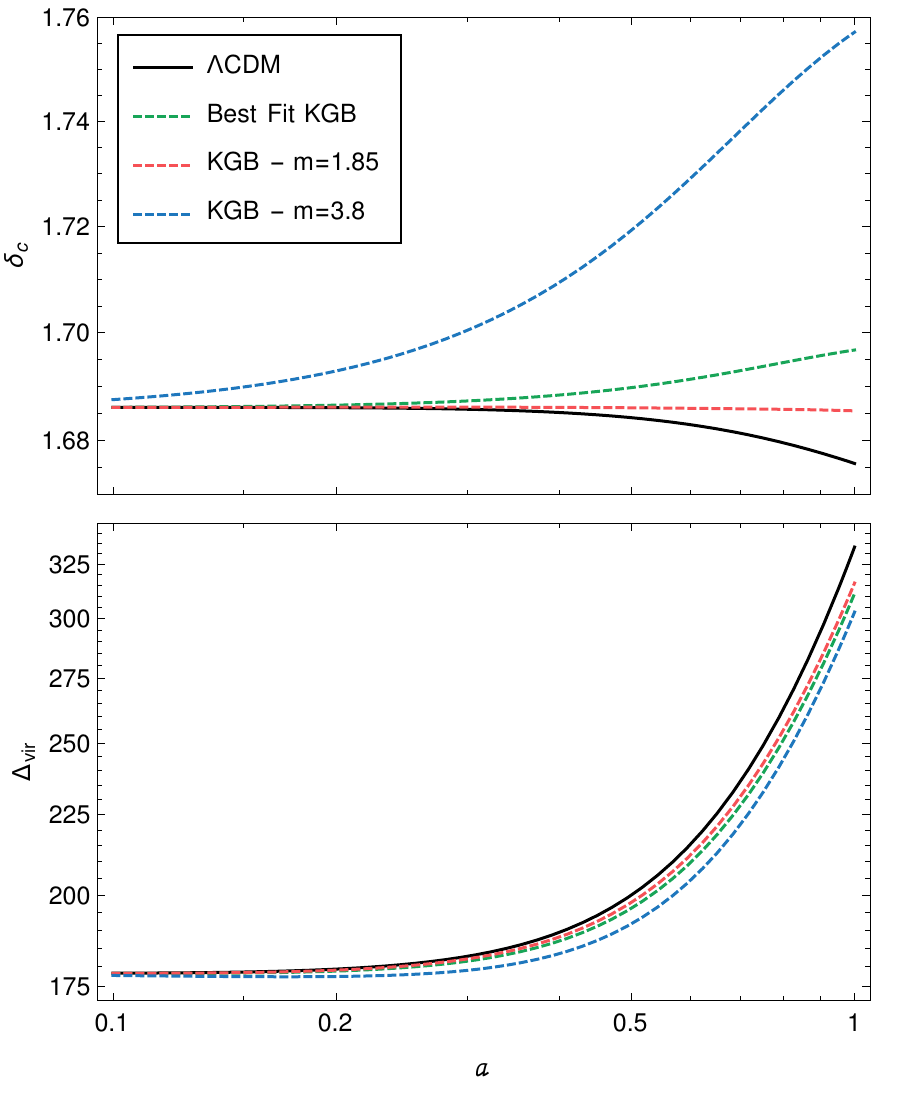}
\caption{Evolution of the critical density contrast $\delta_{\rm c}$ (top panel) and of the virial overdensity $\Delta_{\rm vir}$ (bottom panel) as a function of the scale factor $a$ for the three KGB-mimic models with varying $m$. The $\Lambda$CDM model (black solid line) is shown for comparison.}
\label{fig:deltasKGBm}
\end{figure}

We then discuss the values of $\delta_{\rm c}$ and $\Delta_{\rm vir}$ at present time as well as $a_{\rm vir}$ in \autoref{tab:KGBmquantities}, and we plot the time evolution of the linear critical density and virial overdensity in \autoref{fig:deltasKGBm}. In the top panel, we observe that models with large $m$, such as $m=3.8$, have a critical density larger than the Einstein-de Sitter value already at earlier times ($a\sim 0.1$) and then it monotonically increases up to the current value $\delta_{\rm c}(a=1)=1.76$. Decreasing the value of $m$ leads to a $\delta_{\rm c}$ close to the $\Lambda$CDM value at early times, while starting to monotonically grow only at later times. We can relate this behaviour with the evolution of the linear gravitational coupling: an earlier deviation of $\mu^{\rm L}$ from unity translates into an earlier departure of the critical density from $\Lambda$CDM. Furthermore, a larger deviation of $\mu^{\rm L}$ from the GR limit at present time corresponds to a larger value of $\delta_{\rm c}$ at $a=1$, as quantitatively reported in \autoref{tab:KGBmquantities}.

The virial overdensity $\Delta_{\rm vir}$ (bottom panel of \autoref{fig:deltasKGBm}) shows that the KGB-mimic cases have smaller values than the standard model at all times. The suppression in $\Delta_{\rm vir}$ is larger for increasing values of $m$. The model with the smallest value of this parameter, $m=1.85$, has $\Delta_{\rm vir} (a=1) = 316.0$, which is the closest value to the $\Lambda$CDM. Alternatively, for the model $m=3.8$ we find $\Delta_{\rm vir}(a=1)=302.8$, which is the smallest among all the KGB models.

\begin{figure}[t!]
\includegraphics[width=.48\textwidth]{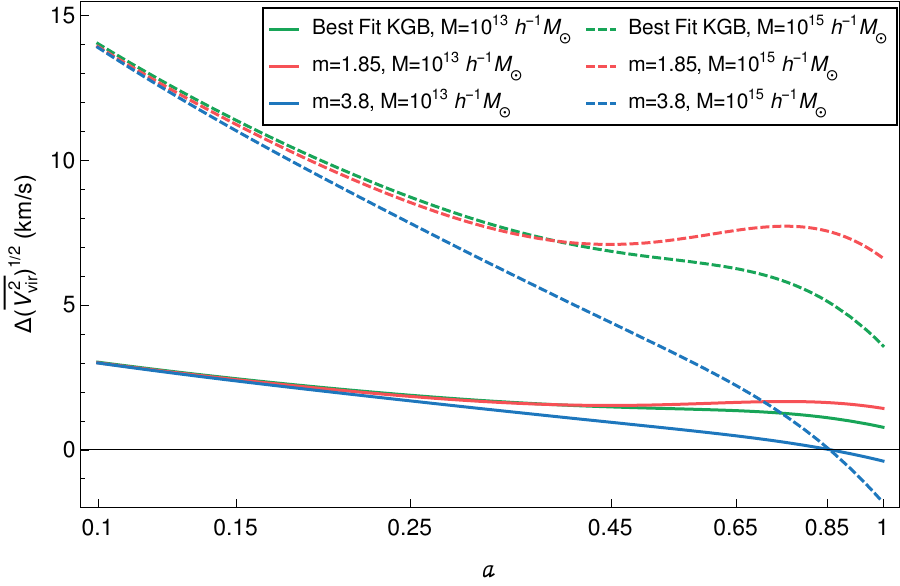}
\caption{Time evolution of the difference between the root-mean-square virial velocity $(\overline{V_\mathrm{vir}^2})^{1/2}$ of three KGB-mimic models with varying $m$ and the corresponding value for the $\Lambda$CDM model considering halos with masses $M = 10^{13} \, h^{-1} \mathrm{M}_\odot$ (solid lines) and $M = 10^{15} \, h^{-1} \mathrm{M}_\odot$ (dashed lines).}
\label{fig:deltaVvirKGBm}
\end{figure}

The root-mean-square virial velocity $(\overline{V_\mathrm{vir}})^{1/2}$ as function of time and mass is quantitatively and qualitatively very similar to that of models considered in the previous section. As shown in \autoref{fig:deltaVvirKGBm}, it varies by 1 to 15~km~s$^{-1}$ depending on epoch and mass with respect to $\Lambda$CDM.

In conclusion, we find that varying $m$ while keeping all other parameters fixed has a large impact on the magnitude of the different physical quantities, but not on their time evolution. The larger the value of $m$, the biggest the deviations we find from the standard $\Lambda$CDM model. Similarly to the case where we vary $\alpha_{B,0}$, small values of $m$ seem to bring the KGB-mimic closer to $\Lambda$CDM.

\section{The halo mass function} \label{Sec:V}

The study of the mass function is critical for precision cosmology. Theoretical predictions for the halo mass function are key ingredients to model the galaxy formation \cite{White:1991mr,Somerville:2014ika}, constrain cosmological parameters using the abundance of galaxy clusters \cite{WangSteinhardt1998,Majumdar:2002hd,Allen:2011zs,Planck:2013lkt,DES:2020mlx}, estimate the halo merger rate \cite{Lacey:1994su,Cohn:2000cm,Giocoli:2007gf,Ali-Haimoud:2017rtz,Fakhry:2020plg}, and explore the $\sigma_8$-tension \cite{Gu:2023jef} using  gravitational lensing \cite{WeinbergKamionkowski2003,Bartelmann:2010fz,Massey:2010hh}. To this purpose, we study the effects of shift-symmetric Horndeski models on the abundance of bound structures, or halos. Similar analyses have been already performed \cite{Kimura:2010di,Bellini:2012qn,Barreira:2013eea,Frusciante:2020zfs} but for specific shift-symmetric models; here we use the general parametrization introduced in Eq.~\eqref{eq:alphaKGB}.
 
For this study, we adopt the Sheth-Tormen prescription \cite{Sheth:1998ew,Sheth:1999su,Sheth:2001dp,Murray:2013qza}, according to which the differential mass function is defined by
\begin{eqnarray}
    \frac{\mathrm{d}n}{\mathrm{d}M} = &-& \sqrt{\frac{2 \tilde{a}}{\pi}} A \left[ 1 + \left( \frac{\tilde{a}\, \delta_{\rm c}^2}{ \sigma_{M}^2} \right)^{-p} \right] \frac{\rho_m}{M^2} \frac{\delta_{\rm c}}{ \sigma_M} \nonumber \\
    &\times& \frac{d \ln{\sigma_M}}{d \ln{M}} \, \exp {\left( - \frac{\tilde{a}\, \delta_{\rm c}^2}{2 \sigma_M^2}\right)} \,, \label{eq:massfunction}
\end{eqnarray}
where $\tilde{a} = 0.707$, $A=0.2162$ and $p=0.3$ are numerical factors and $\sigma_M$ is the variance of the linear matter power spectrum defined as
\begin{equation}
 \sigma_M^2 = \frac{1}{2 \pi^2} \int_0^{\infty} \mathrm{d}k \, k^2 W^2( k R ) P^{\rm L}(k, z) \,.
\end{equation}
The linear matter power spectrum $P^{\rm L} (k,z)$ is obtained from a modified version of the publicly available Einstein-Boltzmann solver \texttt{EFTCAMB}\footnote{Web page: \url{http://www.eftcamb.org}}  \cite{Hu:2013twa}, and the window function $W(x)=3x^{-3}(\sin x-x\cos x)$ is the Fourier transform of the top-hat filter. Finally, $R$ denotes the comoving radius enclosing the mass $M = (4\pi/3) \rho_{{\rm m},0} R^3$. Note that in this case the mass of the halo is the observable and it is assumed to be the same for all the models. Nevertheless, since we evaluate the virial overdensity within the formalism of the spherical collapse model, this mass corresponds to the virial mass of the halo and properly accounts for the effects of KGB.

\begin{figure}[t!]
  \centering
  \includegraphics[scale=0.56]{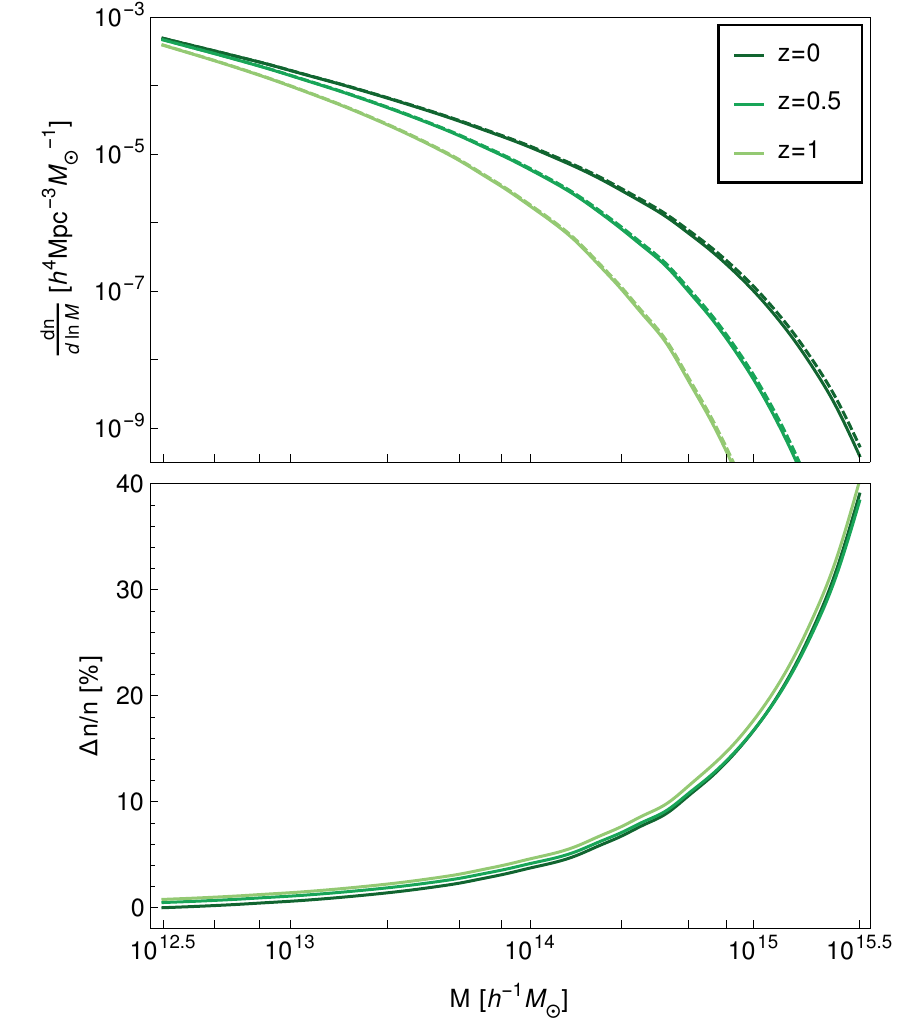}
\caption{Top panel: differential mass function at redshift $z=0$, $z=0.5$ and $z=1$, for the Best Fit KGB model (dashed lines) and $\Lambda$CDM (solid lines in top panel). Bottom panel: relative difference between the Best Fit KGB model and $\Lambda$CDM for the same values of $z$.}
\label{fig:MFforLKGBdiffz}
\end{figure}

\begin{figure*}[t!]
  \centering
  \includegraphics[scale=0.56]{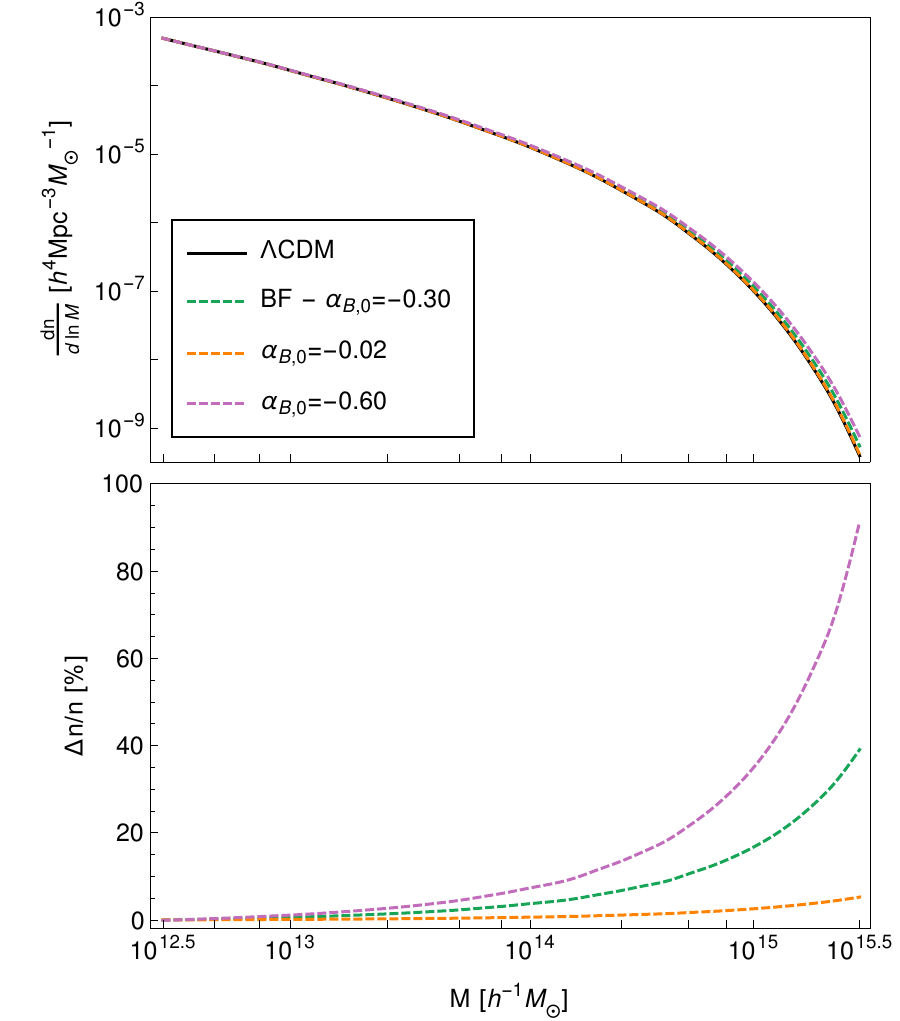}
  \includegraphics[scale=0.56]{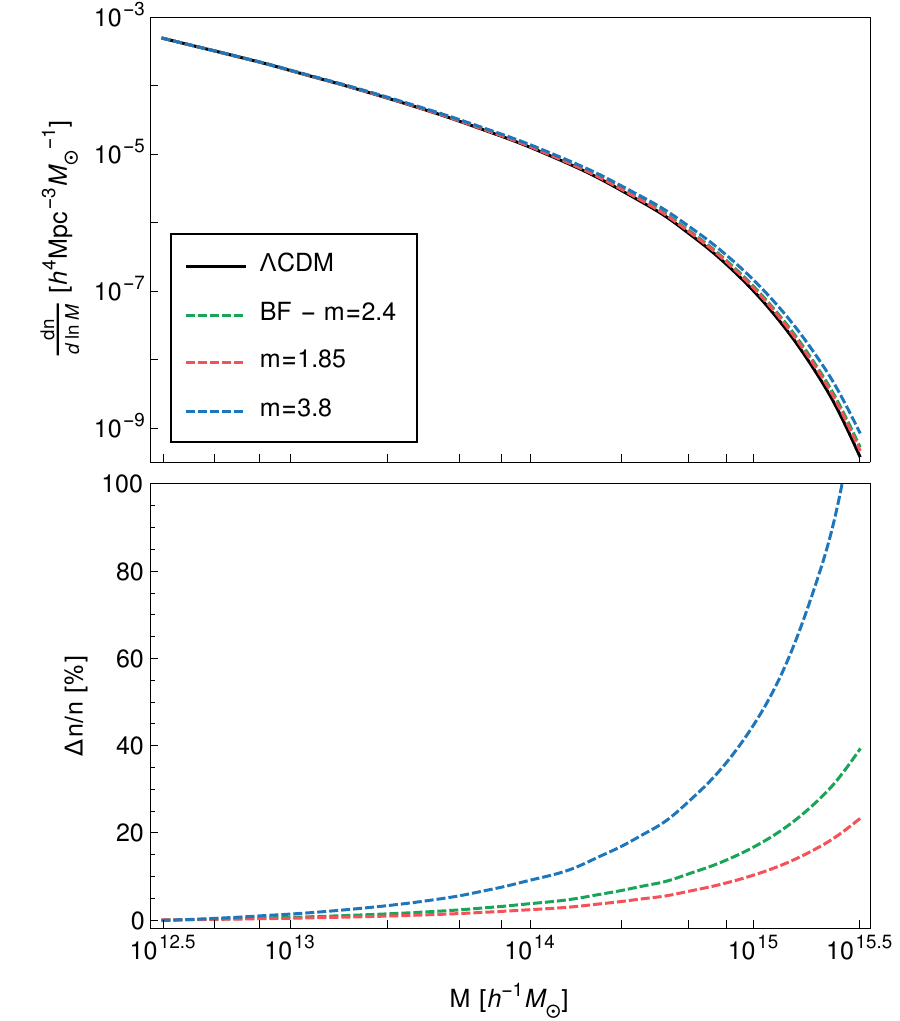}
\caption{Top: differential mass function as a function of the halo mass $M$ at $z=0$ for the KGB-mimic models with varying $\alpha_{B,0}$ (left panels) and varying $m$ (right panels), including the Best Fit KGB model (quoted as BF) and compared to $\Lambda$CDM (black solid line). Bottom: relative difference between KGB-mimic models and $\Lambda$CDM (same line style).}
\label{fig:MFforvaralpha-varm}
\end{figure*}

The differential mass function is shown in \autoref{fig:MFforLKGBdiffz} for the Best Fit KGB model at three redshifts, $z=0$, $z=0.5$ and $z=1$, and in Figure~\ref{fig:MFforvaralpha-varm} for the other KGB models with varying $\alpha_{B,0}$ and $m$ at $z=0$. In all these figures the curves are compared with the $\Lambda$CDM model, including the relative differences $\Delta n/n$ (bottom panels). The halo mass range $10^{12.5}~h^{-1} \mathrm{M}_\odot <M<10^{15.5}~h^{-1} \mathrm{M}_\odot$ encompasses the typical values from galaxies to clusters of galaxies, where the effects of the KGB-mimic models are potentially more important.

\autoref{fig:MFforLKGBdiffz} shows that the mass function strongly depends on redshift, as expected because of the exponential cut-off depending on $\delta_{\rm c}$. On the other hand, the redshift dependence of the relative difference is very small. However, while for masses smaller than $M \sim 10^{14} \, h^{-1} \mathrm{M}_\odot $ the difference is $\lesssim 5\%$, for higher masses it is in general very large, reaching $39\%$ at $z=0$, $38.4\%$ at $z=0.5$, and $\sim 40\%$ at $z=1$ for halos with $M\sim 10^{15} \, h^{-1} \mathrm{M}_\odot $.

The reason why the KGB model has an enhanced mass function compared to the standard scenario is related to the exponential factor in the mass function, Eq.~\eqref{eq:massfunction}, which depends on $\delta_{\rm c}$ in the numerator and on the mass variance $\sigma_M$ in the denominator. Therefore, a higher $\delta_{\rm c}$ leads to a suppression of the mass function, on the contrary higher values of $\sigma_M$ work towards an enhancement. As discussed in Figures \ref{fig:deltasKGBalpha} and \ref{fig:deltasKGBm}, the critical density for the Best Fit KGB model is enhanced w.r.t.\ $\Lambda$CDM. However, we have verified that the differences w.r.t.\ this model due to $\sigma_M$ are larger than those due to $\delta_{\rm c}$ for all three redshifts as well. Therefore, since $\sigma_M (\text{Best Fit}) > \sigma_M (\Lambda\text{CDM})$, we observe an enhancement in the number of objects.

\begin{figure*}[t!]
  \centering
  \includegraphics[scale=0.51]{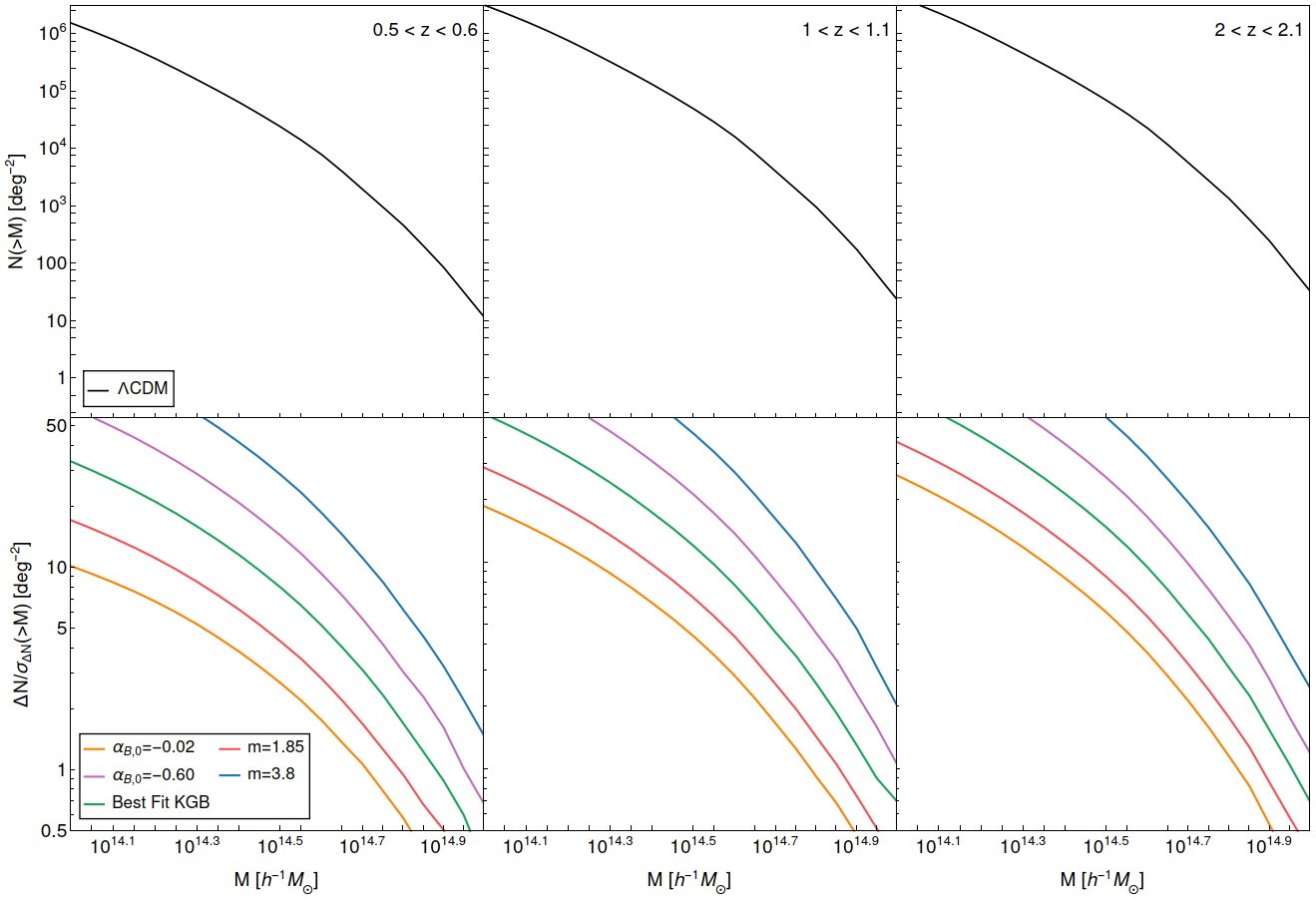}
\caption{Top panels: number of halos with mass larger than $M$ in the $\Lambda$CDM model, per square degree and in three redshift bins as indicated. Bottom panels: difference of counts $N(>M)$ between KGB-mimic models and $\Lambda$CDM in units of the uncertainty of the difference, computed supposing Poisson errors and quadratic propagation.}
\label{fig:MFtailcounts}
\end{figure*}

In Figure~\ref{fig:MFforvaralpha-varm} the differential mass function is shown at $z=0$ for different values of $\alpha_{B,0}$ and $m$ respectively. We find a very similar behavior as the one described for the case of the Best Fit KGB. Note that the larger $|\alpha_{B,0}|$ or $m$, the larger the difference w.r.t.\ $\Lambda$CDM, especially for halos with masses typical of the largest galaxy groups and galaxy clusters, $M\gtrsim  10^{14} \, h^{-1}\mathrm{M}_\odot$. For very massive halos, $M\sim 10^{15.5} \, h^{-1} \mathrm{M}_\odot$, their number density increases by $\sim 91.2\%$ for $|\alpha_{B,0}|=0.6$ and by more than a factor of 2 ($\sim 121.9 \%$) for $m=3.8$.

The large deviation at the high-mass end caused by the exponential tail of the mass function does not indicate, however, that the abundance of very massive halos is a promising probe to distinguish the KGB-mimic models from $\Lambda$CDM. Supposing similar values for counts and Poisson errors, the uncertainty on $\Delta n/n$ is $\sim\sqrt{2/n}$, i.e., exponentially increasing with $M$. To assess this question quantitatively, it is useful to consider the number counts of halos with mass exceeding a given threshold, $N(>M)$. Figure~\ref{fig:MFtailcounts} and Table~\ref{tab:counts} report the results per square degree limited to the mass range $10^{14} \, h^{-1} \mathrm{M}_\odot < M < 10^{15} \, h^{-1} \mathrm{M}_\odot$, in three redshift bins sampling the typical range of existing and forthcoming surveys. Actual counts are shown only for the reference $\Lambda$CDM model (top panels), then reporting the difference of counts $\Delta N$ between the KGB-mimic models and $\Lambda$CDM in units of its error $\sigma_{\Delta N}$ (bottom panels). Poisson errors are considered for simplicity. As expected, the statistical significance of the difference, $\Delta N/\sigma_{\Delta N}$, increases as long as smaller halos are included in the counting process. Focusing on the very massive objects with mass $M\gtrsim10^{15} \, h^{-1} \mathrm{M}_\odot$, which according to the hierarchical scenario are the last formed and then less affected by small-scale instabilities perturbing their spheroidal shape, the statistical significance of counts is marginal, though sufficient to distinguish some of the KGB-mimic models, in particular those with largest value of $\lvert \alpha_{B,0} \rvert$ and $m$, especially at large redshift.

In conclusion, as long as the spherical collapse model provides an accurate description of the nonlinear clustering, not only the extreme KGB-mimic models but also the Best Fit KGB potentially leave significant distinguishable signatures in the halo mass function compared to $\Lambda$CDM. This analysis can therefore be considered as a very promising tool to test the KGB models against the standard cosmology.

\renewcommand{\arraystretch}{1.2}
\begin{table*}\label{tab:counts}
 \caption{Number counts $N(>M)$ per square degree in $\Lambda$CDM (column 2) and difference of counts $\Delta N(>M)$ between KGB-mimic models and $\Lambda$CDM (columns 3-7), in the redshift range $0.5<z<6$ (upper bloc), $1<z<1.1$ (central), and $2<z<2.1$ (lower). Errors are computed assuming Poisson statistics and quadratic propagation (indicated in parenthesis).}
 \centering
\begin{tabular}{|c|c|ccccc|}
\hline
$\log(M/h^{-1} \mathrm{M}_\odot)$ &
$N_{\Lambda{\rm}CDM}$ &
$\Delta N_\mathrm{BF}$ &
$\Delta N_{\alpha_{B,0}=-0.02}$ &
$\Delta N_{\alpha_{B,0}=-0.60}$ &
$\Delta N_{m=1.85}$ &
$\Delta N_{m=3.8}$ \\
\hline\hline
\multicolumn{7}{|c|}{$0.5<z<6$} \\
\hline
 14.0 & 1574813~(1255) & 73636~(1795) & 22244~(1781) & 136702~(1813) & 37562~(1785) & 277343~(1851) \\
 14.1 & 802880~(896) & 42389~(1284) & 13168~(1272) & 78301~(1298) & 21915~(1276) & 158874~(1329) \\
 14.2 & 372909~(611) & 22629~(877) & 7216~(868) & 41610~(888) & 11853~(870) & 84526~(911) \\
 14.3 & 158872~(399) & 11144~(574) & 3630~(567) & 20420~(581) & 5903~(569) & 41579~(600) \\
 14.4 & 64628~(254) & 5147~(366) & 1700~(362) & 9413~(372) & 2748~(363) & 19234~(386) \\
 14.5 & 24192~(156) & 2185~(225) & 729~(222) & 3993~(229) & 1174~(223) & 8196~(238) \\
 14.6 & 7899~(89) & 810~(129) & 272~(127) & 1481~(132) & 438~(127) & 3059~(138) \\
 14.7 & 1964~(44) & 242~(64) & 82~(63) & 442~(66) & 131~(64) & 922~(70) \\
 14.8 & 468~(22) & 66~(32) & 22~(31 & 121~(33) & 36~(31) & 256~(35) \\
 14.9 & 86~(9) & 14~(13) & 5~(13) & 27~(14) & 8~(13) & 57~(15) \\
 15.0 & 12~(3) & 2~(5) & 1~(5) & 4~(5) & 1~(5) & 10~(6) \\
 15.1 & 1~(1) & 0~(1) & 0~(1) & 1~(1) & 0~(1) & 1~(1) \\
 \hline\hline
 \multicolumn{7}{|c|}{$1.0<z<1.1$} \\
 \hline
  14.0 & 3238361~(1800) & 164718~(2578) & 58625~(2557) & 294913~(2603) & 90247~(2563) & 585255~(2657) \\
 14.1 & 1650999~(1285) & 93986~(1843) & 33661~(1826) & 168122~(1863) & 51719~(1831) & 334459~(1907) \\
 14.2 & 766829~(876) & 49725~(1259) & 17905~(1246) & 88910~(1274) & 27478~(1250) & 177505~(1308) \\
 14.3 & 326697~(572) & 24285~(823) & 8774~(814) & 43435~(835) & 13467~(817) & 87117~(861) \\
 14.4 & 132898~(365) & 11146~(527) & 4030~(520) & 19954~(535) & 6195~(522) & 40228~(553) \\
 14.5 & 49747~(223) & 4706~(323) & 1700~(318) & 8438~(328) & 2620~(320) & 17115~(342) \\
 14.6 & 16243~(127) & 1737~(185) & 625~(182) & 3121~(188) & 967~(182) & 6379~(197) \\
 14.7 & 4039~(64) & 515~(93) & 184~(91) & 928~(95) & 287~(92) & 1918~(100) \\
 14.8 & 962~(31) & 141~(45) & 50~(45) & 255~(47) & 78~(45) & 533~(50) \\
 14.9 & 177~(13) & 30~(19) & 11~(19) & 55~(20) & 17~(19) & 118~(21) \\
 15.0 & 24~(5) & 6~(7) & 2~(7) & 10~(8) & 3~(7) & 21~(9) \\
 15.1 & 2~(1) & 1~(2) & 0~(1) & 1~(2) & 0~(1) & 3~(2)
\\
 \hline\hline
 \multicolumn{7}{|c|}{$2.0<z<2.1$} \\
 \hline
  14.0 & 4512344~(2124) & 246019~(3044) & 97674~(3020) & 428064~(3075) & 141889~(3027) & 834035~(3140) \\
 14.1 & 2300508~(1517) & 139421~(2177) & 55072~(2158) & 243081~(2201) & 80321~(2164) & 475663~(2253) \\
 14.2 & 1068503~(1034) & 73245~(1487) & 28753~(1472) & 128036~(1505) & 42139~(1477) & 251914~(1546) \\
 14.3 & 455220~(675) & 35542~(973) & 13852~(962) & 62318~(986) & 20414~(965) & 123396~(1017) \\
 14.4 & 185181~(430) & 16229~(622) & 6278~(614) & 28544~(631) & 9306~(616) & 56893~(653) \\
 14.5 & 69318~(263) & 6821~(381) & 2617~(375) & 12039~(388) & 3904~(378) & 24172~(403) \\
 14.6 & 22633~(150) & 2507~(219) & 953~(215) & 4443~(223) & 1431~(216) & 8998~(233) \\
 14.7 & 5628~(75) & 739~(110) & 277~(107) & 1317~(112) & 420~(108) & 2702~(118) \\
 14.8 & 1341~(37) & 201~(54) & 74~(53) & 360~(55) & 114~(53) & 749~(59) \\
 14.9 & 247~(16) & 43~(23) & 15~(23) & 78~(24) & 24~(23) & 165~(26) \\
 15.0 & 34~(6) & 7~(8) & 3~(8) & 13~(9) & 4~(8) & 29~(10) \\
 15.1 & 3~(2) & 1~(3) & 0~(3) & 1~(3) & 0~(3) & 3~(3) \\
 \hline
\end{tabular}
\end{table*}

\section{Conclusion} \label{Sec:VI}

This study examines the nonlinear growth of bound cosmological structures in a general parametrization of shift-symmetric Galileon models, by means of the spherical collapse model and deducing theoretical predictions for the abundance of halos. The use of a general parametrization for shift-symmetric Galileon models, dubbed KGB, identifies physical quantities characteristic of the spherical collapse that are common to such a large class of models, supporting the prediction and interpretation of observables using a small parameter space that is well-suited to explore deviations from the standard $\Lambda$CDM cosmology. In particular, the parameterized model had four free parameters, two affecting only the background ($w_0,w_a$) and two concerning the perturbations ($\alpha_{B,0},m$). Specifically, we have analyzed only the modifications in the gravitational interaction introduced by $\alpha_{B,0}$ and $m$ by considering as a starting point a Best Fit KGB model very close to $\Lambda$CDM and proved to describe current data from CMB, BAO, RSD, and SnIa. Additionally, we have also presented results for four KGB-mimic models in which $\alpha_{B,0}$ and $m$ assume different values from the best fit. 

We have found that the theoretical predictions in the growth of structures and in the spherical collapse are quite different from the standard scenario, especially for larger values of $|\alpha_{B,0}|$ and $m$. This is particularly evident for the critical density, $\delta_{\rm c}$, and for the virial overdensity, $\Delta_{\rm vir}$, shown in Figures~\ref{fig:deltasKGBalpha}-\ref{fig:deltasKGBm}. For the largest amplitude of the braiding function, $\alpha_{B,0}=-0.60$, at present time $\delta_{\rm c}$ and $\Delta_{\rm vir}$ differ respectively by $\sim 3.2\%$ and $\sim 13\%$ from the standard $\Lambda$CDM model; likewise, for the largest slope of the braiding function, $m=3.8$, the values of $\delta_{\rm c}$ and $\Delta_{\rm vir}$ differ from the standard $\Lambda$CDM model respectively by $\sim 5\%$ and $\sim 9\%$. The following mean-squared virial velocity shown in Figures~\ref{fig:deltaVvirKGBalpha}-\ref{fig:deltaVvirKGBm} is quantitatively similar to the value expected for the $\Lambda$CDM model, with a non-trivial decreasing evolution with time depending on the KGB-mimic model, though predicting variations of no more than 15~km~s$^{-1}$ for the most massive halos at very high redshift.

Following the Sheth-Tormen prescription, we have used the spherical collapse predictions to deduce the halo mass function and the tail distribution of number counts in three specific redshift intervals. As shown in Figure~\ref{fig:MFforvaralpha-varm}, the number density of halos per unit mass in the Best Fit KGB model crucially differs from the $\Lambda$CDM for systems with size typical of galaxy groups and clusters, having mass $M\gtrsim 10^{14} \, h^{-1} \mathrm{M}_\odot$. This difference can be as large as $\sim 40\%$ depending on redshift. For larger values of $|\alpha_{B,0}|$ and $m$ this difference can be even larger, reaching $\sim 90\%$ for $|\alpha_{B,0}|=0.60$ and $\sim 120\%$ for $m=3.8$ at present time. Indeed, if supposing Poisson statistics, the counts of very massive and rare halos are significantly uncertain. Nonetheless, the statistical significance of the survival counts shown in Figure~\ref{fig:MFtailcounts}, suggests that using counts of objects at redshift $z>0.5$ with mass larger than $8 \times 10^{14} \, h^{-1} \mathrm{M}_\odot$, namely large galaxy groups and small-to-intermediate galaxy clusters, one could distinguish not only the more extreme KGB but also the Best Fit KGB model from $\Lambda$CDM.

This study shows very peculiar and measurable features that can help in discriminating between the shift-symmetric model and the standard scenario. These results will be very useful to model the nonlinear matter power spectrum entering all the usual observables including gravitational lensing, which require nonlinear corrections for the screening mechanisms. An upcoming study is quantifying the observational effect of KGB-mimic models by forecasting the number counts of lensing convergence peaks for surveys such as the DESI Legacy Imaging Surveys\footnote{\url{http://legacysurvey.org}} and DES\footnote{\url{https://www.darkenergysurvey.org}} and forthcoming surveys operated by Vera Rubin Observatory \cite{LSST:2008ijt} and Euclid \cite{EUCLID:2011zbd}.

\acknowledgments
The research of I.S.A. has received funding from the Funda\c{c}\~{a}o para a Ci\^{e}ncia e a Tecnologia (FCT) PhD fellowship grant with ref.\ number 2020.07237.BD.
N.F. is supported by the Italian Ministry of University and Research (MUR) through the Rita Levi Montalcini project ``Tests of gravity on cosmic scales" with reference PGR19ILFGP.
F.P. is supported by the INFN grant InDark and the Departments of Excellence grant L.232/2016 of the Italian Ministry of University and Research (MUR).
I.S.A., N.F. and F.P. also acknowledge the FCT project with ref.\ number PTDC/FIS-AST/0054/2021. 
C.S. received support from the French government under the France 2030 investment plan, as part of the Excellence Initiative of Aix-Marseille University - A*MIDEX (AMX-19-IET-008 - IPhU).

This article is based upon work from COST Action CA21136 Addressing observational tensions in cosmology with systematics and fundamental physics (CosmoVerse) supported by COST (European Cooperation in Science and Technology).

\bibliography{Bib_entries}

\begin{thebibliography}{115}%
\makeatletter
\providecommand \@ifxundefined [1]{%
 \@ifx{#1\undefined}
}%
\providecommand \@ifnum [1]{%
 \ifnum #1\expandafter \@firstoftwo
 \else \expandafter \@secondoftwo
 \fi
}%
\providecommand \@ifx [1]{%
 \ifx #1\expandafter \@firstoftwo
 \else \expandafter \@secondoftwo
 \fi
}%
\providecommand \natexlab [1]{#1}%
\providecommand \enquote  [1]{``#1''}%
\providecommand \bibnamefont  [1]{#1}%
\providecommand \bibfnamefont [1]{#1}%
\providecommand \citenamefont [1]{#1}%
\providecommand \href@noop [0]{\@secondoftwo}%
\providecommand \href [0]{\begingroup \@sanitize@url \@href}%
\providecommand \@href[1]{\@@startlink{#1}\@@href}%
\providecommand \@@href[1]{\endgroup#1\@@endlink}%
\providecommand \@sanitize@url [0]{\catcode `\\12\catcode `\$12\catcode
  `\&12\catcode `\#12\catcode `\^12\catcode `\_12\catcode `\%12\relax}%
\providecommand \@@startlink[1]{}%
\providecommand \@@endlink[0]{}%
\providecommand \url  [0]{\begingroup\@sanitize@url \@url }%
\providecommand \@url [1]{\endgroup\@href {#1}{\urlprefix }}%
\providecommand \urlprefix  [0]{URL }%
\providecommand \Eprint [0]{\href }%
\providecommand \doibase [0]{http://dx.doi.org/}%
\providecommand \selectlanguage [0]{\@gobble}%
\providecommand \bibinfo  [0]{\@secondoftwo}%
\providecommand \bibfield  [0]{\@secondoftwo}%
\providecommand \translation [1]{[#1]}%
\providecommand \BibitemOpen [0]{}%
\providecommand \bibitemStop [0]{}%
\providecommand \bibitemNoStop [0]{.\EOS\space}%
\providecommand \EOS [0]{\spacefactor3000\relax}%
\providecommand \BibitemShut  [1]{\csname bibitem#1\endcsname}%
\let\auto@bib@innerbib\@empty
\bibitem [{\citenamefont {Weinberg}(1989)}]{Weinberg:1988cp}%
  \BibitemOpen
  \bibfield  {author} {\bibinfo {author} {\bibfnamefont {S.}~\bibnamefont
  {Weinberg}},\ }\href {\doibase 10.1103/RevModPhys.61.1} {\bibfield  {journal}
  {\bibinfo  {journal} {Rev. Mod. Phys.}\ }\textbf {\bibinfo {volume} {61}},\
  \bibinfo {pages} {1} (\bibinfo {year} {1989})}\BibitemShut {NoStop}%
\bibitem [{\citenamefont {Carroll}(2001)}]{Carroll:2000fy}%
  \BibitemOpen
  \bibfield  {author} {\bibinfo {author} {\bibfnamefont {S.~M.}\ \bibnamefont
  {Carroll}},\ }\href {\doibase 10.12942/lrr-2001-1} {\bibfield  {journal}
  {\bibinfo  {journal} {Living Rev. Rel.}\ }\textbf {\bibinfo {volume} {4}},\
  \bibinfo {pages} {1} (\bibinfo {year} {2001})},\ \Eprint
  {http://arxiv.org/abs/astro-ph/0004075} {arXiv:astro-ph/0004075} \BibitemShut
  {NoStop}%
\bibitem [{\citenamefont {Velten}\ \emph {et~al.}(2014)\citenamefont {Velten},
  \citenamefont {vom Marttens},\ and\ \citenamefont
  {Zimdahl}}]{Velten:2014nra}%
  \BibitemOpen
  \bibfield  {author} {\bibinfo {author} {\bibfnamefont {H.~E.~S.}\
  \bibnamefont {Velten}}, \bibinfo {author} {\bibfnamefont {R.~F.}\
  \bibnamefont {vom Marttens}}, \ and\ \bibinfo {author} {\bibfnamefont
  {W.}~\bibnamefont {Zimdahl}},\ }\href {\doibase
  10.1140/epjc/s10052-014-3160-4} {\bibfield  {journal} {\bibinfo  {journal}
  {Eur. Phys. J. C}\ }\textbf {\bibinfo {volume} {74}},\ \bibinfo {pages}
  {3160} (\bibinfo {year} {2014})},\ \Eprint {http://arxiv.org/abs/1410.2509}
  {arXiv:1410.2509 [astro-ph.CO]} \BibitemShut {NoStop}%
\bibitem [{\citenamefont {Joyce}\ \emph {et~al.}(2015)\citenamefont {Joyce},
  \citenamefont {Jain}, \citenamefont {Khoury},\ and\ \citenamefont
  {Trodden}}]{Joyce:2014kja}%
  \BibitemOpen
  \bibfield  {author} {\bibinfo {author} {\bibfnamefont {A.}~\bibnamefont
  {Joyce}}, \bibinfo {author} {\bibfnamefont {B.}~\bibnamefont {Jain}},
  \bibinfo {author} {\bibfnamefont {J.}~\bibnamefont {Khoury}}, \ and\ \bibinfo
  {author} {\bibfnamefont {M.}~\bibnamefont {Trodden}},\ }\href {\doibase
  10.1016/j.physrep.2014.12.002} {\bibfield  {journal} {\bibinfo  {journal}
  {Phys. Rept.}\ }\textbf {\bibinfo {volume} {568}},\ \bibinfo {pages} {1}
  (\bibinfo {year} {2015})},\ \Eprint {http://arxiv.org/abs/1407.0059}
  {arXiv:1407.0059 [astro-ph.CO]} \BibitemShut {NoStop}%
\bibitem [{\citenamefont {Riess}\ \emph {et~al.}(2019)\citenamefont {Riess},
  \citenamefont {Casertano}, \citenamefont {Yuan}, \citenamefont {Macri},\ and\
  \citenamefont {Scolnic}}]{Riess:2019cxk}%
  \BibitemOpen
  \bibfield  {author} {\bibinfo {author} {\bibfnamefont {A.~G.}\ \bibnamefont
  {Riess}}, \bibinfo {author} {\bibfnamefont {S.}~\bibnamefont {Casertano}},
  \bibinfo {author} {\bibfnamefont {W.}~\bibnamefont {Yuan}}, \bibinfo {author}
  {\bibfnamefont {L.~M.}\ \bibnamefont {Macri}}, \ and\ \bibinfo {author}
  {\bibfnamefont {D.}~\bibnamefont {Scolnic}},\ }\href {\doibase
  10.3847/1538-4357/ab1422} {\bibfield  {journal} {\bibinfo  {journal}
  {Astrophys. J.}\ }\textbf {\bibinfo {volume} {876}},\ \bibinfo {pages} {85}
  (\bibinfo {year} {2019})},\ \Eprint {http://arxiv.org/abs/1903.07603}
  {arXiv:1903.07603 [astro-ph.CO]} \BibitemShut {NoStop}%
\bibitem [{\citenamefont {Wong}\ \emph {et~al.}(2020)\citenamefont {Wong} \emph
  {et~al.}}]{Wong:2019kwg}%
  \BibitemOpen
  \bibfield  {author} {\bibinfo {author} {\bibfnamefont {K.~C.}\ \bibnamefont
  {Wong}} \emph {et~al.},\ }\href {\doibase 10.1093/mnras/stz3094} {\bibfield
  {journal} {\bibinfo  {journal} {Mon. Not. Roy. Astron. Soc.}\ }\textbf
  {\bibinfo {volume} {498}},\ \bibinfo {pages} {1420} (\bibinfo {year}
  {2020})},\ \Eprint {http://arxiv.org/abs/1907.04869} {arXiv:1907.04869
  [astro-ph.CO]} \BibitemShut {NoStop}%
\bibitem [{\citenamefont {Freedman}\ \emph {et~al.}(2019)\citenamefont
  {Freedman} \emph {et~al.}}]{Freedman:2019jwv}%
  \BibitemOpen
  \bibfield  {author} {\bibinfo {author} {\bibfnamefont {W.~L.}\ \bibnamefont
  {Freedman}} \emph {et~al.},\ }\href {\doibase 10.3847/1538-4357/ab2f73}
  {\bibfield  {journal} {\bibinfo  {journal} {Astrophys. J.}\ }\textbf
  {\bibinfo {volume} {882}},\ \bibinfo {pages} {34} (\bibinfo {year} {2019})},\
  \Eprint {http://arxiv.org/abs/1907.05922} {arXiv:1907.05922 [astro-ph.CO]}
  \BibitemShut {NoStop}%
\bibitem [{\citenamefont {Di~Valentino}\ \emph
  {et~al.}(2021{\natexlab{a}})\citenamefont {Di~Valentino} \emph
  {et~al.}}]{DiValentino:2020zio}%
  \BibitemOpen
  \bibfield  {author} {\bibinfo {author} {\bibfnamefont {E.}~\bibnamefont
  {Di~Valentino}} \emph {et~al.},\ }\href {\doibase
  10.1016/j.astropartphys.2021.102605} {\bibfield  {journal} {\bibinfo
  {journal} {Astropart. Phys.}\ }\textbf {\bibinfo {volume} {131}},\ \bibinfo
  {pages} {102605} (\bibinfo {year} {2021}{\natexlab{a}})},\ \Eprint
  {http://arxiv.org/abs/2008.11284} {arXiv:2008.11284 [astro-ph.CO]}
  \BibitemShut {NoStop}%
\bibitem [{\citenamefont {Kuijken}\ \emph {et~al.}(2015)\citenamefont {Kuijken}
  \emph {et~al.}}]{Kuijken:2015vca}%
  \BibitemOpen
  \bibfield  {author} {\bibinfo {author} {\bibfnamefont {K.}~\bibnamefont
  {Kuijken}} \emph {et~al.},\ }\href {\doibase 10.1093/mnras/stv2140}
  {\bibfield  {journal} {\bibinfo  {journal} {Mon. Not. Roy. Astron. Soc.}\
  }\textbf {\bibinfo {volume} {454}},\ \bibinfo {pages} {3500} (\bibinfo {year}
  {2015})},\ \Eprint {http://arxiv.org/abs/1507.00738} {arXiv:1507.00738
  [astro-ph.CO]} \BibitemShut {NoStop}%
\bibitem [{\citenamefont {de~Jong}\ \emph {et~al.}(2015)\citenamefont {de~Jong}
  \emph {et~al.}}]{deJong:2015wca}%
  \BibitemOpen
  \bibfield  {author} {\bibinfo {author} {\bibfnamefont {J.~T.~A.}\
  \bibnamefont {de~Jong}} \emph {et~al.},\ }\href {\doibase
  10.1051/0004-6361/201526601} {\bibfield  {journal} {\bibinfo  {journal}
  {Astron. Astrophys.}\ }\textbf {\bibinfo {volume} {582}},\ \bibinfo {pages}
  {A62} (\bibinfo {year} {2015})},\ \Eprint {http://arxiv.org/abs/1507.00742}
  {arXiv:1507.00742 [astro-ph.CO]} \BibitemShut {NoStop}%
\bibitem [{\citenamefont {Hildebrandt}\ \emph {et~al.}(2017)\citenamefont
  {Hildebrandt} \emph {et~al.}}]{Hildebrandt:2016iqg}%
  \BibitemOpen
  \bibfield  {author} {\bibinfo {author} {\bibfnamefont {H.}~\bibnamefont
  {Hildebrandt}} \emph {et~al.},\ }\href {\doibase 10.1093/mnras/stw2805}
  {\bibfield  {journal} {\bibinfo  {journal} {Mon. Not. Roy. Astron. Soc.}\
  }\textbf {\bibinfo {volume} {465}},\ \bibinfo {pages} {1454} (\bibinfo {year}
  {2017})},\ \Eprint {http://arxiv.org/abs/1606.05338} {arXiv:1606.05338
  [astro-ph.CO]} \BibitemShut {NoStop}%
\bibitem [{\citenamefont {Di~Valentino}\ \emph
  {et~al.}(2021{\natexlab{b}})\citenamefont {Di~Valentino} \emph
  {et~al.}}]{DiValentino:2020vvd}%
  \BibitemOpen
  \bibfield  {author} {\bibinfo {author} {\bibfnamefont {E.}~\bibnamefont
  {Di~Valentino}} \emph {et~al.},\ }\href {\doibase
  10.1016/j.astropartphys.2021.102604} {\bibfield  {journal} {\bibinfo
  {journal} {Astropart. Phys.}\ }\textbf {\bibinfo {volume} {131}},\ \bibinfo
  {pages} {102604} (\bibinfo {year} {2021}{\natexlab{b}})},\ \Eprint
  {http://arxiv.org/abs/2008.11285} {arXiv:2008.11285 [astro-ph.CO]}
  \BibitemShut {NoStop}%
\bibitem [{\citenamefont {Lue}\ \emph {et~al.}(2004)\citenamefont {Lue},
  \citenamefont {Scoccimarro},\ and\ \citenamefont {Starkman}}]{Lue:2004rj}%
  \BibitemOpen
  \bibfield  {author} {\bibinfo {author} {\bibfnamefont {A.}~\bibnamefont
  {Lue}}, \bibinfo {author} {\bibfnamefont {R.}~\bibnamefont {Scoccimarro}}, \
  and\ \bibinfo {author} {\bibfnamefont {G.~D.}\ \bibnamefont {Starkman}},\
  }\href {\doibase 10.1103/PhysRevD.69.124015} {\bibfield  {journal} {\bibinfo
  {journal} {Phys. Rev. D}\ }\textbf {\bibinfo {volume} {69}},\ \bibinfo
  {pages} {124015} (\bibinfo {year} {2004})},\ \Eprint
  {http://arxiv.org/abs/astro-ph/0401515} {arXiv:astro-ph/0401515} \BibitemShut
  {NoStop}%
\bibitem [{\citenamefont {Copeland}\ \emph {et~al.}(2006)\citenamefont
  {Copeland}, \citenamefont {Sami},\ and\ \citenamefont
  {Tsujikawa}}]{Copeland:2006wr}%
  \BibitemOpen
  \bibfield  {author} {\bibinfo {author} {\bibfnamefont {E.~J.}\ \bibnamefont
  {Copeland}}, \bibinfo {author} {\bibfnamefont {M.}~\bibnamefont {Sami}}, \
  and\ \bibinfo {author} {\bibfnamefont {S.}~\bibnamefont {Tsujikawa}},\ }\href
  {\doibase 10.1142/S021827180600942X} {\bibfield  {journal} {\bibinfo
  {journal} {Int. J. Mod. Phys. D}\ }\textbf {\bibinfo {volume} {15}},\
  \bibinfo {pages} {1753} (\bibinfo {year} {2006})},\ \Eprint
  {http://arxiv.org/abs/hep-th/0603057} {arXiv:hep-th/0603057} \BibitemShut
  {NoStop}%
\bibitem [{\citenamefont {Silvestri}\ and\ \citenamefont
  {Trodden}(2009)}]{Silvestri:2009hh}%
  \BibitemOpen
  \bibfield  {author} {\bibinfo {author} {\bibfnamefont {A.}~\bibnamefont
  {Silvestri}}\ and\ \bibinfo {author} {\bibfnamefont {M.}~\bibnamefont
  {Trodden}},\ }\href {\doibase 10.1088/0034-4885/72/9/096901} {\bibfield
  {journal} {\bibinfo  {journal} {Rept. Prog. Phys.}\ }\textbf {\bibinfo
  {volume} {72}},\ \bibinfo {pages} {096901} (\bibinfo {year} {2009})},\
  \Eprint {http://arxiv.org/abs/0904.0024} {arXiv:0904.0024 [astro-ph.CO]}
  \BibitemShut {NoStop}%
\bibitem [{\citenamefont {Nojiri}\ and\ \citenamefont
  {Odintsov}(2011)}]{Nojiri:2010wj}%
  \BibitemOpen
  \bibfield  {author} {\bibinfo {author} {\bibfnamefont {S.}~\bibnamefont
  {Nojiri}}\ and\ \bibinfo {author} {\bibfnamefont {S.~D.}\ \bibnamefont
  {Odintsov}},\ }\href {\doibase 10.1016/j.physrep.2011.04.001} {\bibfield
  {journal} {\bibinfo  {journal} {Phys. Rept.}\ }\textbf {\bibinfo {volume}
  {505}},\ \bibinfo {pages} {59} (\bibinfo {year} {2011})},\ \Eprint
  {http://arxiv.org/abs/1011.0544} {arXiv:1011.0544 [gr-qc]} \BibitemShut
  {NoStop}%
\bibitem [{\citenamefont {Tsujikawa}(2010)}]{Tsujikawa:2010zza}%
  \BibitemOpen
  \bibfield  {author} {\bibinfo {author} {\bibfnamefont {S.}~\bibnamefont
  {Tsujikawa}},\ }\href {\doibase 10.1007/978-3-642-10598-2_3} {\bibfield
  {journal} {\bibinfo  {journal} {Lect. Notes Phys.}\ }\textbf {\bibinfo
  {volume} {800}},\ \bibinfo {pages} {99} (\bibinfo {year} {2010})},\ \Eprint
  {http://arxiv.org/abs/1101.0191} {arXiv:1101.0191 [gr-qc]} \BibitemShut
  {NoStop}%
\bibitem [{\citenamefont {Capozziello}\ and\ \citenamefont
  {De~Laurentis}(2011)}]{Capozziello:2011et}%
  \BibitemOpen
  \bibfield  {author} {\bibinfo {author} {\bibfnamefont {S.}~\bibnamefont
  {Capozziello}}\ and\ \bibinfo {author} {\bibfnamefont {M.}~\bibnamefont
  {De~Laurentis}},\ }\href {\doibase 10.1016/j.physrep.2011.09.003} {\bibfield
  {journal} {\bibinfo  {journal} {Phys. Rept.}\ }\textbf {\bibinfo {volume}
  {509}},\ \bibinfo {pages} {167} (\bibinfo {year} {2011})},\ \Eprint
  {http://arxiv.org/abs/1108.6266} {arXiv:1108.6266 [gr-qc]} \BibitemShut
  {NoStop}%
\bibitem [{\citenamefont {Clifton}\ \emph {et~al.}(2012)\citenamefont
  {Clifton}, \citenamefont {Ferreira}, \citenamefont {Padilla},\ and\
  \citenamefont {Skordis}}]{Clifton:2011jh}%
  \BibitemOpen
  \bibfield  {author} {\bibinfo {author} {\bibfnamefont {T.}~\bibnamefont
  {Clifton}}, \bibinfo {author} {\bibfnamefont {P.~G.}\ \bibnamefont
  {Ferreira}}, \bibinfo {author} {\bibfnamefont {A.}~\bibnamefont {Padilla}}, \
  and\ \bibinfo {author} {\bibfnamefont {C.}~\bibnamefont {Skordis}},\
  }\href@noop {} {\bibfield  {journal} {\bibinfo  {journal} {Phys. Rept.}\
  }\textbf {\bibinfo {volume} {513}},\ \bibinfo {pages} {1} (\bibinfo {year}
  {2012})},\ \Eprint {http://arxiv.org/abs/1106.2476} {arXiv:1106.2476
  [astro-ph.CO]} \BibitemShut {NoStop}%
\bibitem [{\citenamefont {Kobayashi}(2019)}]{Kobayashi:2019hrl}%
  \BibitemOpen
  \bibfield  {author} {\bibinfo {author} {\bibfnamefont {T.}~\bibnamefont
  {Kobayashi}},\ }\href {\doibase 10.1088/1361-6633/ab2429} {\bibfield
  {journal} {\bibinfo  {journal} {Rept. Prog. Phys.}\ }\textbf {\bibinfo
  {volume} {82}},\ \bibinfo {pages} {086901} (\bibinfo {year} {2019})},\
  \Eprint {http://arxiv.org/abs/1901.07183} {arXiv:1901.07183 [gr-qc]}
  \BibitemShut {NoStop}%
\bibitem [{\citenamefont {Frusciante}\ and\ \citenamefont
  {Perenon}(2020)}]{Frusciante:2019xia}%
  \BibitemOpen
  \bibfield  {author} {\bibinfo {author} {\bibfnamefont {N.}~\bibnamefont
  {Frusciante}}\ and\ \bibinfo {author} {\bibfnamefont {L.}~\bibnamefont
  {Perenon}},\ }\href {\doibase 10.1016/j.physrep.2020.02.004} {\bibfield
  {journal} {\bibinfo  {journal} {Phys. Rept.}\ }\textbf {\bibinfo {volume}
  {857}},\ \bibinfo {pages} {1} (\bibinfo {year} {2020})},\ \Eprint
  {http://arxiv.org/abs/1907.03150} {arXiv:1907.03150 [astro-ph.CO]}
  \BibitemShut {NoStop}%
\bibitem [{\citenamefont {Akrami}\ \emph {et~al.}(2021)\citenamefont {Akrami}
  \emph {et~al.}}]{CANTATA:2021ktz}%
  \BibitemOpen
  \bibfield  {author} {\bibinfo {author} {\bibfnamefont {Y.}~\bibnamefont
  {Akrami}} \emph {et~al.} (\bibinfo {collaboration} {CANTATA}),\ }\href
  {\doibase 10.1007/978-3-030-83715-0} {\emph {\bibinfo {title} {{Modified
  Gravity and Cosmology}: {An Update by the CANTATA Network}}}},\ edited by\
  \bibinfo {editor} {\bibfnamefont {E.~N.}\ \bibnamefont {Saridakis}}, \bibinfo
  {editor} {\bibfnamefont {R.}~\bibnamefont {Lazkoz}}, \bibinfo {editor}
  {\bibfnamefont {V.}~\bibnamefont {Salzano}}, \bibinfo {editor} {\bibfnamefont
  {P.}~\bibnamefont {Vargas~Moniz}}, \bibinfo {editor} {\bibfnamefont
  {S.}~\bibnamefont {Capozziello}}, \bibinfo {editor} {\bibfnamefont
  {J.}~\bibnamefont {Beltr\'an~Jim\'enez}}, \bibinfo {editor} {\bibfnamefont
  {M.}~\bibnamefont {De~Laurentis}}, \ and\ \bibinfo {editor} {\bibfnamefont
  {G.~J.}\ \bibnamefont {Olmo}}\ (\bibinfo  {publisher} {Springer},\ \bibinfo
  {year} {2021})\ \Eprint {http://arxiv.org/abs/2105.12582} {arXiv:2105.12582
  [gr-qc]} \BibitemShut {NoStop}%
\bibitem [{\citenamefont {Horndeski}(1974)}]{Horndeski:1974wa}%
  \BibitemOpen
  \bibfield  {author} {\bibinfo {author} {\bibfnamefont {G.~W.}\ \bibnamefont
  {Horndeski}},\ }\href@noop {} {\bibfield  {journal} {\bibinfo  {journal}
  {Int. J. Theor. Phys.}\ }\textbf {\bibinfo {volume} {10}},\ \bibinfo {pages}
  {363} (\bibinfo {year} {1974})}\BibitemShut {NoStop}%
\bibitem [{\citenamefont {Nicolis}\ \emph {et~al.}(2009)\citenamefont
  {Nicolis}, \citenamefont {Rattazzi},\ and\ \citenamefont
  {Trincherini}}]{Nicolis:2008in}%
  \BibitemOpen
  \bibfield  {author} {\bibinfo {author} {\bibfnamefont {A.}~\bibnamefont
  {Nicolis}}, \bibinfo {author} {\bibfnamefont {R.}~\bibnamefont {Rattazzi}}, \
  and\ \bibinfo {author} {\bibfnamefont {E.}~\bibnamefont {Trincherini}},\
  }\href {\doibase 10.1103/PhysRevD.79.064036} {\bibfield  {journal} {\bibinfo
  {journal} {Phys. Rev. D}\ }\textbf {\bibinfo {volume} {79}},\ \bibinfo
  {pages} {064036} (\bibinfo {year} {2009})},\ \Eprint
  {http://arxiv.org/abs/0811.2197} {arXiv:0811.2197 [hep-th]} \BibitemShut
  {NoStop}%
\bibitem [{\citenamefont {Deffayet}\ \emph
  {et~al.}(2009{\natexlab{a}})\citenamefont {Deffayet}, \citenamefont
  {Esposito-Farese},\ and\ \citenamefont {Vikman}}]{Deffayet:2009wt}%
  \BibitemOpen
  \bibfield  {author} {\bibinfo {author} {\bibfnamefont {C.}~\bibnamefont
  {Deffayet}}, \bibinfo {author} {\bibfnamefont {G.}~\bibnamefont
  {Esposito-Farese}}, \ and\ \bibinfo {author} {\bibfnamefont {A.}~\bibnamefont
  {Vikman}},\ }\href {\doibase 10.1103/PhysRevD.79.084003} {\bibfield
  {journal} {\bibinfo  {journal} {Phys. Rev. D}\ }\textbf {\bibinfo {volume}
  {79}},\ \bibinfo {pages} {084003} (\bibinfo {year} {2009}{\natexlab{a}})},\
  \Eprint {http://arxiv.org/abs/0901.1314} {arXiv:0901.1314 [hep-th]}
  \BibitemShut {NoStop}%
\bibitem [{\citenamefont {Deffayet}\ \emph
  {et~al.}(2009{\natexlab{b}})\citenamefont {Deffayet}, \citenamefont {Deser},\
  and\ \citenamefont {Esposito-Farese}}]{Deffayet:2009mn}%
  \BibitemOpen
  \bibfield  {author} {\bibinfo {author} {\bibfnamefont {C.}~\bibnamefont
  {Deffayet}}, \bibinfo {author} {\bibfnamefont {S.}~\bibnamefont {Deser}}, \
  and\ \bibinfo {author} {\bibfnamefont {G.}~\bibnamefont {Esposito-Farese}},\
  }\href {\doibase 10.1103/PhysRevD.80.064015} {\bibfield  {journal} {\bibinfo
  {journal} {Phys. Rev. D}\ }\textbf {\bibinfo {volume} {80}},\ \bibinfo
  {pages} {064015} (\bibinfo {year} {2009}{\natexlab{b}})},\ \Eprint
  {http://arxiv.org/abs/0906.1967} {arXiv:0906.1967 [gr-qc]} \BibitemShut
  {NoStop}%
\bibitem [{\citenamefont {Kobayashi}\ \emph
  {et~al.}(2011{\natexlab{a}})\citenamefont {Kobayashi}, \citenamefont
  {Yamaguchi},\ and\ \citenamefont {Yokoyama}}]{Kobayashi:2011nu}%
  \BibitemOpen
  \bibfield  {author} {\bibinfo {author} {\bibfnamefont {T.}~\bibnamefont
  {Kobayashi}}, \bibinfo {author} {\bibfnamefont {M.}~\bibnamefont
  {Yamaguchi}}, \ and\ \bibinfo {author} {\bibfnamefont {J.}~\bibnamefont
  {Yokoyama}},\ }\href {\doibase 10.1143/PTP.126.511} {\bibfield  {journal}
  {\bibinfo  {journal} {Prog. Theor. Phys.}\ }\textbf {\bibinfo {volume}
  {126}},\ \bibinfo {pages} {511} (\bibinfo {year} {2011}{\natexlab{a}})},\
  \Eprint {http://arxiv.org/abs/1105.5723} {arXiv:1105.5723 [hep-th]}
  \BibitemShut {NoStop}%
\bibitem [{\citenamefont {Burrage}\ \emph {et~al.}(2011)\citenamefont
  {Burrage}, \citenamefont {de~Rham}, \citenamefont {Seery},\ and\
  \citenamefont {Tolley}}]{Burrage:2010cu}%
  \BibitemOpen
  \bibfield  {author} {\bibinfo {author} {\bibfnamefont {C.}~\bibnamefont
  {Burrage}}, \bibinfo {author} {\bibfnamefont {C.}~\bibnamefont {de~Rham}},
  \bibinfo {author} {\bibfnamefont {D.}~\bibnamefont {Seery}}, \ and\ \bibinfo
  {author} {\bibfnamefont {A.~J.}\ \bibnamefont {Tolley}},\ }\href {\doibase
  10.1088/1475-7516/2011/01/014} {\bibfield  {journal} {\bibinfo  {journal}
  {JCAP}\ }\textbf {\bibinfo {volume} {01}},\ \bibinfo {pages} {014} (\bibinfo
  {year} {2011})},\ \Eprint {http://arxiv.org/abs/1009.2497} {arXiv:1009.2497
  [hep-th]} \BibitemShut {NoStop}%
\bibitem [{\citenamefont {Creminelli}\ \emph {et~al.}(2011)\citenamefont
  {Creminelli}, \citenamefont {D'Amico}, \citenamefont {Musso}, \citenamefont
  {Norena},\ and\ \citenamefont {Trincherini}}]{Creminelli:2010qf}%
  \BibitemOpen
  \bibfield  {author} {\bibinfo {author} {\bibfnamefont {P.}~\bibnamefont
  {Creminelli}}, \bibinfo {author} {\bibfnamefont {G.}~\bibnamefont {D'Amico}},
  \bibinfo {author} {\bibfnamefont {M.}~\bibnamefont {Musso}}, \bibinfo
  {author} {\bibfnamefont {J.}~\bibnamefont {Norena}}, \ and\ \bibinfo {author}
  {\bibfnamefont {E.}~\bibnamefont {Trincherini}},\ }\href {\doibase
  10.1088/1475-7516/2011/02/006} {\bibfield  {journal} {\bibinfo  {journal}
  {JCAP}\ }\textbf {\bibinfo {volume} {02}},\ \bibinfo {pages} {006} (\bibinfo
  {year} {2011})},\ \Eprint {http://arxiv.org/abs/1011.3004} {arXiv:1011.3004
  [hep-th]} \BibitemShut {NoStop}%
\bibitem [{\citenamefont {Renaux-Petel}\ \emph {et~al.}(2011)\citenamefont
  {Renaux-Petel}, \citenamefont {Mizuno},\ and\ \citenamefont
  {Koyama}}]{Renaux-Petel:2011rmu}%
  \BibitemOpen
  \bibfield  {author} {\bibinfo {author} {\bibfnamefont {S.}~\bibnamefont
  {Renaux-Petel}}, \bibinfo {author} {\bibfnamefont {S.}~\bibnamefont
  {Mizuno}}, \ and\ \bibinfo {author} {\bibfnamefont {K.}~\bibnamefont
  {Koyama}},\ }\href {\doibase 10.1088/1475-7516/2011/11/042} {\bibfield
  {journal} {\bibinfo  {journal} {JCAP}\ }\textbf {\bibinfo {volume} {11}},\
  \bibinfo {pages} {042} (\bibinfo {year} {2011})},\ \Eprint
  {http://arxiv.org/abs/1108.0305} {arXiv:1108.0305 [astro-ph.CO]} \BibitemShut
  {NoStop}%
\bibitem [{\citenamefont {Kamada}\ \emph {et~al.}(2011)\citenamefont {Kamada},
  \citenamefont {Kobayashi}, \citenamefont {Yamaguchi},\ and\ \citenamefont
  {Yokoyama}}]{Kamada:2010qe}%
  \BibitemOpen
  \bibfield  {author} {\bibinfo {author} {\bibfnamefont {K.}~\bibnamefont
  {Kamada}}, \bibinfo {author} {\bibfnamefont {T.}~\bibnamefont {Kobayashi}},
  \bibinfo {author} {\bibfnamefont {M.}~\bibnamefont {Yamaguchi}}, \ and\
  \bibinfo {author} {\bibfnamefont {J.}~\bibnamefont {Yokoyama}},\ }\href
  {\doibase 10.1103/PhysRevD.83.083515} {\bibfield  {journal} {\bibinfo
  {journal} {Phys. Rev. D}\ }\textbf {\bibinfo {volume} {83}},\ \bibinfo
  {pages} {083515} (\bibinfo {year} {2011})},\ \Eprint
  {http://arxiv.org/abs/1012.4238} {arXiv:1012.4238 [astro-ph.CO]} \BibitemShut
  {NoStop}%
\bibitem [{\citenamefont {Kobayashi}\ \emph
  {et~al.}(2011{\natexlab{b}})\citenamefont {Kobayashi}, \citenamefont
  {Yamaguchi},\ and\ \citenamefont {Yokoyama}}]{Kobayashi:2011pc}%
  \BibitemOpen
  \bibfield  {author} {\bibinfo {author} {\bibfnamefont {T.}~\bibnamefont
  {Kobayashi}}, \bibinfo {author} {\bibfnamefont {M.}~\bibnamefont
  {Yamaguchi}}, \ and\ \bibinfo {author} {\bibfnamefont {J.}~\bibnamefont
  {Yokoyama}},\ }\href {\doibase 10.1103/PhysRevD.83.103524} {\bibfield
  {journal} {\bibinfo  {journal} {Phys. Rev. D}\ }\textbf {\bibinfo {volume}
  {83}},\ \bibinfo {pages} {103524} (\bibinfo {year} {2011}{\natexlab{b}})},\
  \Eprint {http://arxiv.org/abs/1103.1740} {arXiv:1103.1740 [hep-th]}
  \BibitemShut {NoStop}%
\bibitem [{\citenamefont {Frusciante}\ \emph {et~al.}(2013)\citenamefont
  {Frusciante}, \citenamefont {Zhou},\ and\ \citenamefont
  {Sotiriou}}]{Frusciante:2013haa}%
  \BibitemOpen
  \bibfield  {author} {\bibinfo {author} {\bibfnamefont {N.}~\bibnamefont
  {Frusciante}}, \bibinfo {author} {\bibfnamefont {S.-Y.}\ \bibnamefont
  {Zhou}}, \ and\ \bibinfo {author} {\bibfnamefont {T.~P.}\ \bibnamefont
  {Sotiriou}},\ }\href {\doibase 10.1088/1475-7516/2013/07/020} {\bibfield
  {journal} {\bibinfo  {journal} {JCAP}\ }\textbf {\bibinfo {volume} {07}},\
  \bibinfo {pages} {020} (\bibinfo {year} {2013})},\ \Eprint
  {http://arxiv.org/abs/1303.6628} {arXiv:1303.6628 [astro-ph.CO]} \BibitemShut
  {NoStop}%
\bibitem [{\citenamefont {Deffayet}\ \emph {et~al.}(2010)\citenamefont
  {Deffayet}, \citenamefont {Pujolas}, \citenamefont {Sawicki},\ and\
  \citenamefont {Vikman}}]{Deffayet:2010qz}%
  \BibitemOpen
  \bibfield  {author} {\bibinfo {author} {\bibfnamefont {C.}~\bibnamefont
  {Deffayet}}, \bibinfo {author} {\bibfnamefont {O.}~\bibnamefont {Pujolas}},
  \bibinfo {author} {\bibfnamefont {I.}~\bibnamefont {Sawicki}}, \ and\
  \bibinfo {author} {\bibfnamefont {A.}~\bibnamefont {Vikman}},\ }\href
  {\doibase 10.1088/1475-7516/2010/10/026} {\bibfield  {journal} {\bibinfo
  {journal} {JCAP}\ }\textbf {\bibinfo {volume} {10}},\ \bibinfo {pages} {026}
  (\bibinfo {year} {2010})},\ \Eprint {http://arxiv.org/abs/1008.0048}
  {arXiv:1008.0048 [hep-th]} \BibitemShut {NoStop}%
\bibitem [{\citenamefont {Kobayashi}\ \emph {et~al.}(2010)\citenamefont
  {Kobayashi}, \citenamefont {Yamaguchi},\ and\ \citenamefont
  {Yokoyama}}]{Kobayashi:2010cm}%
  \BibitemOpen
  \bibfield  {author} {\bibinfo {author} {\bibfnamefont {T.}~\bibnamefont
  {Kobayashi}}, \bibinfo {author} {\bibfnamefont {M.}~\bibnamefont
  {Yamaguchi}}, \ and\ \bibinfo {author} {\bibfnamefont {J.}~\bibnamefont
  {Yokoyama}},\ }\href {\doibase 10.1103/PhysRevLett.105.231302} {\bibfield
  {journal} {\bibinfo  {journal} {Phys. Rev. Lett.}\ }\textbf {\bibinfo
  {volume} {105}},\ \bibinfo {pages} {231302} (\bibinfo {year} {2010})},\
  \Eprint {http://arxiv.org/abs/1008.0603} {arXiv:1008.0603 [hep-th]}
  \BibitemShut {NoStop}%
\bibitem [{\citenamefont {Nesseris}\ \emph {et~al.}(2010)\citenamefont
  {Nesseris}, \citenamefont {De~Felice},\ and\ \citenamefont
  {Tsujikawa}}]{Nesseris:2010pc}%
  \BibitemOpen
  \bibfield  {author} {\bibinfo {author} {\bibfnamefont {S.}~\bibnamefont
  {Nesseris}}, \bibinfo {author} {\bibfnamefont {A.}~\bibnamefont {De~Felice}},
  \ and\ \bibinfo {author} {\bibfnamefont {S.}~\bibnamefont {Tsujikawa}},\
  }\href {\doibase 10.1103/PhysRevD.82.124054} {\bibfield  {journal} {\bibinfo
  {journal} {Phys. Rev. D}\ }\textbf {\bibinfo {volume} {82}},\ \bibinfo
  {pages} {124054} (\bibinfo {year} {2010})},\ \Eprint
  {http://arxiv.org/abs/1010.0407} {arXiv:1010.0407 [astro-ph.CO]} \BibitemShut
  {NoStop}%
\bibitem [{\citenamefont {Charmousis}\ \emph {et~al.}(2012)\citenamefont
  {Charmousis}, \citenamefont {Copeland}, \citenamefont {Padilla},\ and\
  \citenamefont {Saffin}}]{Charmousis:2011bf}%
  \BibitemOpen
  \bibfield  {author} {\bibinfo {author} {\bibfnamefont {C.}~\bibnamefont
  {Charmousis}}, \bibinfo {author} {\bibfnamefont {E.~J.}\ \bibnamefont
  {Copeland}}, \bibinfo {author} {\bibfnamefont {A.}~\bibnamefont {Padilla}}, \
  and\ \bibinfo {author} {\bibfnamefont {P.~M.}\ \bibnamefont {Saffin}},\
  }\href {\doibase 10.1103/PhysRevLett.108.051101} {\bibfield  {journal}
  {\bibinfo  {journal} {Phys. Rev. Lett.}\ }\textbf {\bibinfo {volume} {108}},\
  \bibinfo {pages} {051101} (\bibinfo {year} {2012})},\ \Eprint
  {http://arxiv.org/abs/1106.2000} {arXiv:1106.2000 [hep-th]} \BibitemShut
  {NoStop}%
\bibitem [{\citenamefont {Barreira}\ \emph
  {et~al.}(2013{\natexlab{a}})\citenamefont {Barreira}, \citenamefont {Li},
  \citenamefont {Baugh},\ and\ \citenamefont {Pascoli}}]{Barreira:2013xea}%
  \BibitemOpen
  \bibfield  {author} {\bibinfo {author} {\bibfnamefont {A.}~\bibnamefont
  {Barreira}}, \bibinfo {author} {\bibfnamefont {B.}~\bibnamefont {Li}},
  \bibinfo {author} {\bibfnamefont {C.~M.}\ \bibnamefont {Baugh}}, \ and\
  \bibinfo {author} {\bibfnamefont {S.}~\bibnamefont {Pascoli}},\ }\href
  {\doibase 10.1088/1475-7516/2013/11/056} {\bibfield  {journal} {\bibinfo
  {journal} {JCAP}\ }\textbf {\bibinfo {volume} {11}},\ \bibinfo {pages} {056}
  (\bibinfo {year} {2013}{\natexlab{a}})},\ \Eprint
  {http://arxiv.org/abs/1308.3699} {arXiv:1308.3699 [astro-ph.CO]} \BibitemShut
  {NoStop}%
\bibitem [{\citenamefont {Barreira}\ \emph {et~al.}(2014)\citenamefont
  {Barreira}, \citenamefont {Li}, \citenamefont {Baugh},\ and\ \citenamefont
  {Pascoli}}]{Barreira:2014jha}%
  \BibitemOpen
  \bibfield  {author} {\bibinfo {author} {\bibfnamefont {A.}~\bibnamefont
  {Barreira}}, \bibinfo {author} {\bibfnamefont {B.}~\bibnamefont {Li}},
  \bibinfo {author} {\bibfnamefont {C.}~\bibnamefont {Baugh}}, \ and\ \bibinfo
  {author} {\bibfnamefont {S.}~\bibnamefont {Pascoli}},\ }\href {\doibase
  10.1088/1475-7516/2014/08/059} {\bibfield  {journal} {\bibinfo  {journal}
  {JCAP}\ }\textbf {\bibinfo {volume} {08}},\ \bibinfo {pages} {059} (\bibinfo
  {year} {2014})},\ \Eprint {http://arxiv.org/abs/1406.0485} {arXiv:1406.0485
  [astro-ph.CO]} \BibitemShut {NoStop}%
\bibitem [{\citenamefont {Renk}\ \emph {et~al.}(2017)\citenamefont {Renk},
  \citenamefont {Zumalac\'arregui}, \citenamefont {Montanari},\ and\
  \citenamefont {Barreira}}]{Renk:2017rzu}%
  \BibitemOpen
  \bibfield  {author} {\bibinfo {author} {\bibfnamefont {J.}~\bibnamefont
  {Renk}}, \bibinfo {author} {\bibfnamefont {M.}~\bibnamefont
  {Zumalac\'arregui}}, \bibinfo {author} {\bibfnamefont {F.}~\bibnamefont
  {Montanari}}, \ and\ \bibinfo {author} {\bibfnamefont {A.}~\bibnamefont
  {Barreira}},\ }\href {\doibase 10.1088/1475-7516/2017/10/020} {\bibfield
  {journal} {\bibinfo  {journal} {JCAP}\ }\textbf {\bibinfo {volume} {10}},\
  \bibinfo {pages} {020} (\bibinfo {year} {2017})},\ \Eprint
  {http://arxiv.org/abs/1707.02263} {arXiv:1707.02263 [astro-ph.CO]}
  \BibitemShut {NoStop}%
\bibitem [{\citenamefont {Peirone}\ \emph {et~al.}(2018)\citenamefont
  {Peirone}, \citenamefont {Frusciante}, \citenamefont {Hu}, \citenamefont
  {Raveri},\ and\ \citenamefont {Silvestri}}]{Peirone:2017vcq}%
  \BibitemOpen
  \bibfield  {author} {\bibinfo {author} {\bibfnamefont {S.}~\bibnamefont
  {Peirone}}, \bibinfo {author} {\bibfnamefont {N.}~\bibnamefont {Frusciante}},
  \bibinfo {author} {\bibfnamefont {B.}~\bibnamefont {Hu}}, \bibinfo {author}
  {\bibfnamefont {M.}~\bibnamefont {Raveri}}, \ and\ \bibinfo {author}
  {\bibfnamefont {A.}~\bibnamefont {Silvestri}},\ }\href {\doibase
  10.1103/PhysRevD.97.063518} {\bibfield  {journal} {\bibinfo  {journal} {Phys.
  Rev. D}\ }\textbf {\bibinfo {volume} {97}},\ \bibinfo {pages} {063518}
  (\bibinfo {year} {2018})},\ \Eprint {http://arxiv.org/abs/1711.04760}
  {arXiv:1711.04760 [astro-ph.CO]} \BibitemShut {NoStop}%
\bibitem [{\citenamefont {Kase}\ and\ \citenamefont
  {Tsujikawa}(2018)}]{Kase:2018iwp}%
  \BibitemOpen
  \bibfield  {author} {\bibinfo {author} {\bibfnamefont {R.}~\bibnamefont
  {Kase}}\ and\ \bibinfo {author} {\bibfnamefont {S.}~\bibnamefont
  {Tsujikawa}},\ }\href {\doibase 10.1103/PhysRevD.97.103501} {\bibfield
  {journal} {\bibinfo  {journal} {Phys. Rev. D}\ }\textbf {\bibinfo {volume}
  {97}},\ \bibinfo {pages} {103501} (\bibinfo {year} {2018})},\ \Eprint
  {http://arxiv.org/abs/1802.02728} {arXiv:1802.02728 [gr-qc]} \BibitemShut
  {NoStop}%
\bibitem [{\citenamefont {Frusciante}\ \emph {et~al.}(2018)\citenamefont
  {Frusciante}, \citenamefont {Kase}, \citenamefont {Nunes},\ and\
  \citenamefont {Tsujikawa}}]{Frusciante:2018aew}%
  \BibitemOpen
  \bibfield  {author} {\bibinfo {author} {\bibfnamefont {N.}~\bibnamefont
  {Frusciante}}, \bibinfo {author} {\bibfnamefont {R.}~\bibnamefont {Kase}},
  \bibinfo {author} {\bibfnamefont {N.~J.}\ \bibnamefont {Nunes}}, \ and\
  \bibinfo {author} {\bibfnamefont {S.}~\bibnamefont {Tsujikawa}},\ }\href
  {\doibase 10.1103/PhysRevD.98.123517} {\bibfield  {journal} {\bibinfo
  {journal} {Phys. Rev. D}\ }\textbf {\bibinfo {volume} {98}},\ \bibinfo
  {pages} {123517} (\bibinfo {year} {2018})},\ \Eprint
  {http://arxiv.org/abs/1810.07957} {arXiv:1810.07957 [gr-qc]} \BibitemShut
  {NoStop}%
\bibitem [{\citenamefont {Albuquerque}\ \emph {et~al.}(2018)\citenamefont
  {Albuquerque}, \citenamefont {Frusciante}, \citenamefont {Nunes},\ and\
  \citenamefont {Tsujikawa}}]{Albuquerque:2018ymr}%
  \BibitemOpen
  \bibfield  {author} {\bibinfo {author} {\bibfnamefont {I.~S.}\ \bibnamefont
  {Albuquerque}}, \bibinfo {author} {\bibfnamefont {N.}~\bibnamefont
  {Frusciante}}, \bibinfo {author} {\bibfnamefont {N.~J.}\ \bibnamefont
  {Nunes}}, \ and\ \bibinfo {author} {\bibfnamefont {S.}~\bibnamefont
  {Tsujikawa}},\ }\href {\doibase 10.1103/PhysRevD.98.064038} {\bibfield
  {journal} {\bibinfo  {journal} {Phys. Rev. D}\ }\textbf {\bibinfo {volume}
  {98}},\ \bibinfo {pages} {064038} (\bibinfo {year} {2018})},\ \Eprint
  {http://arxiv.org/abs/1807.09800} {arXiv:1807.09800 [gr-qc]} \BibitemShut
  {NoStop}%
\bibitem [{\citenamefont {Frusciante}\ \emph {et~al.}(2020)\citenamefont
  {Frusciante}, \citenamefont {Peirone}, \citenamefont {Atayde},\ and\
  \citenamefont {De~Felice}}]{Frusciante:2019puu}%
  \BibitemOpen
  \bibfield  {author} {\bibinfo {author} {\bibfnamefont {N.}~\bibnamefont
  {Frusciante}}, \bibinfo {author} {\bibfnamefont {S.}~\bibnamefont {Peirone}},
  \bibinfo {author} {\bibfnamefont {L.}~\bibnamefont {Atayde}}, \ and\ \bibinfo
  {author} {\bibfnamefont {A.}~\bibnamefont {De~Felice}},\ }\href {\doibase
  10.1103/PhysRevD.101.064001} {\bibfield  {journal} {\bibinfo  {journal}
  {Phys. Rev. D}\ }\textbf {\bibinfo {volume} {101}},\ \bibinfo {pages}
  {064001} (\bibinfo {year} {2020})},\ \Eprint
  {http://arxiv.org/abs/1912.07586} {arXiv:1912.07586 [astro-ph.CO]}
  \BibitemShut {NoStop}%
\bibitem [{\citenamefont {Peirone}\ \emph {et~al.}(2019)\citenamefont
  {Peirone}, \citenamefont {Benevento}, \citenamefont {Frusciante},\ and\
  \citenamefont {Tsujikawa}}]{Peirone:2019aua}%
  \BibitemOpen
  \bibfield  {author} {\bibinfo {author} {\bibfnamefont {S.}~\bibnamefont
  {Peirone}}, \bibinfo {author} {\bibfnamefont {G.}~\bibnamefont {Benevento}},
  \bibinfo {author} {\bibfnamefont {N.}~\bibnamefont {Frusciante}}, \ and\
  \bibinfo {author} {\bibfnamefont {S.}~\bibnamefont {Tsujikawa}},\ }\href
  {\doibase 10.1103/PhysRevD.100.063540} {\bibfield  {journal} {\bibinfo
  {journal} {Phys. Rev. D}\ }\textbf {\bibinfo {volume} {100}},\ \bibinfo
  {pages} {063540} (\bibinfo {year} {2019})},\ \Eprint
  {http://arxiv.org/abs/1905.05166} {arXiv:1905.05166 [astro-ph.CO]}
  \BibitemShut {NoStop}%
\bibitem [{\citenamefont {Albuquerque}\ \emph {et~al.}(2022)\citenamefont
  {Albuquerque}, \citenamefont {Frusciante},\ and\ \citenamefont
  {Martinelli}}]{Albuquerque:2021grl}%
  \BibitemOpen
  \bibfield  {author} {\bibinfo {author} {\bibfnamefont {I.~S.}\ \bibnamefont
  {Albuquerque}}, \bibinfo {author} {\bibfnamefont {N.}~\bibnamefont
  {Frusciante}}, \ and\ \bibinfo {author} {\bibfnamefont {M.}~\bibnamefont
  {Martinelli}},\ }\href {\doibase 10.1103/PhysRevD.105.044056} {\bibfield
  {journal} {\bibinfo  {journal} {Phys. Rev. D}\ }\textbf {\bibinfo {volume}
  {105}},\ \bibinfo {pages} {044056} (\bibinfo {year} {2022})},\ \Eprint
  {http://arxiv.org/abs/2112.06892} {arXiv:2112.06892 [astro-ph.CO]}
  \BibitemShut {NoStop}%
\bibitem [{\citenamefont {Abbott}\ \emph
  {et~al.}(2017{\natexlab{a}})\citenamefont {Abbott} \emph
  {et~al.}}]{LIGOScientific:2017zic}%
  \BibitemOpen
  \bibfield  {author} {\bibinfo {author} {\bibfnamefont {B.~P.}\ \bibnamefont
  {Abbott}} \emph {et~al.} (\bibinfo {collaboration} {LIGO Scientific, Virgo,
  Fermi-GBM, INTEGRAL}),\ }\href {\doibase 10.3847/2041-8213/aa920c} {\bibfield
   {journal} {\bibinfo  {journal} {Astrophys. J. Lett.}\ }\textbf {\bibinfo
  {volume} {848}},\ \bibinfo {pages} {L13} (\bibinfo {year}
  {2017}{\natexlab{a}})},\ \Eprint {http://arxiv.org/abs/1710.05834}
  {arXiv:1710.05834 [astro-ph.HE]} \BibitemShut {NoStop}%
\bibitem [{\citenamefont {Creminelli}\ and\ \citenamefont
  {Vernizzi}(2017)}]{Creminelli:2017sry}%
  \BibitemOpen
  \bibfield  {author} {\bibinfo {author} {\bibfnamefont {P.}~\bibnamefont
  {Creminelli}}\ and\ \bibinfo {author} {\bibfnamefont {F.}~\bibnamefont
  {Vernizzi}},\ }\href {\doibase 10.1103/PhysRevLett.119.251302} {\bibfield
  {journal} {\bibinfo  {journal} {Phys. Rev. Lett.}\ }\textbf {\bibinfo
  {volume} {119}},\ \bibinfo {pages} {251302} (\bibinfo {year} {2017})},\
  \Eprint {http://arxiv.org/abs/1710.05877} {arXiv:1710.05877 [astro-ph.CO]}
  \BibitemShut {NoStop}%
\bibitem [{\citenamefont {Baker}\ \emph {et~al.}(2017)\citenamefont {Baker},
  \citenamefont {Bellini}, \citenamefont {Ferreira}, \citenamefont {Lagos},
  \citenamefont {Noller},\ and\ \citenamefont {Sawicki}}]{Baker:2017hug}%
  \BibitemOpen
  \bibfield  {author} {\bibinfo {author} {\bibfnamefont {T.}~\bibnamefont
  {Baker}}, \bibinfo {author} {\bibfnamefont {E.}~\bibnamefont {Bellini}},
  \bibinfo {author} {\bibfnamefont {P.~G.}\ \bibnamefont {Ferreira}}, \bibinfo
  {author} {\bibfnamefont {M.}~\bibnamefont {Lagos}}, \bibinfo {author}
  {\bibfnamefont {J.}~\bibnamefont {Noller}}, \ and\ \bibinfo {author}
  {\bibfnamefont {I.}~\bibnamefont {Sawicki}},\ }\href {\doibase
  10.1103/PhysRevLett.119.251301} {\bibfield  {journal} {\bibinfo  {journal}
  {Phys. Rev. Lett.}\ }\textbf {\bibinfo {volume} {119}},\ \bibinfo {pages}
  {251301} (\bibinfo {year} {2017})},\ \Eprint
  {http://arxiv.org/abs/1710.06394} {arXiv:1710.06394 [astro-ph.CO]}
  \BibitemShut {NoStop}%
\bibitem [{\citenamefont {Ezquiaga}\ and\ \citenamefont
  {Zumalac\'arregui}(2017)}]{Ezquiaga:2017ekz}%
  \BibitemOpen
  \bibfield  {author} {\bibinfo {author} {\bibfnamefont {J.~M.}\ \bibnamefont
  {Ezquiaga}}\ and\ \bibinfo {author} {\bibfnamefont {M.}~\bibnamefont
  {Zumalac\'arregui}},\ }\href {\doibase 10.1103/PhysRevLett.119.251304}
  {\bibfield  {journal} {\bibinfo  {journal} {Phys. Rev. Lett.}\ }\textbf
  {\bibinfo {volume} {119}},\ \bibinfo {pages} {251304} (\bibinfo {year}
  {2017})},\ \Eprint {http://arxiv.org/abs/1710.05901} {arXiv:1710.05901
  [astro-ph.CO]} \BibitemShut {NoStop}%
\bibitem [{\citenamefont {Gomes}\ and\ \citenamefont
  {Amendola}(2014)}]{Gomes:2013ema}%
  \BibitemOpen
  \bibfield  {author} {\bibinfo {author} {\bibfnamefont {A.~R.}\ \bibnamefont
  {Gomes}}\ and\ \bibinfo {author} {\bibfnamefont {L.}~\bibnamefont
  {Amendola}},\ }\href {\doibase 10.1088/1475-7516/2014/03/041} {\bibfield
  {journal} {\bibinfo  {journal} {JCAP}\ }\textbf {\bibinfo {volume} {03}},\
  \bibinfo {pages} {041} (\bibinfo {year} {2014})},\ \Eprint
  {http://arxiv.org/abs/1306.3593} {arXiv:1306.3593 [astro-ph.CO]} \BibitemShut
  {NoStop}%
\bibitem [{\citenamefont {Kimura}\ and\ \citenamefont
  {Yamamoto}(2011)}]{Kimura:2010di}%
  \BibitemOpen
  \bibfield  {author} {\bibinfo {author} {\bibfnamefont {R.}~\bibnamefont
  {Kimura}}\ and\ \bibinfo {author} {\bibfnamefont {K.}~\bibnamefont
  {Yamamoto}},\ }\href {\doibase 10.1088/1475-7516/2011/04/025} {\bibfield
  {journal} {\bibinfo  {journal} {JCAP}\ }\textbf {\bibinfo {volume} {04}},\
  \bibinfo {pages} {025} (\bibinfo {year} {2011})},\ \Eprint
  {http://arxiv.org/abs/1011.2006} {arXiv:1011.2006 [astro-ph.CO]} \BibitemShut
  {NoStop}%
\bibitem [{\citenamefont {Kimura}\ \emph
  {et~al.}(2012{\natexlab{a}})\citenamefont {Kimura}, \citenamefont
  {Kobayashi},\ and\ \citenamefont {Yamamoto}}]{Kimura:2011td}%
  \BibitemOpen
  \bibfield  {author} {\bibinfo {author} {\bibfnamefont {R.}~\bibnamefont
  {Kimura}}, \bibinfo {author} {\bibfnamefont {T.}~\bibnamefont {Kobayashi}}, \
  and\ \bibinfo {author} {\bibfnamefont {K.}~\bibnamefont {Yamamoto}},\ }\href
  {\doibase 10.1103/PhysRevD.85.123503} {\bibfield  {journal} {\bibinfo
  {journal} {Phys. Rev. D}\ }\textbf {\bibinfo {volume} {85}},\ \bibinfo
  {pages} {123503} (\bibinfo {year} {2012}{\natexlab{a}})},\ \Eprint
  {http://arxiv.org/abs/1110.3598} {arXiv:1110.3598 [astro-ph.CO]} \BibitemShut
  {NoStop}%
\bibitem [{\citenamefont {Bellini}\ \emph {et~al.}(2012)\citenamefont
  {Bellini}, \citenamefont {Bartolo},\ and\ \citenamefont
  {Matarrese}}]{Bellini:2012qn}%
  \BibitemOpen
  \bibfield  {author} {\bibinfo {author} {\bibfnamefont {E.}~\bibnamefont
  {Bellini}}, \bibinfo {author} {\bibfnamefont {N.}~\bibnamefont {Bartolo}}, \
  and\ \bibinfo {author} {\bibfnamefont {S.}~\bibnamefont {Matarrese}},\ }\href
  {\doibase 10.1088/1475-7516/2012/06/019} {\bibfield  {journal} {\bibinfo
  {journal} {JCAP}\ }\textbf {\bibinfo {volume} {06}},\ \bibinfo {pages} {019}
  (\bibinfo {year} {2012})},\ \Eprint {http://arxiv.org/abs/1202.2712}
  {arXiv:1202.2712 [astro-ph.CO]} \BibitemShut {NoStop}%
\bibitem [{\citenamefont {Barreira}\ \emph
  {et~al.}(2013{\natexlab{b}})\citenamefont {Barreira}, \citenamefont {Li},
  \citenamefont {Hellwing}, \citenamefont {Baugh},\ and\ \citenamefont
  {Pascoli}}]{Barreira:2013eea}%
  \BibitemOpen
  \bibfield  {author} {\bibinfo {author} {\bibfnamefont {A.}~\bibnamefont
  {Barreira}}, \bibinfo {author} {\bibfnamefont {B.}~\bibnamefont {Li}},
  \bibinfo {author} {\bibfnamefont {W.~A.}\ \bibnamefont {Hellwing}}, \bibinfo
  {author} {\bibfnamefont {C.~M.}\ \bibnamefont {Baugh}}, \ and\ \bibinfo
  {author} {\bibfnamefont {S.}~\bibnamefont {Pascoli}},\ }\href {\doibase
  10.1088/1475-7516/2013/10/027} {\bibfield  {journal} {\bibinfo  {journal}
  {JCAP}\ }\textbf {\bibinfo {volume} {10}},\ \bibinfo {pages} {027} (\bibinfo
  {year} {2013}{\natexlab{b}})},\ \Eprint {http://arxiv.org/abs/1306.3219}
  {arXiv:1306.3219 [astro-ph.CO]} \BibitemShut {NoStop}%
\bibitem [{\citenamefont {Frusciante}\ and\ \citenamefont
  {Pace}(2020)}]{Frusciante:2020zfs}%
  \BibitemOpen
  \bibfield  {author} {\bibinfo {author} {\bibfnamefont {N.}~\bibnamefont
  {Frusciante}}\ and\ \bibinfo {author} {\bibfnamefont {F.}~\bibnamefont
  {Pace}},\ }\href {\doibase 10.1016/j.dark.2020.100686} {\bibfield  {journal}
  {\bibinfo  {journal} {Phys. Dark Univ.}\ }\textbf {\bibinfo {volume} {30}},\
  \bibinfo {pages} {100686} (\bibinfo {year} {2020})},\ \Eprint
  {http://arxiv.org/abs/2004.11881} {arXiv:2004.11881 [astro-ph.CO]}
  \BibitemShut {NoStop}%
\bibitem [{\citenamefont {Vainshtein}(1972)}]{Vainshtein:1972sx}%
  \BibitemOpen
  \bibfield  {author} {\bibinfo {author} {\bibfnamefont {A.~I.}\ \bibnamefont
  {Vainshtein}},\ }\href {\doibase 10.1016/0370-2693(72)90147-5} {\bibfield
  {journal} {\bibinfo  {journal} {Phys. Lett. B}\ }\textbf {\bibinfo {volume}
  {39}},\ \bibinfo {pages} {393} (\bibinfo {year} {1972})}\BibitemShut
  {NoStop}%
\bibitem [{\citenamefont {Kimura}\ \emph
  {et~al.}(2012{\natexlab{b}})\citenamefont {Kimura}, \citenamefont
  {Kobayashi},\ and\ \citenamefont {Yamamoto}}]{Kimura:2011dc}%
  \BibitemOpen
  \bibfield  {author} {\bibinfo {author} {\bibfnamefont {R.}~\bibnamefont
  {Kimura}}, \bibinfo {author} {\bibfnamefont {T.}~\bibnamefont {Kobayashi}}, \
  and\ \bibinfo {author} {\bibfnamefont {K.}~\bibnamefont {Yamamoto}},\ }\href
  {\doibase 10.1103/PhysRevD.85.024023} {\bibfield  {journal} {\bibinfo
  {journal} {Phys. Rev. D}\ }\textbf {\bibinfo {volume} {85}},\ \bibinfo
  {pages} {024023} (\bibinfo {year} {2012}{\natexlab{b}})},\ \Eprint
  {http://arxiv.org/abs/1111.6749} {arXiv:1111.6749 [astro-ph.CO]} \BibitemShut
  {NoStop}%
\bibitem [{\citenamefont {Babichev}\ and\ \citenamefont
  {Deffayet}(2013)}]{Babichev:2013usa}%
  \BibitemOpen
  \bibfield  {author} {\bibinfo {author} {\bibfnamefont {E.}~\bibnamefont
  {Babichev}}\ and\ \bibinfo {author} {\bibfnamefont {C.}~\bibnamefont
  {Deffayet}},\ }\href {\doibase 10.1088/0264-9381/30/18/184001} {\bibfield
  {journal} {\bibinfo  {journal} {Class. Quant. Grav.}\ }\textbf {\bibinfo
  {volume} {30}},\ \bibinfo {pages} {184001} (\bibinfo {year} {2013})},\
  \Eprint {http://arxiv.org/abs/1304.7240} {arXiv:1304.7240 [gr-qc]}
  \BibitemShut {NoStop}%
\bibitem [{\citenamefont {Koyama}\ \emph {et~al.}(2013)\citenamefont {Koyama},
  \citenamefont {Niz},\ and\ \citenamefont {Tasinato}}]{Koyama:2013paa}%
  \BibitemOpen
  \bibfield  {author} {\bibinfo {author} {\bibfnamefont {K.}~\bibnamefont
  {Koyama}}, \bibinfo {author} {\bibfnamefont {G.}~\bibnamefont {Niz}}, \ and\
  \bibinfo {author} {\bibfnamefont {G.}~\bibnamefont {Tasinato}},\ }\href
  {\doibase 10.1103/PhysRevD.88.021502} {\bibfield  {journal} {\bibinfo
  {journal} {Phys. Rev. D}\ }\textbf {\bibinfo {volume} {88}},\ \bibinfo
  {pages} {021502} (\bibinfo {year} {2013})},\ \Eprint
  {http://arxiv.org/abs/1305.0279} {arXiv:1305.0279 [hep-th]} \BibitemShut
  {NoStop}%
\bibitem [{\citenamefont {Baessler}\ \emph {et~al.}(1999)\citenamefont
  {Baessler}, \citenamefont {Heckel}, \citenamefont {Adelberger}, \citenamefont
  {Gundlach}, \citenamefont {Schmidt},\ and\ \citenamefont
  {Swanson}}]{Baessler:1999iv}%
  \BibitemOpen
  \bibfield  {author} {\bibinfo {author} {\bibfnamefont {S.}~\bibnamefont
  {Baessler}}, \bibinfo {author} {\bibfnamefont {B.~R.}\ \bibnamefont
  {Heckel}}, \bibinfo {author} {\bibfnamefont {E.~G.}\ \bibnamefont
  {Adelberger}}, \bibinfo {author} {\bibfnamefont {J.~H.}\ \bibnamefont
  {Gundlach}}, \bibinfo {author} {\bibfnamefont {U.}~\bibnamefont {Schmidt}}, \
  and\ \bibinfo {author} {\bibfnamefont {H.~E.}\ \bibnamefont {Swanson}},\
  }\href {\doibase 10.1103/PhysRevLett.83.003585} {\bibfield  {journal}
  {\bibinfo  {journal} {Phys. Rev. Lett.}\ }\textbf {\bibinfo {volume} {83}},\
  \bibinfo {pages} {3585} (\bibinfo {year} {1999})}\BibitemShut {NoStop}%
\bibitem [{\citenamefont {Will}(2006)}]{Will:2005va}%
  \BibitemOpen
  \bibfield  {author} {\bibinfo {author} {\bibfnamefont {C.~M.}\ \bibnamefont
  {Will}},\ }\href {\doibase 10.12942/lrr-2006-3} {\bibfield  {journal}
  {\bibinfo  {journal} {Living Rev. Rel.}\ }\textbf {\bibinfo {volume} {9}},\
  \bibinfo {pages} {3} (\bibinfo {year} {2006})},\ \Eprint
  {http://arxiv.org/abs/gr-qc/0510072} {arXiv:gr-qc/0510072} \BibitemShut
  {NoStop}%
\bibitem [{\citenamefont {Uzan}(2011)}]{Uzan:2010pm}%
  \BibitemOpen
  \bibfield  {author} {\bibinfo {author} {\bibfnamefont {J.-P.}\ \bibnamefont
  {Uzan}},\ }\href {\doibase 10.12942/lrr-2011-2} {\bibfield  {journal}
  {\bibinfo  {journal} {Living Rev. Rel.}\ }\textbf {\bibinfo {volume} {14}},\
  \bibinfo {pages} {2} (\bibinfo {year} {2011})},\ \Eprint
  {http://arxiv.org/abs/1009.5514} {arXiv:1009.5514 [astro-ph.CO]} \BibitemShut
  {NoStop}%
\bibitem [{\citenamefont {Gunn}\ and\ \citenamefont
  {Gott}(1972)}]{Gunn:1972sv}%
  \BibitemOpen
  \bibfield  {author} {\bibinfo {author} {\bibfnamefont {J.~E.}\ \bibnamefont
  {Gunn}}\ and\ \bibinfo {author} {\bibfnamefont {J.~R.}\ \bibnamefont {Gott},
  \bibfnamefont {III}},\ }\href {\doibase 10.1086/151605} {\bibfield  {journal}
  {\bibinfo  {journal} {Astrophys. J.}\ }\textbf {\bibinfo {volume} {176}},\
  \bibinfo {pages} {1} (\bibinfo {year} {1972})}\BibitemShut {NoStop}%
\bibitem [{\citenamefont {Cataneo}\ \emph {et~al.}(2019)\citenamefont
  {Cataneo}, \citenamefont {Lombriser}, \citenamefont {Heymans}, \citenamefont
  {Mead}, \citenamefont {Barreira}, \citenamefont {Bose},\ and\ \citenamefont
  {Li}}]{Cataneo:2018cic}%
  \BibitemOpen
  \bibfield  {author} {\bibinfo {author} {\bibfnamefont {M.}~\bibnamefont
  {Cataneo}}, \bibinfo {author} {\bibfnamefont {L.}~\bibnamefont {Lombriser}},
  \bibinfo {author} {\bibfnamefont {C.}~\bibnamefont {Heymans}}, \bibinfo
  {author} {\bibfnamefont {A.}~\bibnamefont {Mead}}, \bibinfo {author}
  {\bibfnamefont {A.}~\bibnamefont {Barreira}}, \bibinfo {author}
  {\bibfnamefont {S.}~\bibnamefont {Bose}}, \ and\ \bibinfo {author}
  {\bibfnamefont {B.}~\bibnamefont {Li}},\ }\href {\doibase
  10.1093/mnras/stz1836} {\bibfield  {journal} {\bibinfo  {journal} {Mon. Not.
  Roy. Astron. Soc.}\ }\textbf {\bibinfo {volume} {488}},\ \bibinfo {pages}
  {2121} (\bibinfo {year} {2019})},\ \Eprint {http://arxiv.org/abs/1812.05594}
  {arXiv:1812.05594 [astro-ph.CO]} \BibitemShut {NoStop}%
\bibitem [{\citenamefont {Ivezi\'c}\ \emph {et~al.}(2019)\citenamefont
  {Ivezi\'c} \emph {et~al.}}]{LSST:2008ijt}%
  \BibitemOpen
  \bibfield  {author} {\bibinfo {author} {\bibfnamefont {v.}~\bibnamefont
  {Ivezi\'c}} \emph {et~al.} (\bibinfo {collaboration} {LSST}),\ }\href
  {\doibase 10.3847/1538-4357/ab042c} {\bibfield  {journal} {\bibinfo
  {journal} {Astrophys. J.}\ }\textbf {\bibinfo {volume} {873}},\ \bibinfo
  {pages} {111} (\bibinfo {year} {2019})},\ \Eprint
  {http://arxiv.org/abs/0805.2366} {arXiv:0805.2366 [astro-ph]} \BibitemShut
  {NoStop}%
\bibitem [{\citenamefont {Aghamousa}\ \emph {et~al.}(2016)\citenamefont
  {Aghamousa} \emph {et~al.}}]{DESI:2016fyo}%
  \BibitemOpen
  \bibfield  {author} {\bibinfo {author} {\bibfnamefont {A.}~\bibnamefont
  {Aghamousa}} \emph {et~al.} (\bibinfo {collaboration} {DESI}),\ }\href@noop
  {} {\  (\bibinfo {year} {2016})},\ \Eprint {http://arxiv.org/abs/1611.00036}
  {arXiv:1611.00036 [astro-ph.IM]} \BibitemShut {NoStop}%
\bibitem [{\citenamefont {Laureijs}\ \emph {et~al.}(2011)\citenamefont
  {Laureijs} \emph {et~al.}}]{EUCLID:2011zbd}%
  \BibitemOpen
  \bibfield  {author} {\bibinfo {author} {\bibfnamefont {R.}~\bibnamefont
  {Laureijs}} \emph {et~al.} (\bibinfo {collaboration} {EUCLID}),\ }\href@noop
  {} {\  (\bibinfo {year} {2011})},\ \Eprint {http://arxiv.org/abs/1110.3193}
  {arXiv:1110.3193 [astro-ph.CO]} \BibitemShut {NoStop}%
\bibitem [{\citenamefont {Weltman}\ \emph {et~al.}(2020)\citenamefont {Weltman}
  \emph {et~al.}}]{Weltman:2018zrl}%
  \BibitemOpen
  \bibfield  {author} {\bibinfo {author} {\bibfnamefont {A.}~\bibnamefont
  {Weltman}} \emph {et~al.},\ }\href {\doibase 10.1017/pasa.2019.42} {\bibfield
   {journal} {\bibinfo  {journal} {Publ. Astron. Soc. Austral.}\ }\textbf
  {\bibinfo {volume} {37}},\ \bibinfo {pages} {e002} (\bibinfo {year}
  {2020})},\ \Eprint {http://arxiv.org/abs/1810.02680} {arXiv:1810.02680
  [astro-ph.CO]} \BibitemShut {NoStop}%
\bibitem [{\citenamefont {Gubitosi}\ \emph {et~al.}(2013)\citenamefont
  {Gubitosi}, \citenamefont {Piazza},\ and\ \citenamefont
  {Vernizzi}}]{Gubitosi:2012hu}%
  \BibitemOpen
  \bibfield  {author} {\bibinfo {author} {\bibfnamefont {G.}~\bibnamefont
  {Gubitosi}}, \bibinfo {author} {\bibfnamefont {F.}~\bibnamefont {Piazza}}, \
  and\ \bibinfo {author} {\bibfnamefont {F.}~\bibnamefont {Vernizzi}},\ }\href
  {\doibase 10.1088/1475-7516/2013/02/032} {\bibfield  {journal} {\bibinfo
  {journal} {JCAP}\ }\textbf {\bibinfo {volume} {02}},\ \bibinfo {pages} {032}
  (\bibinfo {year} {2013})},\ \Eprint {http://arxiv.org/abs/1210.0201}
  {arXiv:1210.0201 [hep-th]} \BibitemShut {NoStop}%
\bibitem [{\citenamefont {Bloomfield}\ \emph {et~al.}(2013)\citenamefont
  {Bloomfield}, \citenamefont {Flanagan}, \citenamefont {Park},\ and\
  \citenamefont {Watson}}]{Bloomfield:2012ff}%
  \BibitemOpen
  \bibfield  {author} {\bibinfo {author} {\bibfnamefont {J.~K.}\ \bibnamefont
  {Bloomfield}}, \bibinfo {author} {\bibfnamefont {E.~E.}\ \bibnamefont
  {Flanagan}}, \bibinfo {author} {\bibfnamefont {M.}~\bibnamefont {Park}}, \
  and\ \bibinfo {author} {\bibfnamefont {S.}~\bibnamefont {Watson}},\ }\href
  {\doibase 10.1088/1475-7516/2013/08/010} {\bibfield  {journal} {\bibinfo
  {journal} {JCAP}\ }\textbf {\bibinfo {volume} {08}},\ \bibinfo {pages} {010}
  (\bibinfo {year} {2013})},\ \Eprint {http://arxiv.org/abs/1211.7054}
  {arXiv:1211.7054 [astro-ph.CO]} \BibitemShut {NoStop}%
\bibitem [{\citenamefont {Traykova}\ \emph {et~al.}(2021)\citenamefont
  {Traykova}, \citenamefont {Bellini}, \citenamefont {Ferreira}, \citenamefont
  {Garc\'\i{}a-Garc\'\i{}a}, \citenamefont {Noller},\ and\ \citenamefont
  {Zumalac\'arregui}}]{Traykova:2021hbr}%
  \BibitemOpen
  \bibfield  {author} {\bibinfo {author} {\bibfnamefont {D.}~\bibnamefont
  {Traykova}}, \bibinfo {author} {\bibfnamefont {E.}~\bibnamefont {Bellini}},
  \bibinfo {author} {\bibfnamefont {P.~G.}\ \bibnamefont {Ferreira}}, \bibinfo
  {author} {\bibfnamefont {C.}~\bibnamefont {Garc\'\i{}a-Garc\'\i{}a}},
  \bibinfo {author} {\bibfnamefont {J.}~\bibnamefont {Noller}}, \ and\ \bibinfo
  {author} {\bibfnamefont {M.}~\bibnamefont {Zumalac\'arregui}},\ }\href
  {\doibase 10.1103/PhysRevD.104.083502} {\bibfield  {journal} {\bibinfo
  {journal} {Phys. Rev. D}\ }\textbf {\bibinfo {volume} {104}},\ \bibinfo
  {pages} {083502} (\bibinfo {year} {2021})},\ \Eprint
  {http://arxiv.org/abs/2103.11195} {arXiv:2103.11195 [astro-ph.CO]}
  \BibitemShut {NoStop}%
\bibitem [{\citenamefont {Abbott}\ \emph
  {et~al.}(2017{\natexlab{b}})\citenamefont {Abbott} \emph
  {et~al.}}]{LIGOScientific:2017vwq}%
  \BibitemOpen
  \bibfield  {author} {\bibinfo {author} {\bibfnamefont {B.~P.}\ \bibnamefont
  {Abbott}} \emph {et~al.} (\bibinfo {collaboration} {LIGO Scientific,
  Virgo}),\ }\href {\doibase 10.1103/PhysRevLett.119.161101} {\bibfield
  {journal} {\bibinfo  {journal} {Phys. Rev. Lett.}\ }\textbf {\bibinfo
  {volume} {119}},\ \bibinfo {pages} {161101} (\bibinfo {year}
  {2017}{\natexlab{b}})},\ \Eprint {http://arxiv.org/abs/1710.05832}
  {arXiv:1710.05832 [gr-qc]} \BibitemShut {NoStop}%
\bibitem [{\citenamefont {Bean}\ and\ \citenamefont
  {Tangmatitham}(2010)}]{Bean:2010zq}%
  \BibitemOpen
  \bibfield  {author} {\bibinfo {author} {\bibfnamefont {R.}~\bibnamefont
  {Bean}}\ and\ \bibinfo {author} {\bibfnamefont {M.}~\bibnamefont
  {Tangmatitham}},\ }\href {\doibase 10.1103/PhysRevD.81.083534} {\bibfield
  {journal} {\bibinfo  {journal} {Phys. Rev. D}\ }\textbf {\bibinfo {volume}
  {81}},\ \bibinfo {pages} {083534} (\bibinfo {year} {2010})},\ \Eprint
  {http://arxiv.org/abs/1002.4197} {arXiv:1002.4197 [astro-ph.CO]} \BibitemShut
  {NoStop}%
\bibitem [{\citenamefont {Silvestri}\ \emph {et~al.}(2013)\citenamefont
  {Silvestri}, \citenamefont {Pogosian},\ and\ \citenamefont
  {Buniy}}]{Silvestri:2013ne}%
  \BibitemOpen
  \bibfield  {author} {\bibinfo {author} {\bibfnamefont {A.}~\bibnamefont
  {Silvestri}}, \bibinfo {author} {\bibfnamefont {L.}~\bibnamefont {Pogosian}},
  \ and\ \bibinfo {author} {\bibfnamefont {R.~V.}\ \bibnamefont {Buniy}},\
  }\href {\doibase 10.1103/PhysRevD.87.104015} {\bibfield  {journal} {\bibinfo
  {journal} {Phys. Rev. D}\ }\textbf {\bibinfo {volume} {87}},\ \bibinfo
  {pages} {104015} (\bibinfo {year} {2013})},\ \Eprint
  {http://arxiv.org/abs/1302.1193} {arXiv:1302.1193 [astro-ph.CO]} \BibitemShut
  {NoStop}%
\bibitem [{\citenamefont {Pogosian}\ \emph {et~al.}(2010)\citenamefont
  {Pogosian}, \citenamefont {Silvestri}, \citenamefont {Koyama},\ and\
  \citenamefont {Zhao}}]{Pogosian:2010tj}%
  \BibitemOpen
  \bibfield  {author} {\bibinfo {author} {\bibfnamefont {L.}~\bibnamefont
  {Pogosian}}, \bibinfo {author} {\bibfnamefont {A.}~\bibnamefont {Silvestri}},
  \bibinfo {author} {\bibfnamefont {K.}~\bibnamefont {Koyama}}, \ and\ \bibinfo
  {author} {\bibfnamefont {G.-B.}\ \bibnamefont {Zhao}},\ }\href {\doibase
  10.1103/PhysRevD.81.104023} {\bibfield  {journal} {\bibinfo  {journal} {Phys.
  Rev. D}\ }\textbf {\bibinfo {volume} {81}},\ \bibinfo {pages} {104023}
  (\bibinfo {year} {2010})},\ \Eprint {http://arxiv.org/abs/1002.2382}
  {arXiv:1002.2382 [astro-ph.CO]} \BibitemShut {NoStop}%
\bibitem [{\citenamefont {Boisseau}\ \emph {et~al.}(2000)\citenamefont
  {Boisseau}, \citenamefont {Esposito-Farese}, \citenamefont {Polarski},\ and\
  \citenamefont {Starobinsky}}]{Boisseau:2000pr}%
  \BibitemOpen
  \bibfield  {author} {\bibinfo {author} {\bibfnamefont {B.}~\bibnamefont
  {Boisseau}}, \bibinfo {author} {\bibfnamefont {G.}~\bibnamefont
  {Esposito-Farese}}, \bibinfo {author} {\bibfnamefont {D.}~\bibnamefont
  {Polarski}}, \ and\ \bibinfo {author} {\bibfnamefont {A.~A.}\ \bibnamefont
  {Starobinsky}},\ }\href {\doibase 10.1103/PhysRevLett.85.2236} {\bibfield
  {journal} {\bibinfo  {journal} {Phys. Rev. Lett.}\ }\textbf {\bibinfo
  {volume} {85}},\ \bibinfo {pages} {2236} (\bibinfo {year} {2000})},\ \Eprint
  {http://arxiv.org/abs/gr-qc/0001066} {arXiv:gr-qc/0001066} \BibitemShut
  {NoStop}%
\bibitem [{\citenamefont {De~Felice}\ \emph {et~al.}(2011)\citenamefont
  {De~Felice}, \citenamefont {Kobayashi},\ and\ \citenamefont
  {Tsujikawa}}]{DeFelice:2011hq}%
  \BibitemOpen
  \bibfield  {author} {\bibinfo {author} {\bibfnamefont {A.}~\bibnamefont
  {De~Felice}}, \bibinfo {author} {\bibfnamefont {T.}~\bibnamefont
  {Kobayashi}}, \ and\ \bibinfo {author} {\bibfnamefont {S.}~\bibnamefont
  {Tsujikawa}},\ }\href {\doibase 10.1016/j.physletb.2011.11.028} {\bibfield
  {journal} {\bibinfo  {journal} {Phys. Lett. B}\ }\textbf {\bibinfo {volume}
  {706}},\ \bibinfo {pages} {123} (\bibinfo {year} {2011})},\ \Eprint
  {http://arxiv.org/abs/1108.4242} {arXiv:1108.4242 [gr-qc]} \BibitemShut
  {NoStop}%
\bibitem [{\citenamefont {Sawicki}\ and\ \citenamefont
  {Bellini}(2015)}]{Sawicki:2015zya}%
  \BibitemOpen
  \bibfield  {author} {\bibinfo {author} {\bibfnamefont {I.}~\bibnamefont
  {Sawicki}}\ and\ \bibinfo {author} {\bibfnamefont {E.}~\bibnamefont
  {Bellini}},\ }\href {\doibase 10.1103/PhysRevD.92.084061} {\bibfield
  {journal} {\bibinfo  {journal} {Phys. Rev. D}\ }\textbf {\bibinfo {volume}
  {92}},\ \bibinfo {pages} {084061} (\bibinfo {year} {2015})},\ \Eprint
  {http://arxiv.org/abs/1503.06831} {arXiv:1503.06831 [astro-ph.CO]}
  \BibitemShut {NoStop}%
\bibitem [{\citenamefont {Pogosian}\ and\ \citenamefont
  {Silvestri}(2016)}]{Pogosian:2016pwr}%
  \BibitemOpen
  \bibfield  {author} {\bibinfo {author} {\bibfnamefont {L.}~\bibnamefont
  {Pogosian}}\ and\ \bibinfo {author} {\bibfnamefont {A.}~\bibnamefont
  {Silvestri}},\ }\href {\doibase 10.1103/PhysRevD.94.104014} {\bibfield
  {journal} {\bibinfo  {journal} {Phys. Rev. D}\ }\textbf {\bibinfo {volume}
  {94}},\ \bibinfo {pages} {104014} (\bibinfo {year} {2016})},\ \Eprint
  {http://arxiv.org/abs/1606.05339} {arXiv:1606.05339 [astro-ph.CO]}
  \BibitemShut {NoStop}%
\bibitem [{\citenamefont {Bellini}\ and\ \citenamefont
  {Sawicki}(2014)}]{Bellini:2014fua}%
  \BibitemOpen
  \bibfield  {author} {\bibinfo {author} {\bibfnamefont {E.}~\bibnamefont
  {Bellini}}\ and\ \bibinfo {author} {\bibfnamefont {I.}~\bibnamefont
  {Sawicki}},\ }\href {\doibase 10.1088/1475-7516/2014/07/050} {\bibfield
  {journal} {\bibinfo  {journal} {JCAP}\ }\textbf {\bibinfo {volume} {07}},\
  \bibinfo {pages} {050} (\bibinfo {year} {2014})},\ \Eprint
  {http://arxiv.org/abs/1404.3713} {arXiv:1404.3713 [astro-ph.CO]} \BibitemShut
  {NoStop}%
\bibitem [{\citenamefont {Weinberg}\ and\ \citenamefont
  {Kamionkowski}(2003)}]{Weinberg:2002rd}%
  \BibitemOpen
  \bibfield  {author} {\bibinfo {author} {\bibfnamefont {N.~N.}\ \bibnamefont
  {Weinberg}}\ and\ \bibinfo {author} {\bibfnamefont {M.}~\bibnamefont
  {Kamionkowski}},\ }\href {\doibase 10.1046/j.1365-8711.2003.06421.x}
  {\bibfield  {journal} {\bibinfo  {journal} {Mon. Not. Roy. Astron. Soc.}\
  }\textbf {\bibinfo {volume} {341}},\ \bibinfo {pages} {251} (\bibinfo {year}
  {2003})},\ \Eprint {http://arxiv.org/abs/astro-ph/0210134}
  {arXiv:astro-ph/0210134} \BibitemShut {NoStop}%
\bibitem [{\citenamefont {Schmidt}\ \emph {et~al.}(2010)\citenamefont
  {Schmidt}, \citenamefont {Hu},\ and\ \citenamefont {Lima}}]{Schmidt:2009yj}%
  \BibitemOpen
  \bibfield  {author} {\bibinfo {author} {\bibfnamefont {F.}~\bibnamefont
  {Schmidt}}, \bibinfo {author} {\bibfnamefont {W.}~\bibnamefont {Hu}}, \ and\
  \bibinfo {author} {\bibfnamefont {M.}~\bibnamefont {Lima}},\ }\href {\doibase
  10.1103/PhysRevD.81.063005} {\bibfield  {journal} {\bibinfo  {journal} {Phys.
  Rev. D}\ }\textbf {\bibinfo {volume} {81}},\ \bibinfo {pages} {063005}
  (\bibinfo {year} {2010})},\ \Eprint {http://arxiv.org/abs/0911.5178}
  {arXiv:0911.5178 [astro-ph.CO]} \BibitemShut {NoStop}%
\bibitem [{\citenamefont {{Wang}}\ and\ \citenamefont
  {{Steinhardt}}(1998)}]{WangSteinhardt1998}%
  \BibitemOpen
  \bibfield  {author} {\bibinfo {author} {\bibfnamefont {L.}~\bibnamefont
  {{Wang}}}\ and\ \bibinfo {author} {\bibfnamefont {P.~J.}\ \bibnamefont
  {{Steinhardt}}},\ }\href {\doibase 10.1086/306436} {\bibfield  {journal}
  {\bibinfo  {journal} {\apj}\ }\textbf {\bibinfo {volume} {508}},\ \bibinfo
  {pages} {483} (\bibinfo {year} {1998})},\ \Eprint
  {http://arxiv.org/abs/astro-ph/9804015} {arXiv:astro-ph/9804015 [astro-ph]}
  \BibitemShut {NoStop}%
\bibitem [{\citenamefont {Hassani}\ and\ \citenamefont
  {Lombriser}(2020)}]{Hassani:2020rxd}%
  \BibitemOpen
  \bibfield  {author} {\bibinfo {author} {\bibfnamefont {F.}~\bibnamefont
  {Hassani}}\ and\ \bibinfo {author} {\bibfnamefont {L.}~\bibnamefont
  {Lombriser}},\ }\href {\doibase 10.1093/mnras/staa2083} {\bibfield  {journal}
  {\bibinfo  {journal} {Mon. Not. Roy. Astron. Soc.}\ }\textbf {\bibinfo
  {volume} {497}},\ \bibinfo {pages} {1885} (\bibinfo {year} {2020})},\ \Eprint
  {http://arxiv.org/abs/2003.05927} {arXiv:2003.05927 [astro-ph.CO]}
  \BibitemShut {NoStop}%
\bibitem [{\citenamefont {Bose}\ \emph {et~al.}(2022)\citenamefont {Bose},
  \citenamefont {Tsedrik}, \citenamefont {Kennedy}, \citenamefont {Lombriser},
  \citenamefont {Pourtsidou},\ and\ \citenamefont {Taylor}}]{Bose:2022vwi}%
  \BibitemOpen
  \bibfield  {author} {\bibinfo {author} {\bibfnamefont {B.}~\bibnamefont
  {Bose}}, \bibinfo {author} {\bibfnamefont {M.}~\bibnamefont {Tsedrik}},
  \bibinfo {author} {\bibfnamefont {J.}~\bibnamefont {Kennedy}}, \bibinfo
  {author} {\bibfnamefont {L.}~\bibnamefont {Lombriser}}, \bibinfo {author}
  {\bibfnamefont {A.}~\bibnamefont {Pourtsidou}}, \ and\ \bibinfo {author}
  {\bibfnamefont {A.}~\bibnamefont {Taylor}},\ }\href@noop {} {\  (\bibinfo
  {year} {2022})},\ \Eprint {http://arxiv.org/abs/2210.01094} {arXiv:2210.01094
  [astro-ph.CO]} \BibitemShut {NoStop}%
\bibitem [{\citenamefont {Chevallier}\ and\ \citenamefont
  {Polarski}(2001)}]{Chevallier:2000qy}%
  \BibitemOpen
  \bibfield  {author} {\bibinfo {author} {\bibfnamefont {M.}~\bibnamefont
  {Chevallier}}\ and\ \bibinfo {author} {\bibfnamefont {D.}~\bibnamefont
  {Polarski}},\ }\href {\doibase 10.1142/S0218271801000822} {\bibfield
  {journal} {\bibinfo  {journal} {Int. J. Mod. Phys. D}\ }\textbf {\bibinfo
  {volume} {10}},\ \bibinfo {pages} {213} (\bibinfo {year} {2001})},\ \Eprint
  {http://arxiv.org/abs/gr-qc/0009008} {arXiv:gr-qc/0009008} \BibitemShut
  {NoStop}%
\bibitem [{\citenamefont {Linder}(2003)}]{Linder:2002et}%
  \BibitemOpen
  \bibfield  {author} {\bibinfo {author} {\bibfnamefont {E.~V.}\ \bibnamefont
  {Linder}},\ }\href {\doibase 10.1103/PhysRevLett.90.091301} {\bibfield
  {journal} {\bibinfo  {journal} {Phys. Rev. Lett.}\ }\textbf {\bibinfo
  {volume} {90}},\ \bibinfo {pages} {091301} (\bibinfo {year} {2003})},\
  \Eprint {http://arxiv.org/abs/astro-ph/0208512} {arXiv:astro-ph/0208512}
  \BibitemShut {NoStop}%
\bibitem [{\citenamefont {Bellini}\ \emph {et~al.}(2016)\citenamefont
  {Bellini}, \citenamefont {Cuesta}, \citenamefont {Jimenez},\ and\
  \citenamefont {Verde}}]{Bellini:2015xja}%
  \BibitemOpen
  \bibfield  {author} {\bibinfo {author} {\bibfnamefont {E.}~\bibnamefont
  {Bellini}}, \bibinfo {author} {\bibfnamefont {A.~J.}\ \bibnamefont {Cuesta}},
  \bibinfo {author} {\bibfnamefont {R.}~\bibnamefont {Jimenez}}, \ and\
  \bibinfo {author} {\bibfnamefont {L.}~\bibnamefont {Verde}},\ }\href
  {\doibase 10.1088/1475-7516/2016/06/E01} {\bibfield  {journal} {\bibinfo
  {journal} {JCAP}\ }\textbf {\bibinfo {volume} {02}},\ \bibinfo {pages} {053}
  (\bibinfo {year} {2016})},\ \bibinfo {note} {[Erratum: JCAP 06, E01
  (2016)]},\ \Eprint {http://arxiv.org/abs/1509.07816} {arXiv:1509.07816
  [astro-ph.CO]} \BibitemShut {NoStop}%
\bibitem [{\citenamefont {Frusciante}\ \emph {et~al.}(2019)\citenamefont
  {Frusciante}, \citenamefont {Peirone}, \citenamefont {Casas},\ and\
  \citenamefont {Lima}}]{Frusciante:2018jzw}%
  \BibitemOpen
  \bibfield  {author} {\bibinfo {author} {\bibfnamefont {N.}~\bibnamefont
  {Frusciante}}, \bibinfo {author} {\bibfnamefont {S.}~\bibnamefont {Peirone}},
  \bibinfo {author} {\bibfnamefont {S.}~\bibnamefont {Casas}}, \ and\ \bibinfo
  {author} {\bibfnamefont {N.~A.}\ \bibnamefont {Lima}},\ }\href {\doibase
  10.1103/PhysRevD.99.063538} {\bibfield  {journal} {\bibinfo  {journal} {Phys.
  Rev. D}\ }\textbf {\bibinfo {volume} {99}},\ \bibinfo {pages} {063538}
  (\bibinfo {year} {2019})},\ \Eprint {http://arxiv.org/abs/1810.10521}
  {arXiv:1810.10521 [astro-ph.CO]} \BibitemShut {NoStop}%
\bibitem [{\citenamefont {Pace}\ \emph {et~al.}(2010)\citenamefont {Pace},
  \citenamefont {Waizmann},\ and\ \citenamefont {Bartelmann}}]{Pace:2010sn}%
  \BibitemOpen
  \bibfield  {author} {\bibinfo {author} {\bibfnamefont {F.}~\bibnamefont
  {Pace}}, \bibinfo {author} {\bibfnamefont {J.~C.}\ \bibnamefont {Waizmann}},
  \ and\ \bibinfo {author} {\bibfnamefont {M.}~\bibnamefont {Bartelmann}},\
  }\href {\doibase 10.1111/j.1365-2966.2010.16841.x} {\bibfield  {journal}
  {\bibinfo  {journal} {Mon. Not. Roy. Astron. Soc.}\ }\textbf {\bibinfo
  {volume} {406}},\ \bibinfo {pages} {1865} (\bibinfo {year} {2010})},\ \Eprint
  {http://arxiv.org/abs/1005.0233} {arXiv:1005.0233 [astro-ph.CO]} \BibitemShut
  {NoStop}%
\bibitem [{\citenamefont {Pace}\ \emph {et~al.}(2012)\citenamefont {Pace},
  \citenamefont {Fedeli}, \citenamefont {Moscardini},\ and\ \citenamefont
  {Bartelmann}}]{Pace:2011kb}%
  \BibitemOpen
  \bibfield  {author} {\bibinfo {author} {\bibfnamefont {F.}~\bibnamefont
  {Pace}}, \bibinfo {author} {\bibfnamefont {C.}~\bibnamefont {Fedeli}},
  \bibinfo {author} {\bibfnamefont {L.}~\bibnamefont {Moscardini}}, \ and\
  \bibinfo {author} {\bibfnamefont {M.}~\bibnamefont {Bartelmann}},\ }\href
  {\doibase 10.1111/j.1365-2966.2012.20692.x} {\bibfield  {journal} {\bibinfo
  {journal} {Mon. Not. Roy. Astron. Soc.}\ }\textbf {\bibinfo {volume} {422}},\
  \bibinfo {pages} {1186} (\bibinfo {year} {2012})},\ \Eprint
  {http://arxiv.org/abs/1111.1556} {arXiv:1111.1556 [astro-ph.CO]} \BibitemShut
  {NoStop}%
\bibitem [{\citenamefont {Pace}\ \emph {et~al.}(2014)\citenamefont {Pace},
  \citenamefont {Moscardini}, \citenamefont {Crittenden}, \citenamefont
  {Bartelmann},\ and\ \citenamefont {Pettorino}}]{Pace:2013pea}%
  \BibitemOpen
  \bibfield  {author} {\bibinfo {author} {\bibfnamefont {F.}~\bibnamefont
  {Pace}}, \bibinfo {author} {\bibfnamefont {L.}~\bibnamefont {Moscardini}},
  \bibinfo {author} {\bibfnamefont {R.}~\bibnamefont {Crittenden}}, \bibinfo
  {author} {\bibfnamefont {M.}~\bibnamefont {Bartelmann}}, \ and\ \bibinfo
  {author} {\bibfnamefont {V.}~\bibnamefont {Pettorino}},\ }\href {\doibase
  10.1093/mnras/stt1907} {\bibfield  {journal} {\bibinfo  {journal} {Mon. Not.
  Roy. Astron. Soc.}\ }\textbf {\bibinfo {volume} {437}},\ \bibinfo {pages}
  {547} (\bibinfo {year} {2014})},\ \Eprint {http://arxiv.org/abs/1307.7026}
  {arXiv:1307.7026 [astro-ph.CO]} \BibitemShut {NoStop}%
\bibitem [{\citenamefont {Nazari-Pooya}\ \emph {et~al.}(2016)\citenamefont
  {Nazari-Pooya}, \citenamefont {Malekjani}, \citenamefont {Pace},\ and\
  \citenamefont {Jassur}}]{Nazari-Pooya:2016bra}%
  \BibitemOpen
  \bibfield  {author} {\bibinfo {author} {\bibfnamefont {N.}~\bibnamefont
  {Nazari-Pooya}}, \bibinfo {author} {\bibfnamefont {M.}~\bibnamefont
  {Malekjani}}, \bibinfo {author} {\bibfnamefont {F.}~\bibnamefont {Pace}}, \
  and\ \bibinfo {author} {\bibfnamefont {D.~M.-Z.}\ \bibnamefont {Jassur}},\
  }\href {\doibase 10.1093/mnras/stw582} {\bibfield  {journal} {\bibinfo
  {journal} {Mon. Not. Roy. Astron. Soc.}\ }\textbf {\bibinfo {volume} {458}},\
  \bibinfo {pages} {3795} (\bibinfo {year} {2016})},\ \Eprint
  {http://arxiv.org/abs/1601.04593} {arXiv:1601.04593 [gr-qc]} \BibitemShut
  {NoStop}%
\bibitem [{\citenamefont {White}\ and\ \citenamefont
  {Frenk}(1991)}]{White:1991mr}%
  \BibitemOpen
  \bibfield  {author} {\bibinfo {author} {\bibfnamefont {S.~D.~M.}\
  \bibnamefont {White}}\ and\ \bibinfo {author} {\bibfnamefont {C.~S.}\
  \bibnamefont {Frenk}},\ }\href {\doibase 10.1086/170483} {\bibfield
  {journal} {\bibinfo  {journal} {Astrophys. J.}\ }\textbf {\bibinfo {volume}
  {379}},\ \bibinfo {pages} {52} (\bibinfo {year} {1991})}\BibitemShut
  {NoStop}%
\bibitem [{\citenamefont {Somerville}\ and\ \citenamefont
  {Dav\'e}(2015)}]{Somerville:2014ika}%
  \BibitemOpen
  \bibfield  {author} {\bibinfo {author} {\bibfnamefont {R.~S.}\ \bibnamefont
  {Somerville}}\ and\ \bibinfo {author} {\bibfnamefont {R.}~\bibnamefont
  {Dav\'e}},\ }\href {\doibase 10.1146/annurev-astro-082812-140951} {\bibfield
  {journal} {\bibinfo  {journal} {Ann. Rev. Astron. Astrophys.}\ }\textbf
  {\bibinfo {volume} {53}},\ \bibinfo {pages} {51} (\bibinfo {year} {2015})},\
  \Eprint {http://arxiv.org/abs/1412.2712} {arXiv:1412.2712 [astro-ph.GA]}
  \BibitemShut {NoStop}%
\bibitem [{\citenamefont {Majumdar}\ and\ \citenamefont
  {Mohr}(2003)}]{Majumdar:2002hd}%
  \BibitemOpen
  \bibfield  {author} {\bibinfo {author} {\bibfnamefont {S.}~\bibnamefont
  {Majumdar}}\ and\ \bibinfo {author} {\bibfnamefont {J.~J.}\ \bibnamefont
  {Mohr}},\ }\href {\doibase 10.1086/346179} {\bibfield  {journal} {\bibinfo
  {journal} {Astrophys. J.}\ }\textbf {\bibinfo {volume} {585}},\ \bibinfo
  {pages} {603} (\bibinfo {year} {2003})},\ \Eprint
  {http://arxiv.org/abs/astro-ph/0208002} {arXiv:astro-ph/0208002} \BibitemShut
  {NoStop}%
\bibitem [{\citenamefont {Allen}\ \emph {et~al.}(2011)\citenamefont {Allen},
  \citenamefont {Evrard},\ and\ \citenamefont {Mantz}}]{Allen:2011zs}%
  \BibitemOpen
  \bibfield  {author} {\bibinfo {author} {\bibfnamefont {S.~W.}\ \bibnamefont
  {Allen}}, \bibinfo {author} {\bibfnamefont {A.~E.}\ \bibnamefont {Evrard}}, \
  and\ \bibinfo {author} {\bibfnamefont {A.~B.}\ \bibnamefont {Mantz}},\ }\href
  {\doibase 10.1146/annurev-astro-081710-102514} {\bibfield  {journal}
  {\bibinfo  {journal} {Ann. Rev. Astron. Astrophys.}\ }\textbf {\bibinfo
  {volume} {49}},\ \bibinfo {pages} {409} (\bibinfo {year} {2011})},\ \Eprint
  {http://arxiv.org/abs/1103.4829} {arXiv:1103.4829 [astro-ph.CO]} \BibitemShut
  {NoStop}%
\bibitem [{\citenamefont {Ade}\ \emph {et~al.}(2014)\citenamefont {Ade} \emph
  {et~al.}}]{Planck:2013lkt}%
  \BibitemOpen
  \bibfield  {author} {\bibinfo {author} {\bibfnamefont {P.~A.~R.}\
  \bibnamefont {Ade}} \emph {et~al.} (\bibinfo {collaboration} {Planck}),\
  }\href {\doibase 10.1051/0004-6361/201321521} {\bibfield  {journal} {\bibinfo
   {journal} {Astron. Astrophys.}\ }\textbf {\bibinfo {volume} {571}},\
  \bibinfo {pages} {A20} (\bibinfo {year} {2014})},\ \Eprint
  {http://arxiv.org/abs/1303.5080} {arXiv:1303.5080 [astro-ph.CO]} \BibitemShut
  {NoStop}%
\bibitem [{\citenamefont {To}\ \emph {et~al.}(2021)\citenamefont {To} \emph
  {et~al.}}]{DES:2020mlx}%
  \BibitemOpen
  \bibfield  {author} {\bibinfo {author} {\bibfnamefont {C.}~\bibnamefont {To}}
  \emph {et~al.} (\bibinfo {collaboration} {DES}),\ }\href {\doibase
  10.1103/PhysRevLett.126.141301} {\bibfield  {journal} {\bibinfo  {journal}
  {Phys. Rev. Lett.}\ }\textbf {\bibinfo {volume} {126}},\ \bibinfo {pages}
  {141301} (\bibinfo {year} {2021})},\ \Eprint
  {http://arxiv.org/abs/2010.01138} {arXiv:2010.01138 [astro-ph.CO]}
  \BibitemShut {NoStop}%
\bibitem [{\citenamefont {Lacey}\ and\ \citenamefont
  {Cole}(1994)}]{Lacey:1994su}%
  \BibitemOpen
  \bibfield  {author} {\bibinfo {author} {\bibfnamefont {C.~G.}\ \bibnamefont
  {Lacey}}\ and\ \bibinfo {author} {\bibfnamefont {S.}~\bibnamefont {Cole}},\
  }\href {\doibase 10.1093/mnras/271.3.676} {\bibfield  {journal} {\bibinfo
  {journal} {Mon. Not. Roy. Astron. Soc.}\ }\textbf {\bibinfo {volume} {271}},\
  \bibinfo {pages} {676} (\bibinfo {year} {1994})},\ \Eprint
  {http://arxiv.org/abs/astro-ph/9402069} {arXiv:astro-ph/9402069} \BibitemShut
  {NoStop}%
\bibitem [{\citenamefont {Cohn}\ \emph {et~al.}(2001)\citenamefont {Cohn},
  \citenamefont {Bagla},\ and\ \citenamefont {White}}]{Cohn:2000cm}%
  \BibitemOpen
  \bibfield  {author} {\bibinfo {author} {\bibfnamefont {J.~D.}\ \bibnamefont
  {Cohn}}, \bibinfo {author} {\bibfnamefont {J.~S.}\ \bibnamefont {Bagla}}, \
  and\ \bibinfo {author} {\bibfnamefont {M.~J.}\ \bibnamefont {White}},\ }\href
  {\doibase 10.1046/j.1365-8711.2001.04509.x} {\bibfield  {journal} {\bibinfo
  {journal} {Mon. Not. Roy. Astron. Soc.}\ }\textbf {\bibinfo {volume} {325}},\
  \bibinfo {pages} {1053} (\bibinfo {year} {2001})},\ \Eprint
  {http://arxiv.org/abs/astro-ph/0009381} {arXiv:astro-ph/0009381} \BibitemShut
  {NoStop}%
\bibitem [{\citenamefont {Giocoli}\ \emph {et~al.}(2008)\citenamefont
  {Giocoli}, \citenamefont {Pieri},\ and\ \citenamefont
  {Tormen}}]{Giocoli:2007gf}%
  \BibitemOpen
  \bibfield  {author} {\bibinfo {author} {\bibfnamefont {C.}~\bibnamefont
  {Giocoli}}, \bibinfo {author} {\bibfnamefont {L.}~\bibnamefont {Pieri}}, \
  and\ \bibinfo {author} {\bibfnamefont {G.}~\bibnamefont {Tormen}},\ }\href
  {\doibase 10.1111/j.1365-2966.2008.13283.x} {\bibfield  {journal} {\bibinfo
  {journal} {Mon. Not. Roy. Astron. Soc.}\ }\textbf {\bibinfo {volume} {387}},\
  \bibinfo {pages} {689} (\bibinfo {year} {2008})},\ \Eprint
  {http://arxiv.org/abs/0712.1476} {arXiv:0712.1476 [astro-ph]} \BibitemShut
  {NoStop}%
\bibitem [{\citenamefont {Ali-Ha\"\i{}moud}\ \emph {et~al.}(2017)\citenamefont
  {Ali-Ha\"\i{}moud}, \citenamefont {Kovetz},\ and\ \citenamefont
  {Kamionkowski}}]{Ali-Haimoud:2017rtz}%
  \BibitemOpen
  \bibfield  {author} {\bibinfo {author} {\bibfnamefont {Y.}~\bibnamefont
  {Ali-Ha\"\i{}moud}}, \bibinfo {author} {\bibfnamefont {E.~D.}\ \bibnamefont
  {Kovetz}}, \ and\ \bibinfo {author} {\bibfnamefont {M.}~\bibnamefont
  {Kamionkowski}},\ }\href {\doibase 10.1103/PhysRevD.96.123523} {\bibfield
  {journal} {\bibinfo  {journal} {Phys. Rev. D}\ }\textbf {\bibinfo {volume}
  {96}},\ \bibinfo {pages} {123523} (\bibinfo {year} {2017})},\ \Eprint
  {http://arxiv.org/abs/1709.06576} {arXiv:1709.06576 [astro-ph.CO]}
  \BibitemShut {NoStop}%
\bibitem [{\citenamefont {Fakhry}\ \emph {et~al.}(2021)\citenamefont {Fakhry},
  \citenamefont {Firouzjaee},\ and\ \citenamefont {Farhoudi}}]{Fakhry:2020plg}%
  \BibitemOpen
  \bibfield  {author} {\bibinfo {author} {\bibfnamefont {S.}~\bibnamefont
  {Fakhry}}, \bibinfo {author} {\bibfnamefont {J.~T.}\ \bibnamefont
  {Firouzjaee}}, \ and\ \bibinfo {author} {\bibfnamefont {M.}~\bibnamefont
  {Farhoudi}},\ }\href {\doibase 10.1103/PhysRevD.103.123014} {\bibfield
  {journal} {\bibinfo  {journal} {Phys. Rev. D}\ }\textbf {\bibinfo {volume}
  {103}},\ \bibinfo {pages} {123014} (\bibinfo {year} {2021})},\ \Eprint
  {http://arxiv.org/abs/2012.03211} {arXiv:2012.03211 [astro-ph.CO]}
  \BibitemShut {NoStop}%
\bibitem [{\citenamefont {Gu}\ \emph {et~al.}(2023)\citenamefont {Gu},
  \citenamefont {Dor}, \citenamefont {van Waerbeke}, \citenamefont {Asgari},
  \citenamefont {Mead}, \citenamefont {Tr\"oster},\ and\ \citenamefont
  {Yan}}]{Gu:2023jef}%
  \BibitemOpen
  \bibfield  {author} {\bibinfo {author} {\bibfnamefont {S.}~\bibnamefont
  {Gu}}, \bibinfo {author} {\bibfnamefont {M.-A.}\ \bibnamefont {Dor}},
  \bibinfo {author} {\bibfnamefont {L.}~\bibnamefont {van Waerbeke}}, \bibinfo
  {author} {\bibfnamefont {M.}~\bibnamefont {Asgari}}, \bibinfo {author}
  {\bibfnamefont {A.}~\bibnamefont {Mead}}, \bibinfo {author} {\bibfnamefont
  {T.}~\bibnamefont {Tr\"oster}}, \ and\ \bibinfo {author} {\bibfnamefont
  {Z.}~\bibnamefont {Yan}},\ }\href@noop {} {\  (\bibinfo {year} {2023})},\
  \Eprint {http://arxiv.org/abs/2302.00780} {arXiv:2302.00780 [astro-ph.CO]}
  \BibitemShut {NoStop}%
\bibitem [{\citenamefont {{Weinberg}}\ and\ \citenamefont
  {{Kamionkowski}}(2003)}]{WeinbergKamionkowski2003}%
  \BibitemOpen
  \bibfield  {author} {\bibinfo {author} {\bibfnamefont {N.~N.}\ \bibnamefont
  {{Weinberg}}}\ and\ \bibinfo {author} {\bibfnamefont {M.}~\bibnamefont
  {{Kamionkowski}}},\ }\href {\doibase 10.1046/j.1365-8711.2003.06421.x}
  {\bibfield  {journal} {\bibinfo  {journal} {\mnras}\ }\textbf {\bibinfo
  {volume} {341}},\ \bibinfo {pages} {251} (\bibinfo {year} {2003})},\ \Eprint
  {http://arxiv.org/abs/astro-ph/0210134} {arXiv:astro-ph/0210134 [astro-ph]}
  \BibitemShut {NoStop}%
\bibitem [{\citenamefont {Bartelmann}(2010)}]{Bartelmann:2010fz}%
  \BibitemOpen
  \bibfield  {author} {\bibinfo {author} {\bibfnamefont {M.}~\bibnamefont
  {Bartelmann}},\ }\href {\doibase 10.1088/0264-9381/27/23/233001} {\bibfield
  {journal} {\bibinfo  {journal} {Class. Quant. Grav.}\ }\textbf {\bibinfo
  {volume} {27}},\ \bibinfo {pages} {233001} (\bibinfo {year} {2010})},\
  \Eprint {http://arxiv.org/abs/1010.3829} {arXiv:1010.3829 [astro-ph.CO]}
  \BibitemShut {NoStop}%
\bibitem [{\citenamefont {Massey}\ \emph {et~al.}(2010)\citenamefont {Massey},
  \citenamefont {Kitching},\ and\ \citenamefont {Richard}}]{Massey:2010hh}%
  \BibitemOpen
  \bibfield  {author} {\bibinfo {author} {\bibfnamefont {R.}~\bibnamefont
  {Massey}}, \bibinfo {author} {\bibfnamefont {T.}~\bibnamefont {Kitching}}, \
  and\ \bibinfo {author} {\bibfnamefont {J.}~\bibnamefont {Richard}},\ }\href
  {\doibase 10.1088/0034-4885/73/8/086901} {\bibfield  {journal} {\bibinfo
  {journal} {Rept. Prog. Phys.}\ }\textbf {\bibinfo {volume} {73}},\ \bibinfo
  {pages} {086901} (\bibinfo {year} {2010})},\ \Eprint
  {http://arxiv.org/abs/1001.1739} {arXiv:1001.1739 [astro-ph.CO]} \BibitemShut
  {NoStop}%
\bibitem [{\citenamefont {Sheth}(1998)}]{Sheth:1998ew}%
  \BibitemOpen
  \bibfield  {author} {\bibinfo {author} {\bibfnamefont {R.~K.}\ \bibnamefont
  {Sheth}},\ }\href {\doibase 10.1046/j.1365-8711.1998.01976.x} {\bibfield
  {journal} {\bibinfo  {journal} {Mon. Not. Roy. Astron. Soc.}\ }\textbf
  {\bibinfo {volume} {300}},\ \bibinfo {pages} {1057} (\bibinfo {year}
  {1998})},\ \Eprint {http://arxiv.org/abs/astro-ph/9805319}
  {arXiv:astro-ph/9805319} \BibitemShut {NoStop}%
\bibitem [{\citenamefont {Sheth}\ \emph {et~al.}(2001)\citenamefont {Sheth},
  \citenamefont {Mo},\ and\ \citenamefont {Tormen}}]{Sheth:1999su}%
  \BibitemOpen
  \bibfield  {author} {\bibinfo {author} {\bibfnamefont {R.~K.}\ \bibnamefont
  {Sheth}}, \bibinfo {author} {\bibfnamefont {H.~J.}\ \bibnamefont {Mo}}, \
  and\ \bibinfo {author} {\bibfnamefont {G.}~\bibnamefont {Tormen}},\ }\href
  {\doibase 10.1046/j.1365-8711.2001.04006.x} {\bibfield  {journal} {\bibinfo
  {journal} {Mon. Not. Roy. Astron. Soc.}\ }\textbf {\bibinfo {volume} {323}},\
  \bibinfo {pages} {1} (\bibinfo {year} {2001})},\ \Eprint
  {http://arxiv.org/abs/astro-ph/9907024} {arXiv:astro-ph/9907024} \BibitemShut
  {NoStop}%
\bibitem [{\citenamefont {Sheth}\ and\ \citenamefont
  {Tormen}(2002)}]{Sheth:2001dp}%
  \BibitemOpen
  \bibfield  {author} {\bibinfo {author} {\bibfnamefont {R.~K.}\ \bibnamefont
  {Sheth}}\ and\ \bibinfo {author} {\bibfnamefont {G.}~\bibnamefont {Tormen}},\
  }\href {\doibase 10.1046/j.1365-8711.2002.04950.x} {\bibfield  {journal}
  {\bibinfo  {journal} {Mon. Not. Roy. Astron. Soc.}\ }\textbf {\bibinfo
  {volume} {329}},\ \bibinfo {pages} {61} (\bibinfo {year} {2002})},\ \Eprint
  {http://arxiv.org/abs/astro-ph/0105113} {arXiv:astro-ph/0105113} \BibitemShut
  {NoStop}%
\bibitem [{\citenamefont {Murray}\ \emph {et~al.}(2013)\citenamefont {Murray},
  \citenamefont {Power},\ and\ \citenamefont {Robotham}}]{Murray:2013qza}%
  \BibitemOpen
  \bibfield  {author} {\bibinfo {author} {\bibfnamefont {S.}~\bibnamefont
  {Murray}}, \bibinfo {author} {\bibfnamefont {C.}~\bibnamefont {Power}}, \
  and\ \bibinfo {author} {\bibfnamefont {A.~S.~G.}\ \bibnamefont {Robotham}},\
  }\href {\doibase 10.1016/j.ascom.2013.11.001} {\bibfield  {journal} {\bibinfo
   {journal} {Astron. Comput.}\ }\textbf {\bibinfo {volume} {3-4}},\ \bibinfo
  {pages} {23} (\bibinfo {year} {2013})},\ \Eprint
  {http://arxiv.org/abs/1306.6721} {arXiv:1306.6721 [astro-ph.CO]} \BibitemShut
  {NoStop}%
\bibitem [{\citenamefont {Hu}\ \emph {et~al.}(2014)\citenamefont {Hu},
  \citenamefont {Raveri}, \citenamefont {Frusciante},\ and\ \citenamefont
  {Silvestri}}]{Hu:2013twa}%
  \BibitemOpen
  \bibfield  {author} {\bibinfo {author} {\bibfnamefont {B.}~\bibnamefont
  {Hu}}, \bibinfo {author} {\bibfnamefont {M.}~\bibnamefont {Raveri}}, \bibinfo
  {author} {\bibfnamefont {N.}~\bibnamefont {Frusciante}}, \ and\ \bibinfo
  {author} {\bibfnamefont {A.}~\bibnamefont {Silvestri}},\ }\href {\doibase
  10.1103/PhysRevD.89.103530} {\bibfield  {journal} {\bibinfo  {journal} {Phys.
  Rev. D}\ }\textbf {\bibinfo {volume} {89}},\ \bibinfo {pages} {103530}
  (\bibinfo {year} {2014})},\ \Eprint {http://arxiv.org/abs/1312.5742}
  {arXiv:1312.5742 [astro-ph.CO]} \BibitemShut {NoStop}%
\end{thebibliography}%

\end{document}